

\input ioppreprint.sty
\input verbatim

%
%
\def\LaTeX{L\kern-.26em \raise.6ex\hfil\breakox{\fiverm A}%
   \kern-.15em\TeX}%

\def\AmSTeX{%
{$\cal{A}$}\kern-.1667em\lower.5ex\hfil\breakox{%
 $\cal{M}$}\kern-.125em{$\cal{S}$}-\TeX}

\def\gsim{\ \rlap{\raise 2pt \hbox{$>$}}{\lower 2pt \hbox{$\sim$}}\ }
\def\lesssim{\ \rlap{\raise 2pt \hbox{$<$}}{\lower 2pt \hbox{$\sim$}}\ }

\rightline{UCLA/94/TEP/36}
\rightline{November 1994}

\title{Neutrino Masses\footnote{$^*$}{Review for Reports
of Progress in Physics.}}

\author{G Gelmini\dag\ and E Roulet\ddag\ }

\address{\dag\ Department of Physics, UCLA}
\address{\ddag\ CERN TH Div, 1211 Geneva 23, Switzerland}


\beginabstract

Even if neutrino masses are unknown, we know neutrinos are much lighter than
the other fermions we know, and we do not have a good explanation
for it. In the Standard Model of elementary particles neutrinos are
exactly massless, although this is not
insured by any basic principle.
Non-zero neutrino masses arise in many extensions of the Standard
Model. Massive neutrinos and their
associated properties, such as the Dirac or Majorana character of neutrinos,
their mixings, lifetimes and magnetic or electric moments, may have very
important consequences in astrophysics, cosmology and particle physics.
Here we explore these consequences and the constraints they already
impose on neutrino properties, as well as the large body of experimental
and observational efforts currently devoted to elucidate the mystery of
neutrino masses. Several hints for non-zero masses in solar and atmospheric
neutrinos, that will be confirmed or rejected in the near future, make this
field of research particularly exciting at present.

\endabstract

\contents
\entry{1.}{Introduction}{4}
\entry{2.}{Types of Possible Neutrino Masses}{8}
\entry{3.}{Main Elementary Particle Models for Neutrino Masses}{15}
\subentry{3.1.}{The Standard Model}{15}
\subentry{3.2.}{Plain Dirac Masses}{19}
\subentry{3.3.}{Models Without $\nu_R$}{19}
\subentry{3.4.}{See-Saw Models}{21}
\subentry{3.5.}{Majoron Models}{23}
\subentry{3.6.}{Models with Gauged Lepton-Number}{27}
\entry{4.}{Neutrino Mass Searches}{29}
\subentry{4.1.}{Direct Searches of Neutrino Masses}{29}
\subentry{4.2.}{Double Beta Decay}{31}
\entry{5.}{Neutrino in Cosmology}{34}
\entry{6.}{Neutrino Oscillations}{43}
\subentry{6.1.}{Neutrino Mixing}{43}
\subentry{6.2.}{Oscillations in Vacuum}{43}
\subentry{6.3.}{Oscillation Experiments}{46}
\subentry{6.4.}{Present Situation}{47}
\subentry{6.5.}{Future Prospects}{48}
\entry{7.}{Solar Neutrinos and Oscillations in Matter}{50}
\subentry{7.1.}{The Solar Neutrino Problem}{50}
\subentry{7.2.}{Neutrino Oscillations in Matter}{53}
\subentry{7.3.}{Other Solutions}{58}
\subentry{7.4.}{Future Prospects}{60}
\entry{8.}{Atmospheric Neutrinos}{61}
\entry{9.}{Neutrinos from Supernovae and Other Stars}{64}
\entry{10.}{Concluding Remarks}{71}
\entry{}{Acknowledgements}{72}
\entry{}{References}{73}
\entry{}{Figure Captions}{84}
\vfill\eject

\section{Introduction}

Pauli proposed in 1930 the existence of neutrinos as a ``desperate way out"
to explain the continuous spectrum of electrons emitted in
$\beta$-decay.
This idea was considered at the time almost as revolutionary as the
alternative explanation, the violation of the
principle of energy conservation.
In this way the neutrino became the first particle proposed as the
solution to an elementary particle problem, an idea that proved to be
useful many times later and now commonly used (and even over used).
 Pauli postulated
the existence of a new neutral light fermion.
Even if we now know that neutrinos exists and come in three varieties or
``flavours'' (as charged leptons do) and we have no doubts about their
 neutrality, we are still debating their masses.

In fact, fermions come in families or generations, each one a
repetition of the others, except for their mass. The lighter charged
member of each family is heavier than the heaviest of the previous
one. This is an inter-familial mass hierarchy that has no explanation
so far, since in the Standard Model (SM) of elementary particles all
masses are free parameters (see section 3.1). Why the generations are
three is another mystery. In any case, we know neutrinos come only in
the three known varieties, unless the additional ones are inert, i.e.
do not have weak interactions unlike the known neutrinos, or
they are heavier than $M_Z/2 \simeq 45$ GeV.
This results from  the measurement at the CERN
e$^+$e$^-$ collider LEP of the width of the weak boson $Z^o$ into
``invisible'' particles (neutrinos or other exotic particles that do
not interact within the LEP detectors), $ Z^o \to \nu \nu$.

At present we only know upper bounds on neutrino masses,
obtained in direct mass searches (see section 4.1),
$$
m_{\nu_{e}} < O(10\ {\rm eV}) \quad m_{\nu_{\mu}} < 160\ {\rm keV}
\quad m_{\nu_{\tau}}< 31\ {\rm MeV}~~.
\eqno(1.1)
$$
The bound on $m_{\nu_{e}}$  is actually uncertain because unknown
systematic effects are probably responsible for the
negative experimentally measured
$m_{\nu_{e}}^2$ values.
Even if there is a combined
bound of 5 eV with statistical confidence level of 95\%,
the above mentioned systematic effects
bring the bound at about 10 eV. A preliminary
result lowers the bound on $m_{\nu_{\tau}}$ to  29 MeV.
These bounds show that neutrinos are  much lighter than
their corresponding charged fermion,
$m_e = 0.5$ MeV,  $m_\mu = 105.6$ MeV and $m_\tau = 1.77$ GeV,
 and the other members of their respective family.

Thus neutrinos introduce an intra-familial hierarchy problem: why
neutrinos are much lighter than the other members of each family.
In the SM  this last hierarchy is obtained by avoiding producing
neutrino masses through the mechanism that gives origin to all the
other masses, namely the vacuum expectation value of the standard scalar
Higgs field (see section 3.1). Thus neutrinos are exactly massless in the SM.
However this is a rather ad-hoc choice.
No basic physical principle insures the masslessness of neutrinos (as is
the case for the masslessness of the photon, insured by gauge
invariance).

A large body of experimental and observational efforts are devoted right
now to  elucidate the mystery of neutrino masses.
These are direct mass searches, neutrinoless double beta decay
$(\beta\beta 0\nu)$ (or neutrinoless plus a boson $(\beta\beta
0\nu J)$ decay experiments), oscillation experiments in reactors and
accelerators, solar neutrino observations, atmospheric neutrino
observations and supernova neutrino observations.

While direct mass searches  do not rely on any other possible neutrino
property besides its mass, $\beta\beta 0\nu$ requires neutrinos to be
Majorana particles. Majorana proposed in 1937 that neutrinos,
contrary  to their charged family companions, can be their own
antiparticle (see section 2). Particle and antiparticle carry opposite
charges of any conserved lepton number.
So if neutrinos are Majorana fermions, lepton number is not conserved.
This would induce $\beta\beta 0\nu$ where lepton number is violated
by two units. While double beta decay with emission of two neutrinos
has been observed,
$\beta\beta 0\nu$ has not.
This places an upper bound on an effective Majorana electron neutrino
mass (see section 4.2)
$$
\langle m_{\nu_{e}} \rangle \lesssim 1\ {\rm eV}~~.
\eqno(1.2)
$$
If neutrinos are massive, there would be mixings between neutrino flavours
(see section 3.1). One of the most striking manifestations of neutrino mixing
would be neutrino oscillations (see section 6).
These appear because of the difference between weak interaction neutrino
eigenstates and mass eigenstates, if neutrinos are massive.
When a neutrino is produced it is necessarily in an interaction
eigenstate, whose different mass eigenstate components propagate
differently, according to their mass.
Thus, after some time the propagating neutrino becomes a linear combination
of interaction eigenstates, different from the initial one.
Actually, because neutrino oscillations are an interference effect, they are
sensitive to a combination of mixing angles and mass square
differences, $\Delta m^2$. Oscillations may reveal masses  much smaller
than those testable through  direct searches.
Oscillations could explain the so-called solar neutrino problem
and/or the atmospheric neutrino problem,
and are actively being looked for at dozens of experiments at present.

The solar neutrino problem consists of the deficit of observed neutrinos
emitted from the sun with respect to the theoretically expected amount.
It has been with us for more than 30 yr, but for a long time it was based
on only one experiment, led by R. Davis.
Several other experiments have now also observed solar neutrinos.
One of them, Kamiokande, actually became the first neutrino telescope when
it succeeded in identifying the sun in the sky through its neutrino-image.
The results of the four experiments that have observed solar neutrinos so far,
even if still inconclusive, support the existence of a
solar neutrino deficit, and suggest that the solution lies in new neutrino
physics (as opposed to a modification of our standard solar model).
This is a very exciting field of research and several new experiments that
are under way will allow us to understand fully the solar neutrino
problem within the next few years.
One of the favourite explanations of this problem involves matter
induced oscillations, with very small neutrino mass differences,
 $\Delta m^2 \simeq 10^{-6}$ eV$^2$ (see section 7).

Atmospheric neutrinos produced by cosmic rays hitting the earth
atmosphere show a deficit of $\nu_\mu$ relative to $\nu_e$ with respect
to the expected ratio.
This deficit could be explained if part of the $\nu_\mu$ oscillated
into $\nu_\tau$ (oscillations into $\nu_e$ or a sterile neutrino
are not allowed, see section 8)
within the atmosphere, before reaching the surface
of the earth. This would require $\Delta m^2 \simeq 10^{-2}$ eV$^2$
(see section 8).
New long-baseline oscillations experiments are being proposed to test
experimentally this hypothesis (see section 6). The baselines of these
experiments are actually amazingly long. For example, neutrinos produced
at CERN, in Switzerland, could be detected at the Gran Sasso Laboratory, in
Italy.

Neutrinos are not only important in particle physics but also in
astrophysics (section 9) and cosmology (section 5).
Neutrino emissions are an important, and sometimes dominant, energy
loss mechanism in stars.
So much so that neutrino properties can be constrained on the basis of
the unacceptable changes they would introduce in the evolution of
stars (section 9).
Nowhere is the neutrino energy loss more striking than in a supernova
explosion in which 99\% of the energy released goes into neutrinos.
This is the only case in which the matter densities achieved are so high that
neutrinos are temporarily trapped.
The observation of 19 of the neutrinos emitted by the supernova SN1987A
inaugurated neutrino astronomy outside the solar system, and brought
about an amazing amount of information on neutrinos. For example,
a bound on the $\nu_e$ mass was obtained, $m_{\nu_{e}} < 23$ eV.
We would need the observation of a supernova in our galaxy (as opposed
to SN1987A that happened in one of the Milky Way satellite galaxies, the
Large Magellanic Cloud) to be able to say something about (even measure)
the masses of the other neutrinos (if they are larger than about 25 eV
with the experiments under construction at present).

Turning to cosmology (section 5), the theory of
nucleosynthesis as well as  the dark matter problem and its related
issue of structure formation in the universe, would be greatly affected
by new neutrino properties.
The primordial nucleosynthesis of light elements (D + $^3$He, $^4$He and
${}^7$Li) is one of the fundamental pieces of evidence on which the Big
Bang model relies.
Some neutrino properties, such as a finite mass, lifetime or
 magnetic moment may considerably
change the outcome of nucleosynthesis and, in fact, this
is used at present to constrain neutrino properties.
Most of the mass content of the universe, 90 to 99\%, is in a form of
matter that does not emit or absorb light in any observable way.
This is the dark matter (DM).
Neutrinos of mass in the range of a few eV to a few tens of eV could be an
important component of the DM.
Structure formation arguments (namely the formation of galaxies and
clusters of galaxies) seem to require these light neutrinos not to
constitute the bulk of the DM.
These neutrinos would be hot DM (HDM, i.e. relativistic when galaxies
should start forming, namely at a temperature of approximately 1 keV) while
structure formation prefers the bulk of the DM to be cold (CDM).
However, CDM does not seem to account by itself for all the observations,
and  an admixture of something else seems to be required.
This could be some neutrinos as HDM, with mass of a few eV, or heavier
unstable neutrinos with a tuned combination of masses and lifetimes.

In this brief introduction we see that many important effects in physics,
astrophysics, and
cosmology depend on neutrino properties, all associated with their mass.
We have not yet mentioned one of the major motivations that particle
physicists have to find neutrino masses.
This motivation has to do with the triumph and the tragedy of the standard
model of elementary particles, both joined in the fact that it works extremely
well. So well that at this point any sign of failure and a consequent
indication to go beyond this model will be more welcome than a new
confirmation. This is what any non-zero neutrino mass would be.
A potentially rich window towards physics beyond the standard model.
Moreover, the expectation of this opening towards new physics is intense at
present because the solar and atmospheric neutrino problems
seem to be giving hints of non-zero masses, that will be confirmed or rejected
in the near future.

This review is organized in the following manner.
We start by explaining the different neutrino masses corresponding to
different type of neutrinos, Dirac or Majorana, in section 2.
In section 3 we discuss the main elementary particle models for neutrino
masses, starting with an explanation of the origin of the masslessness
of neutrinos in the SM.
We present then the main types of proposed extensions of the SM that
incorporate neutrino masses (plain Dirac masses, left-handed neutrino
Majorana masses, see-saw models) and their distinctive phenomenological
consequences.

In section 4 we describe the status of direct mass searches and Majorana
mass searches in neutrinoless double beta decays.
In section 5 we go over the many cosmological implications of, and
constraints on, neutrino properties, mainly masses and lifetimes.
Sections 6, 7 and 8 review neutrino oscillations, the solar neutrino
problem and the atmospheric neutrino problem, their implications for
neutrino masses and the current and future experiments with which these
problems
will be clarified in the near future.
We explain oscillations in vacuum in section 6 and oscillations in
matter in section 7 (as well as other possible solutions to the solar
neutrino problem).
Section 9 describes stars, mainly SN1987A, as a laboratory for neutrino
physics and summarizes the main bounds they impose on neutrino masses,
lifetimes and magnetic and electric moments.
A few concluding remarks follow.

\section{Types of Possible Neutrino Masses}

Because the electroweak interactions of the Standard Model (SM) violate
parity $P$ (and charge-conjugation $C$) maximally, they distinguish fermions
of different chirality. These are eigenstates of the Dirac matrix
$\gamma_5$, with eigenvalue $+1$ for right-handed chirality and $-1$
for left-handed chirality. Chirality eigenstates are called Weyl spinors.
 Starting from a Dirac spinor $\psi$, that has
four independent complex components, we obtain two orthogonal  two-component
Weyl spinors of definite left and right-handed chirality,
$\psi_L$ and $\psi_R$, by means of the projectors $P_L$ and $P_R$,
$$
\psi_L = P_L \psi \equiv ({1 - \gamma_5\over 2} ) \psi~~,~~ \psi_R =
P_R \psi \equiv ({1 + \gamma_5\over 2}) \psi~~. \eqno(2.1)
$$
The property $\gamma^2_5 = 1$ insures that $P_L$ and $P_R$ are, in fact,
projectors onto orthogonal components, i.e. $P^2_L = 1$, $P^2_R = 1$ and
$P_R P_L = P_L P_R = 0$.   The decomposition,
$$ \psi = \psi_L + \psi_R \eqno (2.2)$$
 is invariant under the homogeneous Lorentz group
 (i.e. the continuous transformations of the Lorentz group,
rotations and boosts), because $\psi_L$ and $\psi_R$ transform
as separate irreducible representations. However,
a discrete parity ($P$) transformation transforms left and right-handed
Weyl spinors into each other.  This is why a $P$-invariant theory must treat
both chiralities identically (equal couplings) and conversely, when only one
chirality is present, the violation of parity is maximal. The experimentally
determined $V-A$ (vector minus axial vector) coupling in charged current
weak interactions means that only left-handed fermions interact with the
gauge bosons of the group $SU_L(2)$ and $P$-violation is maximal.
For example, the leptonic charged current term in the Lagrangian of the
SM is
$$ {\cal{L}}^{CC} = {g\over \sqrt{2}} \sum_{\ell} \bar{\ell}
\gamma^{\mu} {(1 - \gamma_5)\over 2} \nu_{\ell} W^-_{\mu} + h.c.
=  {g\over \sqrt{2}}J^{-{\mu}}_{\ell} W^-_{\mu}+ h.c. ~~.
\eqno(2.3)
$$
Here $g$ is the weak coupling constant, $\ell = e, \mu, \tau$ are the
charged leptons of the
known generations,
${\nu}_{\ell}$ are the associated neutrinos
 and $W^-_{\mu}$ is the charged heavy gauge boson.
Notice that
$$
 {1\over 2} {\overline{\ell}} \gamma^{\mu}(1 - \gamma_5)\nu_{\ell}
 = {\overline{\ell}} \gamma^{\mu} P_L \nu_{\ell}
 = {\overline{{\ell_L}}} \gamma^{\mu} \nu_{\ell {L}} ,
 \eqno(2.4)
$$
because
$\gamma_5$ anticommutes with all the other Dirac matrices $\gamma_{\mu}
= \gamma_0, \gamma_1, \gamma_2, \gamma_3$, i.e. $\gamma_5 \gamma_{\mu} =
- \gamma_{\mu} \gamma_5$, and, thus
$ {\overline{\psi_R}} = {\overline{\psi}} P_L$, ${\overline{\psi_L}} =
{\overline{\psi}} P_R$.

A Weyl spinor can describe only free massless particles, for which chirality
coincides with helicity (the spin projection in the direction of
motion), that is a good quantum number. This is
so because mass terms mix chiralities. Thus, if a fermion is massive,
the helicity eigenstates, the eigenstates of a freely propagating fermion,
do not coincide with the chirality eigenstates, the eigenstates of weak
 interactions. A massive fermion can be  either a Dirac or a Majorana
 particle, that are described respectively by a Dirac and a Majorana spinor.

A Dirac spinor, such as the one describing
an electron, has four independent complex
components, corresponding to
particle and antiparticle, both of left- and right-handed helicity.
It was actually  Dirac's equation for the electron that led to the concept of
particles and antiparticles and to the definition of charge-conjugation.
 Antiparticle spinors are obtained from particle
spinors through charge-conjugation, under which $\psi\to\eta_c\psi^c$,
with $\eta_c$ a phase and
$$
\psi^c = C {\overline{\psi}}^T = C \gamma_0 \psi^{\ast} .
 \eqno(2.5)
$$
Here $C$ is a unitary matrix that is defined by the property $ C^{-1}
\gamma_{\mu} C = - \gamma^T_{\mu}$. If we now apply $P_L$ or  $P_R$ to the
Dirac spinor and  its charge-conjugate spinor, we see that under the
physical charge conjugation operation, that should preserve
chirality\footnote{$^*$}{It should preserve helicity, that is
the actual physical property of a particle,  but thinking of the limit
in which chirality and helicity coincide,  it is obvious that
chirality must  also be preserved.} (and ignoring the phase $\eta_c$
hereafter),    $\psi_L \to (\psi^c)_L$ and
$\psi_R \to (\psi^c)_R$ (see, e.g., Langacker 1981). Hence, under this
operation,  a Dirac particle in a given state of momentum and helicity
is changed into its antiparticle in the same state of momentum and
helicity.

Notice that one does not obtain the charge-conjugate of the
Weyl spinor components of a Dirac spinor by applying the conjugation
operation (2.5) directly to them.
In fact, the operations of conjugation (2.5) and projection
onto chirality components do not commute,
$(\psi_L)^c \equiv C {\overline{\psi_L}}^T$ is  a $R$-handed
spinor, it is the right component of $\psi^c$ in (2.5),
$(\psi_L)^c =(\psi^c)_R$. Hence, in this operation the chirality
is reversed, while, as discussed above, under charge conjugation
$\psi_L\to(\psi^c)_L=(\psi_R)^c$ (thus, an
interaction containing only $L$-handed fields violates
charge-conjugation maximally).
 Many authors then understandably
avoid  calling $(\psi_L)^c$ the
charge-conjugate of $\psi_L$.
They call it simply ``conjugate" (see e.g.
Mohapatra and Pal 1991), or even loosely ``CP-partner" (see
e.g. Langacker 1981 and 1992), because of the additional change of
chirality, even if it is not the CP-conjugate\footnote{$^{**}$}{Actually $CP$
conjugation is a different
operation, involving an additional multiplication by $\gamma_0$ and a
change of $\vec{x}$ into $-\vec{x}$, $\psi^{CP}(t,\vec{x})= \eta_{CP}
\gamma_0 \psi^c(t,-\vec{x})$, where $\eta_{CP}$ is  a phase.}.
We will then call $c$-conjugation the operation defined in (2.5) and we
call $\psi^c$ the $c$-conjugate of $\psi$ for any kind of field
$\psi$, Dirac or Weyl (or later Majorana).
The operation $c$-conjugation coincides, up to a phase, with
charge-conjugation $C$ for a Dirac field (and for a Majorana field)
but not for a Weyl spinor component of a Dirac field.
It is also trivial to check
that using $(\psi_L)^c=(\psi^c)_R$ and $(\psi_R)^c=(\psi^c)_L$
or using the $C$-transformation
$\psi_L \to (\psi^c)_L$ and $\psi_R \to (\psi^c)_R$ yields the
same result when applied to $\psi_L+\psi_R$ (we only exchange which
component of $\psi^c$ we call the conjugate of which component of $\psi$).

 In the language of relativistic quantum field theory, a Weyl field, say
$\psi_L$, can  annihilate a left-handed (L) particle or create a right-handed
(R) antiparticle, while
$(\psi^c)_R = C {\overline{\psi_L}}^T$, can annihilate a R-antiparticle or
create a
L-particle. For $\psi_R$ the roles of right and left are exchanged. Notice
that $\psi_L$ and $(\psi^c)_R$ are not independent fields, they represent the
same two degrees of freedom. The same is valid for $\psi_R$ and $(\psi^c)_L$.
For Dirac fermions $\psi_L$
and $\psi^c_L$ represent different degrees of freedom but for Majorana fermions
they coincide. Since the fields $\psi_L$ and $(\psi^c)_L$ must have
opposite values of all additive quantum numbers, Majorana fermions
cannot carry any conserved charge.

Electrons e and positrons e$^c$ are clearly different particles, since
they have opposite electric charges. However, we
have examples of neutral particles, such as the $\pi^0$, that coincide
with their antiparticles.  Majorana (Majorana 1937) first proposed that a
neutral fermion could have this property. While a Dirac fermion is
 always different from its
antiparticle, a Majorana fermion is
such that $(\psi^M)^c = \psi^M $, or actually,
$$
(\psi^M)^c = e^{i \Theta} \psi^M \eqno(2.6)
$$
since a phase, $e^{i \Theta}$ (called sometimes Majorana
creation phase factor, Kayser 1984), can always be incorporated in the
definition of the fermion $\psi^M$.  Instead of four independent components,
a Majorana spinor has only two, since particle and antiparticle coincide.

Is $\nu^c$ different or identical to $\nu$?

Only neutrinos of left-handed chirality and  antineutrinos of right
-handed chirality (present in the
hermitian conjugated terms of (2.3)), $\nu_L$ and $(\nu^c)_R$, interact
in the SM.
If they exist, the components of opposite chirality,
$\nu_R$ and $(\nu^c)_L$, are ``inert"  since they do not participate in
weak interactions. If neutrinos are massless
they are described by Weyl spinors,
chirality and helicity coincide and we can never produce neutrinos
and antineutrinos of the same helicity in weak processes. ``Neutrino" and
``antineutrino" are just names at this point, for states that interact
differently due to their different helicity. For massive neutrinos
we can change the helicity by a boost that inverts the momentum.
 We can then produce
neutrinos and antineutrinos of the same helicity and in principle
compare them. If they interact
differently, neutrinos are Dirac particles, if they are
identical, neutrinos  are
Majorana particles (for a more detailed explanation,
 see e.g. Kayser 1985). Although   Majorana and
Dirac neutrinos have different properties, the
differences vanish with the neutrino mass.  The only feasible
experiment that may actually prove the Majorana nature of neutrinos
is the neutrinoless double beta decay (see section 4.2).

 A Dirac mass term, of the form
$$-{\cal{L}}_{\rm mass}= m^D \overline{\psi} \psi =
 m^D {\overline{(\psi_L + \psi_R)}} (\psi_L + \psi_R) =
  m^D ({\overline{\psi}_L} \psi_R + {\overline{\psi}_R} \psi_L)
\eqno(2.7)$$
mixes two different Weyl spinors, of opposite chirality.  This is the
type of mass generated in the SM. Thus, the exclusion in the SM of the
$\nu_R$ (and $(\nu^c)_L) $ components insures ad-hoc that
neutrinos are massless.  A Majorana mass term can be written with only
one Weyl spinor, say $\psi_L$, and its $c$-conjugate
$ (\psi_L)^c = C {\overline{(\psi_L)}}~^T = (\psi^c)_R$,
$$
-{\cal{L}}_{\rm mass}=
{1\over2} m^M \biggl[{\overline{(\psi_L)^c}} \psi_L + {\overline{\psi}}_L
(\psi_L)^c \biggr]
={{1\over 2} m^M {\overline{(\psi_L + (\psi_L)^c)}} (\psi_L +
(\psi_L)^c )}~.
\eqno(2.8)
$$
The mass eigenstate field  $\psi^M$,
$$
 \psi^M \equiv \psi_L + (\psi_L)^c = \psi_L + (\psi^c)_R ~.
 \eqno(2.9)
$$
has the canonical mass term
${1 \over 2} m^M {\overline{\psi^M}} \psi^M$ and is a Majorana field
with creation phase $e^{i\theta}=+1$,
since $(\psi^M)^c = \psi^M$ (notice that $((\psi_L)^c)^c = \psi_L$).
The field
$$
 {\widetilde{\psi}}^M = \psi_L - (\psi_L)^c
 \eqno(2.10)
 $$
is also a Majorana field, as defined in  (2.6), with the phase
$e^{i \theta} = -1$, i.e. $({\widetilde{\psi}}^M)^c = -
{\widetilde{\psi}}^M$. Clearly ${\widetilde{\psi}}^M$ and  $\psi^M$
describe the same degrees of freedom.
It is straightforward to check that if we were
to use ${\widetilde{\psi}^M}$ instead of ${\psi}^M$ in (2.8),
the sign of the mass term  (2.8) would be reversed.

Notice that a Majorana mass term carries two units of fermion number
since ${\overline{(\psi_L)^c}}$ has the same fermion number as $\psi_L$.
A Dirac mass term has instead zero fermion number, since
${\overline{\psi}}_R$ and $\psi_L$ have opposite fermion numbers.  Thus,
Majorana mass terms are forbidden if the fermion number, the lepton
number in the case of neutrinos, is conserved.

When both Majorana and Dirac masses are present, in general the massive
fermions are Majorana particles, and fermion number is not conserved.
The most general mass matrix using the Weyl spinors
$\nu_L$ and $\nu_R$ is:
$$ \eqalign{- {\cal{L}}_{\rm mass} = &  m_D ~({\overline{\nu_L}} \nu_R +
{\overline{\nu_R}} \nu_L ) + \cr
& + {1 \over 2} m_L^M \bigl({\overline{\nu_L}}(\nu_L)^c  +
({\overline{\nu_L)^c}} \nu_L\bigr)
+{1 \over 2} m^M_R \bigl({\overline{(\nu_R)^c}} \nu_R +
{\overline{\nu_R}} (\nu_R)^c\bigr)~~.\cr} \eqno(2.11)
$$
In matricial form it becomes (using ${\overline{\nu_L}} \nu_R =
{\overline{(\nu_R)^c}} (\nu_L)^c$)
$$
 - {\cal{L}}_{\rm mass} = {1\over 2} \bigl({\overline{\nu_L}}~~
{\overline{(\nu_R)^c}} \bigr) \pmatrix{m^M_L & {m_D} \cr
{m_D^T} & m^M_R \cr}~~\pmatrix{(\nu_L)^c\cr  \nu_R \cr}
+ h.c. ~~.
\eqno (2.12)
$$
This expression is also valid for an arbitrary number of flavours,
just taking $\nu_L^T=(\nu_{eL},\ \nu_{\mu L},\dots)$ and similarly for
$\nu_R$.
 Due to the anticommutation properties of the fermion fields, the
mass matrix in (2.12) turns out to be symmetric (see e.g.
 Bilenky and Petcov 1987). We recall that a symmetric matrix $M$ can
generally be brought to a diagonal matrix $M_d$ with positive entries, by
means of a unitary transformation $U$, i.e. $U^TMU=M_d$.

With one neutrino species, $m^M_L$, $m^M_R$ and $m_D$
are numbers (thus, $m_D^T$=$m_D$). Let us consider this case first.
When $m^M_L = 0$ and $m^M_R = 0$, the mass eigenstate is a Dirac
fermion $(\nu_L + \nu_R)$ with mass $m^D$.

 In this case the diagonalization of the matrix of
(2.12) yields two chiral eigenstates  $\psi_{L\pm}\equiv(\nu_L\pm
(\nu_R)^c)/\sqrt{2}$, with
equal and opposite mass eigenvalues, $m_D$ and $-m_D$.
In terms of the associated Majorana fields $\psi_{\pm}\equiv
\psi_{L\pm}+(\psi_{L\pm})^c$, that are self-conjugate by construction,
the mass term reads $-2{\cal L}_{mass}=m_D\overline\psi_+\psi^c_+ -
m_D\overline\psi_-\psi^c_-$.
The negative sign in the mass can however be absorbed into a
redefinition of the
corresponding Majorana eigenstates, taking them to be $\psi_1=\psi_+$
and $\psi_2=\psi_{L-}-(\psi_{L-})^c$, and this last will
consequently transform under
conjugation as the field ${\widetilde{\psi}}^M$ discussed above.
Notice that the original sign of the mass, call it $\lambda_m$,
becomes now the Majorana creation phase (the phase in (2.6)) of the mass
eigenstate field with positive mass.
We also note that Majorana particles, unlike Dirac ones, have a well
defined intrinsic $CP$ parity,
$\tilde\eta_{CP}$\footnote{$^{*}$}{$\tilde\eta_{CP}
=e^{i\theta}\eta_{CP}=\pm i$, where $\eta_{CP}$ is the
phase of the $CP$-transformation (see footnote in previous page)  and
$e^{i\theta}$ is the Majorana creation phase.}.
This is so because $CP$
transforms a particle field into the antiparticle field and while for
Dirac neutrinos there is the freedom to redefine the phase of the
antiparticle field to absorb the $CP$ phase, this is not possible in
the Majorana case.
It is always possible to chose the neutrino fields,
as we did in our example, so that
$\tilde\eta_{CP}=\lambda_mi$
(see e.g. Kayser  1988 and 1984 and Bilenky and Petcov 1987).

 The linear combination of degenerate eigenstates $\nu=
(\psi_1 + \psi_2)/ \sqrt{2}$ is a Dirac neutrino,
 $\nu=(\nu_L + \nu_R)$, and $\nu^c= (\psi_1 - \psi_2)/ \sqrt{2}$).
Thus, a (four-component) Dirac field consists of two (two-component)
 degenerate Majorana fields with opposite values of $\lambda_m$
 (or, in terms of particle states, a Dirac state can be viewed as the
sum of two degenerate Majorana states with opposite $CP$-parities).
The phase $\lambda_m$ is irrelevant for freely propagating neutrinos,
but it will appear in the charged currents involving the $(\nu_L)^c$
field, affecting for instance the neutrinoless double beta decay
(see section 4.2).

These phases appear generically when dealing with Majorana
neutrinos.  The reason is the following. When $CP$ is
conserved, arbitrary phases in the original fields can be chosen so
that the mass matrix is real. Because it is also symmetric, and hence
hermitian, one is tempted to diagonalize it by means of an orthogonal
(not just unitary) transformation $O$. The price to pay is, then, the
possible appearance of negative eigenvalues for self-conjugate fields
or antiself-conjugate fields, if one wants positive masses (as in the
simple example of two degenerate eigenstates presented above).  An
alternative procedure would be to insist in having self-conjugate
fields with positive masses, at the expense of having a non-real
transformation matrix. This is obtained by just replacing the
$\psi_{nL}=\sum O_{nl}\nu_{lL}$ mass eigenstates with negative masses
(here $n$ labels mass eigenstates and $l$ interaction eigenstates)
by the states $\psi'_{nL}=i\psi_{nL}=\sum(iO_{nl})\nu_{lL}$. The
Majorana states $\psi'_n=\psi'_{nL}+(\psi'_{nL})^c$ will be
self-conjugate  and have positive masses, but their mixing matrix (and
their coupling to the charged gauge bosons) will involve the non-real
matrix $U_{nl}=iO_{nl}$. These different approaches are of course
generalizable to the $CP$ violating case.

Going back to (2.12), when either $m^M_L$ or $m^M_R$ or both are non
zero, the mass eigenvectors are two Majorana spinors with different
mass. When the Majorana masses are small compared to the Dirac masses, the
resulting eigenstate is a ``pseudo-" Dirac neutrino (Wolfenstein 1981),
in which the two Majorana fields are not exactly degenerate but almost so.

In order to incorporate the three known generations, we take each of
the spinors to be a vector in generation space,  $\nu_L =
(\nu_{e_{L}},  \nu_{\mu_{L}}, \nu_{\tau_{L}})$, etc. and $m^D$, $
m^M_L$ and $m^M_L$ to be matrices. If there are as many right-handed
neutrino species as left-handed ones, the sub-matrices in (2.12) are
$3 \times 3$.  However, it may be that $\nu_R$ are entirely absent, so
only $m^M_L$ exists, or that the number of $\nu_R$ is different from
three. In the case of several generations there is a special type of
Dirac neutrino that may appear as a mass eigenstate, the ZKM type
(Zeldovich 1952, Konopinski and Mahmoud 1953). Usually a Dirac
neutrino left-handed component  is ``active" (it interacts weakly)
while its right-handed component is  ``sterile" (or inert). Both
components of a ZKM neutrino are active. The right handed  component
is the antiparticle of the left-handed neutrino of a  different
generation, e.g. $\nu = \nu_{eL} + (\nu_{\mu L})^c$. The ZKM
neutrino appears when a linear combination of flavour lepton numbers
different that the usual total lepton number  ($L = L_e + L_{\mu} +
L_{\tau}$  leading to a usual Dirac neutrino) is conserved, in the
example it is  $ L_e - L_{\mu}$. This type of neutrino requires a
complicated Higgs  structure not necessary for usual Dirac neutrinos.

A Dirac neutrino has, in general, a
magnetic dipole moment, and may also have an electric dipole moment
(that violates $CP$).  So, in principle, a torque exerted by an
external magnetic or electric field, $\vec{B}$ or $\vec{E}$, can
change its chirality (since dipole moment interactions  couple Weyl
spinors of opposite chiralities).  Majorana neutrinos can only have
``transition" dipole moments, (Schechter and Valle 1981 and 1982,
Kayser 1982, Nieves 1982), which in a $\vec{B}$ or $\vec{E}$ field
cause the neutrino not only to change chirality, but also to change
flavour (i.e. a neutrino of one generation is changed into another
different neutrino of a different generation), for example
$\nu_{\mu L} \to (\nu_e^c)_R$.  Transition moments can exist for
Dirac neutrinos too, and they imply also the radiative decay of the
heavier neutrino into the lighter one and a photon.

\section{Main Elementary Particle Models for Neutrino masses. }

\subsection{The Standard Model (SM)}
In the SM all charged fermions, quarks and leptons, get Dirac masses
through the Higgs mechanism that breaks spontaneously  the
 $SU_L(2) \times U_Y(1)$ electroweak gauge symmetry
 into  $U_{e.m.}(1)$ of electromagnetism, giving mass to three
of the four gauge vector bosons that mediate electroweak interaction,
the $W^+, W^-$, and $Z^0$, while only the photon stays
massless. The absence of the $\nu_R$
prevents the appearance of a Dirac mass for the neutrinos. The
accidental global symmetry $U_{B-L}(1)$
corresponding to the conservation of $B-L$, Baryon number minus Lepton number,
prevents the appearance of Majorana masses for neutrinos. Thus neutrinos
are massless in the SM.

   All masses in the SM, are proportional to the vacuum expectation value (VEV)
of a complex $SU_L(2)$ doublet scalar field $\phi$ and its conjugate
$\tilde{\phi} = i \sigma^2 \phi^{\ast}$  (where $\sigma^2$ is the second Pauli
matrix), that is also  a doublet (because the representations of $SU(2)$ are
real),
$$
 \phi =  \pmatrix {\phi^0 \cr
\phi^-\cr} ~~,~~ {\tilde{\phi}} = \pmatrix{\phi^+ \cr -\phi^{0\ast}
\cr}~~. \eqno(3.1)
$$
A potential energy  for $\phi$ is introduced, that has its minimum at
$$
\langle\phi \rangle = \pmatrix{ v/\sqrt{2} \cr 0 \cr}~~. \eqno(3.2)
$$
Since the component $\phi^0$ that has non-zero VEV is neutral, $U_{e.m.} (1)$
is preserved as a good gauge symmetry.  Given the masses of the $W$ and $Z$
particles, one obtains
$$
 v \simeq 250~ {\rm GeV}~.  \eqno (3.3)
$$
The matter fields are 15 Weyl spinors, repeated twice at larger masses yielding
three generations or families (see the table 1).  The quarks $q$ and leptons
$\ell$ of the different generations are distinguished by their ``flavour".
There are three flavours of charged leptons, the electron e, the muon $\mu$,
and the $\tau$, and their corresponding neutrinos $\nu_e, \nu_{\mu},
\nu_{\tau}$. Quarks carry also ``colours", the quantum numbers of strong
interactions.  There are three colours for each quark flavour, since quarks
belong to the fundamental representation of $SU_{\rm C}(3)$.   In the table 1
the parentheses show the representations of $SU_C(3)$ and $SU_L(2)$ (both
indicated by their dimensionality), and the value of the weak
hypercharge $Y$, for each set of fields.

The left handed fermions  $f_L$ ($q_L$ and $\ell_L$) have weak isospin $T_L =
1/2$ ($T_{3}$ is $+1/2$ and $-1/2$ for the upper and lower components).  The
right handed fermions $f_R$, have $T_L = 0$, (thus $T_{3L} = 0$).  The charge
$Q$ of each field is given by $ Q = T_{3L} + Y$. The charges of the particles
in each generation sum to zero, $\Sigma_f Q = 0$. This insures that anomalies
cancel within each family and, thus, the SM symmetry is preserved at the
quantum level.
\bigskip
\tabcaption{Matter fields in the Standard Model. Here, $i = 1,2,3$
is the colour index.}

\ialign{#\hfil&&\hglue 2pc plus2pc minus1pc#\hfil\cr
\br
\hfil {\bf Quarks}& \hfil    {\bf Leptons} \cr
\mr
\hfil $q_L :~ (3,2,1/6)$ & \hfil  $\ell_L :~(1,2, -1/2)$ \cr
$\pmatrix{u \cr d\cr}_{Li}~,~~\pmatrix{c \cr s\cr}_{Li}~,~~
\pmatrix{t \cr b\cr}_{Li}$ & $\pmatrix{\nu_e\cr e\cr}_L~,~~
\pmatrix{\nu_{\mu} \cr \mu\cr}_L~,~~ \pmatrix{\nu_{\tau}\cr
\tau\cr}_L$ \cr
&\cr
\hfil $u_R :~ (3,1,2/3)$ & \cr
\hfil $u_{Ri}~,~~ c_{Ri}~,~~ t_{Ri}$ & \cr
& \cr
\hfil $d_R :~ (3, 1, -1/3)$ & \hfil $e_R :~ (1, 1, -1)$ \cr
\hfil $d_{R_i}~,~~ s_{Ri}~,~~ b_{Ri}$ & \hfil
$e_R~,~~ \mu_R~ ,~~ \tau_R$ \cr
\br}
\bigskip

The $SU_L (2) \times U_Y(1)$ assignments of $f_L$ and $f_R$
make it impossible to have gauge invariant (i.e. gauge singlet) mass terms,
i.e.
$$
 m_f \overline{f_L} f_R~~, \eqno(3.4)
 $$
in the Lagrangian. Notice that these Dirac mass terms have a weak isospin
$T_L=1/2$, thus they need to be coupled to a weak doublet to form a
singlet. In fact, the fermion masses come from gauge invariant Yukawa type
couplings with the doublet scalar $\phi = (1,2, -1/2)$,
$$
-{\cal{L}}_{\rm Yukawa} = \sum_{\alpha,~\beta} \left[
(\lambda_u)_{\alpha \beta}~ \overline{q_{L \alpha}} \phi u_{R \beta} +
(\lambda_d)_{\alpha \beta}~ \overline{q_{L \alpha}} \tilde{\phi} d_{R
\beta + } + (\lambda_{\ell} )_{ \alpha \beta}~ \overline{\ell_{L
\alpha}} \tilde{\phi} e _{R \beta} \right]+h.c.~~,\eqno(3.5)
$$
where $\alpha, \beta = 1, 2, 3$ are generation indices and
$(\lambda_f)_{\alpha \beta}$ are coupling constants.
Replacing $\phi$ by its VEV  (3.2), one obtains the fermion mass matrices
$$
(m_f) _{\alpha \beta} = (\lambda_f)_{\alpha \beta} {v\over \sqrt{2}}~~.
\eqno(3.6)
$$
In the SM, the couplings $(\lambda_f)_{\alpha \beta}$ are arbitrary complex
numbers, thus all fermion masses are free parameters. As we mentioned before,
it is only the absence of the right-handed neutrinos $\nu_{eR},
\nu_{\mu R}, \nu_{\tau R}$, that prevents the appearance of Dirac neutrino
masses through Yukawa couplings with the Higgs field.

Therefore, in the SM neutrinos are massless because there are no
right-handed neutrinos $\nu_R$ and there are no other Higgs bosons
besides the doublet (that could generate Majorana masses).
 We could entertain the idea that even in this case neutrinos could
get masses, Majorana masses, due to higher order corrections.
This is not possible in the SM due to the accidental conservation of
$B-L$.
Actually $B$ and the separate flavour lepton numbers $L_e$,
$L_{\mu}$ and $L_{\tau}$ are accidental symmetries,
but they are all anomalous symmetries. Only the
combination $B-L$ is anomaly-free and hence it is
a good global symmetry of
the SM. This symmetry prevents the
appearance of Majorana mass terms also beyond the tree level,
because they would violate $L$ and, therefore,
$B-L$, by two units.

We can see now that models of non-zero neutrino masses necessarily add
to the SM either
fermions, typically $\nu_R$, or bosons, or both, and when
the masses are of the Majorana type,
they introduce a violation of $B-L$, either explicit or spontaneous.
We can also see here that if neutrinos are massive, the leptonic charged
currents in terms of mass eigenstates
include a mixing matrix $K_\ell$ equivalent to the
Cabibbo-Kobayashi-Maskawa (Cabibbo 1963, Kobayashi and Maskawa 1973) mixing
matrix $K_q$ present in the weak charged currents of quarks. The mass matrices
(3.6) must be diagonalized, thus quarks and charged leptons are expressed in
the basis of mass eigenstates
$$
u_L \rightarrow U_L^{u {\dag}} u_L\quad, \quad
d_L \rightarrow U_L^{d {\dag}} d_L \quad, \quad
e_L \rightarrow U_L^{\ell {\dag}} e_L~~.
\eqno(3.7)
$$
Because neutrinos are massless in the SM, $U_L^\ell$ is absorbed in the
redefinition of the neutrinos so that one obtains the coupling in (2.3) and
(3.8) where $K_{\ell}=1$,
while the matrix $K_q = U_L^{u} U_L^{d{\dag}}$ remains
in the charged current of quarks (we ignore colour indices),
$$
J^{+\mu} = (\bar u~ \bar c~ \bar t)_L ~\gamma ^\mu K_q\pmatrix{d\cr s\cr
b\cr}_L + (\bar \nu_e ~\bar \nu_\mu~ \bar\nu_\tau)_L \gamma ^\mu \pmatrix
{e\cr \mu\cr \tau\cr}_L
\eqno (3.8)
$$
Neutrino mass terms in the Lagrangian are always written in the basis of
``current  eigenstates'' (or interaction or flavour eigenstates),
that are those  that diagonalize the lepton weak charged
current with the charged lepton mass eigenstates
(as in (2.3) and (3.8)), namely $\nu_{e}, \nu_{\mu}, \nu_{\tau}$.
Moreover charged lepton current eigenstates are chosen to
coincide with the charged mass eigenstates so that $U^\ell_L=1$.
Then, when neutrinos are massive,
current neutrino eigenstates differ from mass
eigenstates by a  mixing matrix $K_\ell = U_L^{\nu}$. Thus in (3.8)
$$
 (\nu_L^{\rm current})^{\rm T} =
( \nu_{e} ~ \nu_{\mu }~ \nu_{\tau })_L=
 (\nu^{\rm mass}_L)^{\rm T} K_\ell^{*}
= (\nu_1 ~ \nu_2~\nu_3~ ...)_LK_\ell^{*} ~~.
\eqno(3.9)
$$
where the dots indicate that there could be more than three neutrino
mass eigenstates (either heavier than $\simeq$ 45 GeV or mainly
consisting of inert neutrinos).

\subsection{Plain Dirac Masses}
This possibility corresponds to having only $m_D$ non-zero in the general
mass matrix of (2.12).
Right-handed neutrinos can be easily added to the SM, as singlets of the
gauge group
$SU_C(3) \times SU(2)\times U_Y (1)$,   $\nu_R: (1,1,0)$.  Since they are
inert, they do not contribute to anomalies and the renormalizability of
the model is not affected.  We could then add neutrino Yukawa couplings
to  (3.5) (similar to the up-quark terms),
$$
\sum_{\alpha \beta} (\lambda_{\nu})_{\alpha, \beta} ~\overline{\ell_{L
\alpha}}~\phi~ \nu_{R \beta}~~,
\eqno(3.10)
$$
that would yield Dirac neutrino masses as (3.6),
 $(m_{\nu})_{\alpha \beta} = (\lambda_{\nu})_{\alpha \beta}~
v/\sqrt{2}$. This is the simplest way to get non-zero neutrino masses.
However, this mechanism provides no explanation for the smallness of
neutrino masses with respect to the other fermions of the same
generation.  All we could say is that $(\lambda_{\nu})_{\alpha \beta}$,
being arbitrary as all the other Yukawa couplings, happen to be much
smaller than the others.  Other mechanisms provide some insight into
this question.
\subsection{Models Without $\nu_R$}
Counting only with left-handed Weyl spinors, $\nu_L$, we can only have the
Majorana mass terms of $m^M_L$ in  (2.12). These terms have weak isospin
$T_L=1$. Thus, renormalizable interactions at tree level     give origin to
these masses (through a Higgs Mechanism) only  if a triplet Higgs field $\Delta
=(\Delta^{++}, \Delta^+, \Delta^0)$ is added to the SM,  with Yukawa couplings
$$
g_{\alpha \beta} [{\overline{(\ell_{\alpha L})^c}}~ \vec{\sigma}~
\ell_{\beta L}]\cdot
\vec{\Delta}  + h.c.~~.
\eqno(3.11)
$$
Here $\vec{\sigma}=(\sigma_1, \sigma_2, \sigma_3)$ are the Pauli matrices,
$\ell_{\alpha L} = (\nu_{\alpha L},~ e_{\alpha L})$ are lepton
doublets, thus $(\ell_{\alpha L})^c =
 ((e_{\alpha L})^c, - (\nu_{\alpha L})^c)$, since $(\ell_{\alpha L})^c$ is also
the SU(2) conjugate of $\ell_{\alpha L}$ (as $\tilde{\phi}$ is the
conjugate of ${\phi}$ in (3.1)),
 $\alpha, \beta$ are generation indices, $\vec{\Delta} = (\Delta_1,
\Delta_2, \Delta_3)$ and $\Delta^0 =
 (\Delta_1 + i \Delta_2)/\sqrt{2}$, $\Delta^+ =
\Delta_3$ and $\Delta^{++} = (\Delta_1 - i \Delta_2)/\sqrt{2}$.
These Yukawa couplings contain terms
$$
g_{\alpha \beta}~ \sqrt{2}~ {\overline{(\nu_{\alpha L})^c}}
\nu_{\beta L} \Delta^0 + \cdots
\eqno(3.12)
$$
that yield Majorana masses proportional to the VEV $\langle
 \Delta^0 \rangle = v_T/\sqrt{2}~~,$
$$
(m_L^M)_{\alpha \beta} = g_{\alpha \beta} v_T~.
\eqno(3.13)
$$
Here the smallness of the neutrino mass depends on choosing ad-hoc a
small $v_T$, $v_T\ll v$.

When Majorana masses are present in a model there is a choice between
 explicit or spontaneous $L$ violation. Moreover $L$
(actually we should speak of $B-L$, because $L$ is anomalous)
 may be a gauge or a global symmetry. If
 $L$ is a global symmetry, its spontaneous breaking would generate a
 Nambu-Goldstone boson, that is called Majoron.

The lepton number of $\Delta$ is  $L_{\Delta} = -2$, as defined by (3.11).
Therefore, if $L$ is conserved in the Lagrangian, $\langle
 \Delta^0 \rangle \not= 0$
violates $L$-conservation spontaneously and a Goldstone boson results. This
is the basic mechanism of the triplet Majoron model (Gelmini
and Roncadelli 1981), now experimentally rejected (see below).

Given the matter content of the SM (table 1), with no $\nu_R$,
there are only two other scalars, besides the standard doublet $\phi$
and the triplet $\Delta$, that can have $SU_L(2) \times U_Y(1)$ invariant
Yukawa couplings.  These are, a singly charged singlet scalar  field
$h^+$, first proposed by Zee (Zee 1980), that could couple to lepton doublets
$$
f_{\alpha \beta} {\overline{(\ell_{\alpha L} )^c}} \ell_{\beta L}
h^+ = f_{\alpha \beta} ( {\overline{(e^-_{\alpha L})^c}} \nu_{\beta L} -
{\overline{(\nu_{\alpha L})^c}} e^-_{\beta L} )
 h^+~, \eqno(3.14)
$$
and a doubly charged singlet scalar $h^{++}$, that could couple to lepton
singlets
$$
f^{\prime}_{\alpha \beta}
{\overline{(e_{\alpha L})^c}} ~(e_{\beta R}) h^{++}~.
\eqno(3.15)
$$
Here, the couplings $f_{\alpha \beta}$ and $f^{\prime}_{\alpha \beta}$ must be
antisymmetric in $\alpha$ and $\beta$ due to Fermi statistics ($\alpha, \beta$
denote generations).  Thus  $h^+$ and $h^{++}$  couple only to two leptons of
different families.   Neither $h^+$ nor $h^{++}$ can have a non-zero
VEV (otherwise electromagnetism would be a spontaneously broken symmetry).
Thus
$L$ violation has to be introduced in the Higgs potential, by adding  at least
a second doublet Higgs field, and Majorana masses for neutrinos are obtained
through radiative corrections (Zee 1980).  While the Zee model, based on $h^+$,
does not include the $h^{++}$ field, models with $h^{++}$ do require $h^+$
(Babu l988).  In both models neutrino masses are small because they are
generated through radiative corrections, and a distinctive feature is the
presence of a neutrino much lighter than the other two.

Some $L$-violating physics beyond the SM, at a large mass scale $\Lambda$, may
induce  $\nu_L$  Majorana masses  through effective (non-renormalizable)
couplings. The leading non-renormalizable $L$-violating operator that can be
written with the fields of the SM is of dimension four (Barbieri \etal 1980)
$$
{\cal L_{\rm eff}} = C_{\alpha\beta}
\biggl [{\overline{(\ell_{\alpha L} )^c}} \vec\sigma
\ell_{\beta {\rm L}} \biggr ] {(\phi^T \sigma_{2} \vec\sigma\phi)
\over{\Lambda}} ~~.
\eqno(3.16)
$$
When the doublet $\phi$ gets a VEV (3.2), (3.16) gives
Majorana masses
$$
(m_L^M)_{\alpha\beta} \simeq {C_{\alpha\beta} v \over\Lambda}~.
\eqno(3.17)
$$
For $\Lambda\gg v$ one naturally obtains small neutrino masses.
As we will see below, in see-saw models $\Lambda$ is the $\nu_R$
Majorana mass scale, $\Lambda \simeq m_R^M$.

We may not need new physics to have terms as (3.16).
We know that gravity is not incorporated into the SM and it may not
respect any global symmetry (for example, global charges disappear when
falling into a black hole). Quantum gravity effects may violate $L$.
In this case $\Lambda \simeq M_{\rm Planck} \simeq 10^{19}$ GeV
and the resulting masses are $m_L^M \simeq C 10^{-5}$ eV.
\subsection{See-Saw Models}
The ``see-saw" mechanism (Yanagida 1979, Gell-Mann \etal 1979)  consists of
making one particle light at the expense of making another heavy. It assumes a
hierarchy in the values of the different elements of the mass matrix in (2.12),
namely
$$
m^M_R = M \gg m_D \gg m^M_L = \mu~,\eqno(3.18)
$$
with $\mu$ either zero or negligible.

Consider first the case of just one generation. Then, the mass matrix has one
heavy eigenvector, mainly consisting of the inert $\nu_R$, $N \simeq [\nu_R +
(\nu_R)^c] +(m_D/M) [\nu_L + (\nu_L)^c]$, and one light eigenvector, mainly
consisting of the active Weyl spinor $\nu_L$,  $\nu \simeq [\nu_L -
(\nu_L)^c]
+ (m_D/M) [\nu_R - (\nu_R)^c]$, with masses
$$
m_N \simeq M,~~~~~~ m_{\nu} \simeq {m_D^2\over M} \ll m_D~.
\eqno(3.19)
$$
 Thus in the see-saw mechanism the
 larger $M$ the lighter is the light neutrino, explaining the
 intra-familial hierarchy  $m_{\nu} \ll m_D$, while the  Dirac
neutrino masses are naturally similar to the Dirac masses
of the other fermions of the same generation.

 If $\mu$ is not negligible, the light mass eigenvalue becomes\footnote{$^*$}{
When this eigenvalue is negative before taking the modulus,
the  minus sign is incorporated in the definition of the light
eigenstate, that becomes anti-selfconjugate (see section2), as
is the case of $\nu$ above.}
$$
m_{\nu} \simeq  \left\vert \mu - {m_D^2\over M} \right\vert
\eqno(3.20)
$$
Thus, unless $\mu < m_D^2/M$ one
loses the natural explanation of the smallness of $m_{\nu}$.

In the case of three generations, each fermion in  (2.12) becomes
a vector and $M$, $m_D$ and $\mu$ become matrices.
Heavy $(N)$ and light $(\nu)$
eigenvectors are found diagonalizing the matrices
$$
m_N = M, \quad m_\nu = \mu - m_D M^{-1} m_D^T~.
\eqno(3.21)
$$
Here, for the heavy masses we kept only the leading term.
In the usual see-saw models there are as many right- as left-handed
neutrinos, $M$, $m_D$ and $\mu$ are $3 \times 3$ matrices (with $\mu$
negligible or zero (3.8))  and
 the symmetric $6 \times 6 $ mass matrix has six eigenvectors,
(usually) three light and three heavy, with masses
$$
{m_{D1}^2\over M_1}~~,~~ {m_{D2}^2\over M_2}~~,~~ {m_{D3}^2\over
M_3}~~,~~ M_1, M_2, M_3~~,
\eqno(3.22)
$$
where $M_i$, i=1,2,3, are the eigenvalues of the matrix $M$.

There are many different versions of this mechanism.  For example, the Dirac
masses $m_{Di}$ in the eigenvalues (3.22)
 could be similar to the up-quark masses (usual in Grand Unified
models), or to the charged lepton masses of the same generation (or even
something different).  The neutrino mass hierarchy depends on this choice since
$m_t/m_c \simeq 100$ while $m_{\tau}/m_{\mu} \simeq 17$.  In a ``quadratic
see-saw", the three  heavy masses $M_i$ are similar, i.e. $M_i \simeq M$ for i
= 1,2,3. Consequently, the hierarchy of light neutrino masses is that of
$m^2_{Di}$, i.e. $m_{\nu_{1}} : m_{\nu_{2}} : m_{\nu_{3}} \simeq m^2_{u} :
m^2_c : m^2_t$ (or $m^2_e : m^2_{\mu} : m^2_{\tau})$. In a ``linear see-saw",
the hierarchy of the heavy masses $M_i$ coincides with that of
$m_{Di}$, $M_1 : M_2 : M_3 ~\simeq~ m_u : m_c : m_t$
(for
example,  $M_i$ generated through loops in a SO(10) model, Witten
1990).
As a consequence, also the
ratio of light neutrino masses is linear in $m_D$. Vastly different scales are
possible for $M$,  depending on the model, from 1 TeV or lower (for example,
in left-right symmetric models), to $10^{16}$ GeV or higher (in GUT's).  Only
in the simplest GUT models the leptonic mixings coincide with those of quarks,
$K_{\ell}= K_q$ (up to renormalization effects)
in (3.8) and (3.9), but one usually also obtains incorrect
relations  of mass ratios (namely $m_e/m_{\mu}=m_d/m_s$). Otherwise the mixing
angles are entirely model dependent.

There are many other less usual possibilities for see-saw mechanisms. For
example, in the ``incomplete see-saw" (Johnson \etal 1986, 1987, Glashow
1991), a $3 \times 3 $ matrix M but of rank 2, i.e. with one zero eigenvalue,
yields a Dirac neutrino mass eigenstate of mass  $\simeq m_D$, besides two
heavy and two light neutrinos, of masses $ \simeq M$ and $\simeq m_D^2/M$
respectively, as before.

\subsection{Majoron Models}
When a global continuous symmetry is spontaneously broken, a zero mass
boson, called Nambu-Goldstone boson, appears for every broken generator.
One may wonder if such massless bosons would lead to new gravitational
or Coulomb-like $r^{-1}$ potentials, against which there are very
stringent upper limits.  This is not the case.  Goldstone bosons generate only
spin-dependent $r^{-3}$ non-relativistic long-range potentials for which bounds
are much less restrictive (Gelmini \etal 1983).

Majorons are the Goldstone bosons associated with the spontaneous breaking of a
leptonic global symmetry, at a scale $V$, that usually also generates  Majorana
neutrino masses (thus the name Majoron), $m_{\nu} \simeq g_{eff} V$. The
effective coupling $g_{eff}$ is also roughly the coupling of neutrinos with the
 Majoron, $J$. These couplings provide the most important phenomenological
consequences characteristic of these models. In general, given a certain range
of masses $m_{\nu}$, for small values of $V$ (typically $ V < v \simeq 100$
GeV), the couplings $g_{eff}$ are large, neutrinos may decay ($\nu_h \to \nu_l
J$) and/or annihilate ($\nu \nu \to J J$) or interconvert ($ \nu \nu \to
\nu^{\prime} \nu^{\prime}$) very fast (so neutrinos in the early Universe could
not remain to be the Dark Matter). Besides, the Majoron could be emitted in
neutrinoless double beta decays at a level that could be observable soon and
Majorons may play a role in energy loss mechanisms in stars or in the collapse
of a supernova. If instead,  given a certain range of masses $m_{\nu}$, $V$ is
large  (typically $V > v \simeq 100$ GeV), then  $g_{eff}$ is relatively small
and the only important consequence of Majorons is, usually, that neutrinos may
decay with much longer lifetimes than before, shorter or
longer than the lifetime of the universe (so that, for example,
neutrinos could be a relevant part of the Dark Matter).

Majoron models are almost unique in allowing  neutrino masses  in the range
between $\simeq 100$ eV and a few GeV, otherwise forbidden due to
cosmological arguments, through neutrino decays or annihilations into invisible
modes in the early universe. Another important consequence of Majoron models is
that the standard Higgs particle may be ``invisible" itself through decaying
dominantly into the invisible channel $H \to J J$ (Schrock and Suzuki 1892, see
for example Lopez-Fernandez \etal 1993 or Brahmachari \etal 1993 and
references therein).

The main properties of these models are determined by the weak isospin of the
Majoron. The singlet Majoron model (Chikashige, Mohapatra and Peccei
1980, 1981),
adds to the SM
right-handed neutrinos  and a singlet Higgs field $\sigma$ coupled to
them,
$$
h_{\alpha \beta} {\overline{(\nu_{\alpha R})^c}} \nu_{\beta R} \sigma
+ h.c.~.
\eqno(3.23)
$$
The VEV $\langle \sigma \rangle \not= 0$, generates Majorana masses for the
$\nu_R$,
$M_{\alpha \beta} \simeq h_{\alpha \beta} \langle \sigma \rangle$.
Light neutrino
eigenstates $\nu_{\alpha}$ result from the see-saw mechanism ((2.12) with
$m_L^M=0$), if  $\langle \sigma \rangle$ is large enough, $\langle \sigma
\rangle \gsim v$. Because $L$
is chosen to be an exact symmetry of the Lagrangian  and   (3.23) defines a
non-zero lepton number for $\sigma$, $L_{\sigma} = -2$, $L$ is spontaneously
broken. The Majoron in this model is $J = \sqrt{2} Im( \sigma)$, thus
$$
\sigma = {1 \over \sqrt{2}} [V + \rho + i J]~.
\eqno(3.24)
$$
Thus, the Majoron couple to the light neutrinos  only through the small
admixture of $\nu_R$, of order $(m_D/M)$, that they contain. Consequently, the
decay rate of a heavier light neutrino $\nu_h$ into a lighter one $\nu_l$, is
$$
\Gamma (\nu_h \to \nu_l J)\lesssim
({m_D\over M})^4 m_{\nu_{h}}~~.
\eqno(3.25)
$$
Actually, in the simplest form of the singlet model presented here, the leading
$(m_D/M)^2$ terms in the amplitude can be rotated away (Schechter \etal 1982),
so that $\Gamma  \simeq ({m_D / M})^8 m_{\nu_{h}}$. Thus $\nu$ lifetimes are
large with respect to the lifetime of the universe. This is not true in more
complicated versions of the model (Gelmini and Valle 1984, Jungman and
Luty 1991, Babu
1991, Glashow 1991).

Besides its effects on neutrinos, a singlet Majoron  is
practically invisible. It has very small couplings to charged fermions, only
through neutrino loops, and, because the field $\sigma$ is a gauge singlet, $J$
is not coupled to the $Z^0$ boson, thus it does not contribute to its
invisible decay modes.

Non-singlet Majorons have been rejected by the LEP bound on the effective
number of neutrinos, $N_\nu < 2.983 \pm 0.025$ (Review of Particle
Properties 1994). The reason is that in these
models there are additional contributions to the invisible width of the
$Z^0$-boson that would count as extra light neutrinos. The new decay mode is $Z
\rightarrow \rho J$, where $\rho$ is a light boson associated to the Majoron
(usually $\rho$ is the real part of the same combination of fields of which J
is the imaginary part, as in (3.24)). Because non-singlet Majoron models are
less ``invisible'' than singlet Majorons, the scale of $L$-violation $V$ is
phenomenologically required to be $V \ll v \simeq 100 $ GeV, and in this case
$m_\rho \ll V$, and $\rho$ can be emitted in a $Z^0$ decay.

The models rejected are the triplet Majoron (Gelmini and
Roncadelli 1981, Georgi \etal 1981), the doublet Majoron (Bertolini and
Santamaria 1988) and supersymmetric Majoron models where
it is the left-handed scalar
neutrino $\tilde\nu_L$ (the supersymmetric partner of $\nu_L$) VEV that
violates $L$ spontaneously,
$V = \langle \tilde\nu_L\rangle$ (Aulakh and Mohapatra
1983, Ross and Valle 1985). In these three models the additional $Z^0$ width,
$\Gamma (Z^0 \rightarrow \rho J)$, equals 2, 0.5 and 0.5 respectively of its
partial width into a light neutrino species, $\Gamma(Z^0\rightarrow \nu_\alpha
\nu_\alpha)$, while only a few percent is still allowed by the LEP results.
Therefore viable Majorons are singlet (coupled or not to $\nu_R$) or
mostly singlet (even if mixed with non-singlets).
Examples are the original CMP singlet model (Chikashige, Mohapatra and
Peccei 1980,
1981), the singlet-triplet or ``invisible triplet'' (Choi \etal 1989,
Choi and Santamaria 1991, D'Ambrosio and Gelmini 1987) and supersymmetric
models where the right-handed sneutrino (an electroweak singlet as its
supersymmetric partner $\nu_R$) VEV breaks $L$ spontaneously,
$V = \langle \tilde\nu_R\rangle$ (Masiero and Valle 1990).

The ``invisible'' triplet model preserves many of the characteristics
of the original triplet Majoron. This is a variation of the triplet
model presented in section 3.2, which yields a Majoron if $L$ is only
spontaneously violated by $\langle\Delta^0 \rangle \simeq v_T \not =
0$. The triplet Majoron is mainly $J \simeq Im\Delta^0$, thus it
couples at tree level to SM neutrinos. In the ``invisible'' triplet a
singlet $\sigma$ (not coupled to fermions) is added, whose VEV
$\langle\sigma\rangle \simeq v_S\not = 0$ also breaks $L$. If $v_S >
v_T$ the Majoron is now mostly singlet, $J \simeq Im (\sigma +
(v_T/v_S)\Delta^0$), and $\rho$ is the real part of the same
combination. Thus now $\Gamma (Z \rightarrow \rho J) \sim (v_T/v_S)^4$
is reduced to an acceptable level.

Singlet models tend to have large $V$ (and thus small $g_{\rm eff})$.
However, several singlet and mixed models have been proposed
(Berezhiani \etal 1992, Burgess and Cline 1993, 1994, Carone 1993) in
which
$g_{\rm eff}$ is large (and $V$ small), mainly motivated by the possibility
of finding Majoron emission in neutrinoless double beta decay, i.e.
having $(g_{\rm eff})_ { \nu_e \nu_e J}$ not much smaller than its
experimental upper bound of $ 0.7 \times 10^{-4} $ (Moe 1994, see
section 4.2).

Also models in which a leptonic global symmetry group $G$ is
spontaneously broken into a conserved lepton number $\widetilde{L}$, $G \to
U_{{\widetilde{L}}} (1)$, have been considered, and the ensuing Goldstone
bosons are also called Majorons many times.  In this case the Majorons
carry the conserved ${\widetilde{L}}$ number.  ${\widetilde{L}}$ may be
a non-orthodox lepton number (such as $L_e + L_{\mu} - L_{\tau}$, for
example.  The existence of this (almost) conserved lepton number
produces the appearance of at least one (pseudo) Dirac neutrino
mass eigenstate.
Models of this type were produced for the ``17 keV neutrino"
and to explain a possible signature for neutrinoless double beta decay
with emission of a boson (see later).

Finally, interesting models result when an explicit $L$-breaking is
present in Majoron models, transforming the Majoron into a massive
pseudo-Golstone boson.
The explicit breaking may be due to quantum gravity (Akhmedov,
Berezhiani and Senjanovic' 1992, Akhmedov \etal
1993b, Cline \etal 1993), although one
may protect the global $L$-symmetry
with a gauge symmetry (gauge symmetries are respected by gravity,
Rothstein \etal 1993) or  by other mechanisms (Lusignoli \etal 1990).
In this models, Majorons could even be the dark matter in the universe
(Akhmedov \etal 1993a, Rothstein \etal 1993, Berezinsky and Valle
1993).

A brief comment on familons  is in order. Familons $F$ are the Goldstone bosons
associated with the spontaneous breaking of an inter-familial global symmetry
(Reiss 1982, Wilczek 1982, Gelmini, Nussinov and Yanagida 1983).
 The main difference between Majorons and familons
is that Majorons have usually much larger couplings to neutrinos than to other
fermions, allowing for faster neutrino decays, for example. Since the
inter-familial  symmetry rotates in principle whole families together, or at
least whole lepton doublets together (in only leptonic inter-familial
symmetries)
the same coupling allowing $\nu_\mu \to \nu_e F$ or $\nu_\tau \to \nu_e F$
would allow $\mu \to e F$  or $\tau \to e F$. From experimental bounds in
these last modes one gets large lower bounds on the breaking scale,
of order 10$^{10}$ GeV and 10$^7$~GeV respectively
(D'Ambrosio and Gelmini 1987).

\subsection{Models with Gauged Lepton-Number}
 The simplest models of this type are left-right symmetric models, where right
and left handed fermions are assumed to play identical roles, consequently
there is a $\nu_R$ for every $\nu_L$ (see for example Mohapatra and Pal 1991).
The maximal symmetry of these models is $SU_L(2) \times SU_R(2) \times U(1)
\times P$, where $P$ is a sort of parity that exchanges left and right indices
($P$-parity insures also that the coupling constants of the two $SU(2)$ groups
coincide).
In the simplest versions of these models there are only three Higgs
fields, two triplets $\Delta_R\equiv (1, 3, 2)$ and $\Delta_L \equiv
(3, 1, 2)$ and the usual doublet $\phi \equiv (2,2,0)$. The VEV
$\langle \Delta_R\rangle \not = 0$ gives masses to the right gauge
bosons $Z_R$,  $W_R^\pm$ and Majorana masses to the $\nu_R$ of the same
order $M$, breaking
the symmetry into the SM.
In these versions, $\langle \Delta_L\rangle \simeq \lambda
v^2/\langle\Delta_R\rangle$ results from the
minimization of the potential (Mohapatra and Senjanovic 1981)
and the mixed mass matrix in (2.12) is
$$
\pmatrix{f \langle \Delta_L\rangle & m_D\cr
m_D & f\langle\Delta_R\rangle\cr}~~.
\eqno(3.26)
$$
where $m_D \sim v$ and $\lambda$, $ f$ are combinations
of coupling constants. Consequently,
the light neutrino masses are, (see (3.18)),
$$
m_\nu = f \langle \Delta_L\rangle - {m_D f^{-1} m_D^T\over
\langle\Delta_R\rangle}
\eqno(3.27)
$$
and, unless $\lambda$ is extremely small, the first term dominates.
These models have been used, adding an inter-familial (or
``horizontal'') symmetry, to obtain three almost degenerate mass
eigenstates, whose splittings are due to the see-saw mechanism (in
models that try to account for most of the present hints for non-zero
neutrino masses, namely accommodating solar and atmospheric neutrino
deficits, hot dark matter and giving rise to neutrinoless double beta
decay close to the present bounds).

By complicating the left-right models a bit, $P$ may be broken at a larger
scale than the rest of the group and $\langle\Delta_L\rangle$ becomes
negligible.
Then a usual see-saw mechanism provides small neutrino masses.
Phenomenologically $M > 1$ TeV, due to the bound on additional $Z$ bosons
 and on $W_R$ bosons.
With $M \simeq 1$  TeV one obtains the hierarchy of neutrino masses
$\nu_e:\nu_\mu: \nu_\tau \simeq$ eV: keV: MeV.

The solution to the solar neutrino problem may require much smaller masses,
(see section 7)
$10^{-6}$ eV:$10^{-3}$ eV:1 eV, pointing to much larger $\nu_R$
Majorana masses, $M \simeq 10^{10} -
10^{12}$ GeV, precisely in the range of the intermediate mass scales in
$SO(10)$ Grand Unified models, necessary to obtain the unification of the
SM coupling constants after the precision measurements at LEP
(Mohapatra and Parida 1992, Babu and Mohapatra 1993). These precise
measurements of the coupling constants at the electroweak energy scale
showed that these couplings do not converge simultaneously to a single
value at a large energy scale, the Grand Unification scale $M_U$,
(as necessary in Grand Unified models, Langacker 1981, Mohapatra 1986)
in non-supersymmetric
models with one stage of unification. Thus either supersymmetric models
or non-supersymmetric models with several stages of partial unification
are indicated by these measurements (see for example Langacker and
Polonsky 1992).

The natural Grand Unified symmetry that incorporates left-right models
is $SO(10)$.
$SO(10)$ predicts the existence of the $\nu_R$ by incorporating all
Weyl spinors of each generation into a 16-dimensional multiplet.
In models with intermediate mass scales, $SO(10)$ breaks into the SM in
two stages.
The first, at the scale $M_U \simeq 10^{16}$ GeV, brings the
symmetry to a left-right symmetric group, that breaks at
the intermediate scale $M_I$ into the
SM.
In the intermediate stage there is a discrete symmetry, called $D$-parity,
that has the same role of the $P$-parity mentioned above.
$D$-parity needs to be broken at a scale larger than $M_I$ to obtain a
see-saw mechanism with $M \simeq M_I$, otherwise a solution like (3.27)
is obtained
(Caldwell and Mohapatra 1994, Ioannisian and Valle 1994, Bamert and
Burgess 1994, Joshipura 1994, Lee and Mohapatra 1994).
In $SO(10)$ models, $m_D$ are similar to the masses of up-quarks.

Supersymmetric $SO(10)$ also yields unification of the three coupling
constants of the SM and in this case a see-saw model with
$M_V \simeq 10^{16}$ GeV produces masses
$10^{-11}$ eV:$10^{-8}$ eV:$10^{-3}$ eV.
But also in supersymmetric $SO(10)$ intermediate scales can be ``gravity
induced'' (Cvetic and Langacker 1992).
There are still many other possibilities, using other grand unified
groups like $E_6$, technicolour, etc..
\section{Neutrino Mass Searches}
\subsection{Direct Searches of Neutrino Mass}

These searches are based on kinematical arguments and assume only
finite neutrino masses.
(For a detailed review on direct searches until 1988, see
Robertson and Knapp 1988).
Indirect searches require additional criteria such as Majorana masses in
neutrinoless double beta decay and large enough mixing angles in
oscillation experiments (that measure actually not masses, but mass
differences).
Neutrinos were originally proposed by Pauli in 1932, to account for
the continuous spectrum of the emitted electrons in weak nuclear
$\beta$-decays, $(A, Z) \rightarrow (A, Z+1)$ e$^- \nu_e^c$.
If only e$^-$ were emitted, their spectrum would be a $\delta$-function
with energy equal to the Q-value, $M_{Z} - M_{Z+1}$,
the difference in mass between the
initial and final nuclei.
The minimum energy of $\nu_e^c$ is its mass, $m_{\nu_{e}}$ (CPT insures
$\nu_{eL}$ and $(\nu_{eL})^c$  have the same mass).
Thus, the upper end-point $E_o$ of the electron spectrum
$E_0 = M_{Z+1} - M_Z - m_{\nu_{e}}$, and what is more relevant
experimentally, the curvature of the spectrum near the end-point, are
sensitive to $m_{\nu_{e}}$.
The best constraints on $m_{\nu_{e}}$ come from the shape of the
spectrum of electrons emitted in the Tritium decay, ${}^3$H$ \rightarrow
{}^3$He e$^- \nu_e^c$, because the low Q-value of this decay, $E_o =
18.58$ keV, allows to notice the effect in the spectrum of relatively
smaller $m_{\nu_{e}}$.
The first bound using this method, $m_{\nu_{e}} < 1$ keV, was obtained
as early as 1948 (Curran \etal 1949), and by the 1970's the limit
was already 55 eV (Bergkvist 1972).
The claim of a positive detection in 1980 by Lyubimov \etal originated
many new experiments attempting to verify this result
(for a review see Holzschuh 1992).
Six of these experiments (see table 2)
have now upper limits that rule out the revised
Moscow claim  of 17 eV $< m_{\nu_{e}} <$ 40 eV (Boris \etal 1987).

A striking feature of these experiments can be seen in the second column
of the table 2, namely they all find negative $m_{\nu_{e}}$ values.
This means the experiments find the e$^-$ spectra near the end-point
deformed with opposite curvature with respect to what a neutrino mass
would cause.
They use then a statistical analysis method (prescribed by the Particle
Data Group 1986) for dealing with a non-physical result, to take into
account only positive $m_{\nu_{e}}^2$ and obtain the
quoted bounds on $m_{\nu_{e}}$.
However, the negative $m_{\nu_{e}}^2$ values are large enough that
statistical fluctuations as a cause have a very low probability (about
1\%).
It seems at this point that the systematics of these experiments are not
well understood (Wilkerson 1993).
Thus, even if nominally the combined results of the first five entries of the
table 2 imply
$m_{\nu_{e}} < 5$ eV at $95\% C.L.$, due to the systematic
uncertainties it is hard to claim more than an upper limit of about 10
eV. The still unpublished new result from Troitsk involves a new kind
of systematics and its low negative $m_{\nu_{e}}^2$ seems promising.

\bigskip
\tabcaption{Present upper bounds on $m_{\nu_e}$.}
\ialign{#\hfil&&\hglue 2pc plus2pc minus1pc#\hfil\cr
\br
Experiment &  $m_{\nu_{e}}^2 \pm {\sigma_{stat}}  \pm
{\sigma_{syst}}[{\rm eV}^2]$
&  95 $\%$ CL limit
&  Reference \cr
& &  on $m_{\nu_{e}}$ [eV] &  \cr
\mr
Los Alamos &  $-147 \pm 68 \pm  41$ &  9.3
&  Robertson \etal 1991\cr
INS, Tokyo &  $-65 \pm 85 \pm 65 $ & 13
& Kawakami \etal 1991\cr
Zurich &  $-24 \pm 48 \pm 61 $ & 11
& Holzschuh \etal 1992\cr
Mainz & $-39 \pm 34 \pm 15$ & 7.2 & Weinheimer \etal~1993~\cr
Livermore & $-130 \pm 20 \pm 15$ & 8 & Robertson ~~1994~~~~~\cr
Troitsk & $-18 \pm 6$ & 4.5 & Belesev \etal ~~1994~~~~~\cr
\br}
\bigskip

The best bounds on $m_{\nu_{\mu}}$ come from the measurement of the
momentum of $\mu$ produced in the decay of pions at rest, $\pi \to
 \mu \nu_\mu$ (Abela \etal 1984, Jeckelmann \etal 1986).
After several recent objections (Wilkerson 1993) have been cleared up,
the present bound is $m_{\nu_{\mu}} <$ 160 keV at 90\% C. L. (Assamagan
\etal 1994).

The $\nu_\tau$ has only been observed as missing energy in $\tau$-decays,
contrary to the other two neutrinos that have also been observed by
producing their corresponding charged lepton in interactions with matter.
Rare multi-particle semileptonic decays of the $\tau$ provide the best
bounds on $\nu_\tau$ (Weinstein and Stroynowski).
The ARGUS collaboration (Albrech \etal 1992) obtained $m_{\nu_{\tau}}<
31$ MeV at $95\%$C.L., with 20 events, studying the reaction
$\tau^{-} \rightarrow 5\pi~ \nu_\tau$.
The CLEO collaboration (Cinabro \etal 1993), with a much larger sample
of 113 events, obtained $m_{\nu_{\tau}} < 32.6$ MeV at 95\%C.L..
There is already a preliminary upper bound from Beijing of 29 MeV at 95\%C.L.
(Darriulat 1994).
Improvements on this method are not expected to lower the $m_{\nu_{\tau}}$
bound to better than about 10 MeV.

Besides these laboratory searches, several cosmological and
astrophysical arguments considered later in this review
(sections 5 and 9) provide additional direct bounds on neutrino masses.

Here we have called $m_{\nu_{e}}$, $m_{\nu_{\mu}}$ and $m_{\nu_{\tau}}$
the three mass eigenvalues (more properly called $m_1$, $m_2$ and $m_3$)
of the mass eigenstates $\nu_1$, $\nu_2$ and $\nu_3$, and we assume $\nu_1$
consists mainly of $\nu_e$, $\nu_2$ of $\nu_\mu$ and
$\nu_3$ of $\nu_\tau$.
In recent years evidences for a 17 keV neutrino seen in nuclear beta
decay spectra were claimed by various experiments, giving origin to a
quite intense experimental and theoretical activity (for a review see,
for example, Gelmini, Nussinov and Peccei 1992).
This neutrino was seen as a shoulder in the e$^-$ spectrum, at 17 keV of
the end-point, indicating that $\nu_e$ was composed mainly of a very
light (or massless) neutrino  mass eigenstate with an admixture of order
 $1\%$ of the heavy neutrino mass eigenstate.
The existence of this shoulder has been definitely rejected in several
recent experiments (Hime 1993, Bonvicini 1993).
\subsection{Double Beta Decay}

Neutrinoless double beta decay ($\beta\beta 0\nu$) is a process that
provides a very sensitive probe of Majorana neutrino masses.
Unlike the ordinary double beta decay with emission of two neutrinos
($\beta\beta 2\nu$), that is just an allowed but rare transition at
second order in the weak interactions $(A,Z)\to(A,Z+2)+2$e$^-+2\bar\nu_e$,
the $\beta\beta 0\nu$ transition $(A,Z)\to(A,Z+2)+2$e$^-$ is a process
that violates lepton number by two units and hence requires a
departure from the SM.

The simplest scenario in which the $\beta\beta 0\nu$ decay takes place
is precisely in the presence of Majorana neutrino masses, that
convert the $(\nu_{eL})^c$, emitted in association with one of the
electrons, into a $\nu_{eL}$ that can thus be absorbed  in the second
vertex. Although the required chirality flip introduces a strong
suppression in the amplitude (with respect to that of the $\beta\beta
2\nu$), the neutrinoless decay has also a larger available phase
space. Furthermore, it has the much cleaner experimental signature of
producing a single peak in the spectrum of the sum of the two electron
energies. The $\beta\beta 2\nu$ decay leads instead to a continuous
spectrum that is hard to identify above the experimental radioactive
background. These considerations also imply that the chances
of observing a $\beta\beta$ signal increase if
elements with large $Q$-values are used, because of the significant
reduction of background at large energies. Another obvious requirement
is that the single $\beta$ decay of the isotope
be absent or strongly suppressed.

Since the typical lifetimes of $\beta\beta$ emitters are
$10^{19}$--$10^{24}$ yr, the first indirect experimental evidence for
this process came from geochemical searches, studying elements for
which the daughter nucleus is a noble gas (and hence naturally absent
in solid materials). Looking for instance for $^{130}$Xe and
$^{128}$Xe in Te rich rocks or for $^{82}$Kr coming from $^{82}$Se,
the double beta decays of these Te and Se isotopes were established.
These experiments however cannot distinguish
between the $2\nu$ and $0\nu$ channels, with the possible exception of
the ratio of $^{128}$Te and $^{130}$Te lifetimes (Bernatowicz \etal
1992), due to the similarity of the corresponding matrix elements and the
difference of phase space ratios in both types of decays.
A somewhat similar radiochemical study of a $^{238}$U
artificial sample revealed the presence of $^{238}$Pu atoms produced by
$\beta\beta$ decay after the original purification (Turkevich \etal
1991).

In the last few years, due to the significant improvements in
background reduction, detector technology and the use of large amounts
of isotopically enriched samples,
the $\beta\beta 2\nu$ process was directly
observed in the isotopes $^{82}$Se (Elliot \etal 1992), $^{76}$Ge
(Balysh \etal 1994, Avignone \etal 1991), $^{100}$Mo (Elliot \etal
1991, Dassie \etal 1994, Ejiri \etal 1994),
$^{116}$Cd and $^{150}$Nd (for recent experimental reviews see Moe
1993 and Morales 1992). The measured lifetimes are also in satisfactory
agreement with recent theoretical calculations (Wu \etal 1991).
This has given renewed impetus to this field.

Many experiments are also devoted to pursue the more interesting
search of the neutrinoless mode.
The non observation at present of any peak at the endpoint of the two
electron spectrum implies  lower bounds on the lifetime of the
$\beta\beta 0\nu$ decay, $T_{1/2}^{0\nu}$.
The Heidelberg-Moscow 90\% C.L. constraint
$T_{1/2}^{0\nu}(^{76}$Ge$)>1.9\times 10^{24}$ yr (Maier 1994),
provides the most restrictive constraint on the effective Majorana
e-neutrino mass (see below), $\langle m_{\nu_e}\rangle<1.1$ eV
(the precise bound slightly depends on the theoretical matrix element
calculation followed). Other bounds have been obtained for instance
from $T^{0\nu}_{1/2}(^{136}$Xe$)>3.4\times 10^{23}$ yr, which implies
$\langle m_{\nu_e}\rangle<2.8$ eV (Vuilleumier \etal 1993).
In the near future, the experiments
with $^{76}$Ge, and probably also those with $^{136}$Xe,
$^{100}$Mo and $^{116}$Cd, are expected  to reach a sensitivity down to
$\langle m_{\nu_e}\rangle\sim 0.1$ eV.

If the electron neutrino is a Majorana mass eigenstate,
the effective e-neutrino mass $\langle m_{\nu_e}\rangle$
which appears in the transition
amplitude is just $m_{\nu_e}$. If
the electron neutrino is instead
a linear combination  of several
mass eigenstates, $\nu_e=\sum U_{em}\nu_m$,
one has
$$
\langle m_{\nu_e}\rangle
=\left\vert\sum_m U_{em}^2m_{\nu_m}\right\vert
{}~,\eqno(4.1)
$$
 where we use
self-conjugate Majorana fields $\nu_m$ with positive masses $m_{\nu_m}$.
The matrix $U_{em}$ is in general complex. If $CP$ is conserved, the
arbitrary phases in the fields can be chosen so that the matrix $U$
becomes real,
but in so doing, if the Majorana fields are taken to be self-conjugate
their masses may not be all positive, i.e. $m_{\nu_m}=\lambda_m\vert
m_{\nu_m}\vert$, with $\lambda_m=\pm 1$.
The effective mass can be written in the $CP$ conserving case as
$$\langle m_{\nu_e}\rangle =\left\vert\sum_m\lambda_m\vert U_{em}^2\vert
\vert m_{\nu_m}\vert\right\vert .\eqno(4.2)$$
We recall that $\tilde \eta_{CP}=\lambda_m i$ is the intrinsic $CP$ parity of
the
Majorana neutrino $\nu_m$ (see section 2, Wolfenstein 1981,
Kayser 1984, 1988).

Equation (4.2) shows that cancellations can occur between different
states in $\langle m_{\nu_e}\rangle$. If this happens, the actual
neutrino masses can be all larger than the bound on $\langle
m_{\nu_e}\rangle$ implied by the non-observation of $\beta\beta 0\nu$.
In particular, a Dirac $\nu_e$ can be viewed as the limiting case of
two degenerate Majorana states with opposite $CP$ parities (see
section 2), for which the cancellation is complete and no neutrinoless
decay takes place. This is of course also expected from the absence of
lepton number violation in this case.
If $CP$ is violated in the neutrino sector, there will be unremovable
phases in the matrix $U$ and hence cancellations can occur even among
states with equal $CP$ parities.

Eq. (4.1) is actually valid if all the $\nu_m$ states appearing in
$\nu_e$ are light ($m_{\nu_m}<10$ MeV), since otherwise the nuclear
matrix elements involving the heavy neutrino propagator are strongly
suppressed. Hence, very heavy components have an additional
suppression besides that of the small $U_{em}^2$ factor.

In addition to the ``mass mechanism" just discussed, the
$\beta\beta 0\nu$ can also take place in left--right symmetric
theories. In this case, the presence of $\nu_R$ states,
right--handed charged currents as well as a larger Higgs sector
combine to give rise to the
neutrinoless decays in several possible ways
(for a review see e.g. Vergados 1986, Tomoda 1991).

Other interesting extensions of the SM affecting $\beta\beta$
processes are the Majoron models (see section 3.5).
Here the coupling of the Majoron Goldstone boson ($J$)
to neutrinos
$$
{\cal L}=-{(g_{\rm eff})_ { \nu_e \nu_e J}\over
\sqrt{2}}~~\overline{\nu_{e L}}(\nu_e^c)_R J
+h.c.
\eqno(4.3)$$
produces the required chirality flip in the exchanged neutrino line by
the emission of the neutral Majoron\footnote{*}{Since LEP has excluded
both triplet and doublet Majorons, this effective coupling can only be
with singlet or mixed Majorons and could arise from a $\nu_e$ mixing with
singlet neutrinos (see section 3.5).}. Due
to the energy carried away by the undetectable $J$,
the electron spectrum is no longer a peak. $\beta\beta 0\nu J$ is
however still concentrated at larger energies than the $\beta\beta
2\nu$ spectrum. The non-observation of this decay has then resulted in
significant constraints on the coupling
$(g_{\rm eff})_ { \nu_e \nu_e J}$.
In particular, the best direct bound $g_{eff}<7\times 10^{-5}$
comes from $^{150}$Nd (Moe 1994),
even if the lifetime limit is weaker than that from Ge (Beck \etal 1993),
due to the
favorable matrix element and phase space factors of the first.

\section{Neutrinos in Cosmology}

Cosmological and astrophysical arguments are a complement to laboratory
experiments as a probe of neutrino physics and actually provide some of
the most restrictive constraints on certain neutrino properties.

The possible cosmological consequences of neutrinos, and in particular of
their non-vanishing masses,
started to be explored in the seventies. Bounds on the masses
from the contribution of stable neutrinos to the density of the universe were
derived first. Then, bounds on unstable neutrinos were obtained from the
effects of the decay products. These cosmological tests derived for neutrinos
are now routinely applied to any new proposed particle to limit its lifetime,
mass, cosmological density and decay modes.

The hot Bing Bang, the standard model of cosmology, establishes that the
universe is expanding from a state of extremely high temperature and density.
The moment the expansion started is taken as the origin of the lifetime of the
universe  $t$, which is now determined to be $ t_o \simeq 1.3$ to $1.7 \times
10^{10}$ yr (from the oldest globular clusters, Demarque, Deliyannis and
Sarajedini 1991 give 14-17 Gyr, Renzini 1993 gives 13-15 $\pm$3 Gyr). Notice
that in cosmology the subscript zero denotes  present values. The Big Bang
model is based on three major empirical pieces of evidence, namely: the Hubble
expansion, the cosmic blackbody microwave background radiation, CMB, and the
relative abundance of the light elements (up to${}^7$Li). These pieces of
evidence are  very difficult to explain with a model different than the Big
Bang (Peebles \etal 1991).

The Hubble parameter provides the proportionality between the velocity of
recession of far away objects v, and their relative distance d, v $=H$d. Its
present value, the Hubble constant, is known up to a factor of two, $H_o =
100 h$ km/sec Mpc, with $ 0.4 \lesssim h \lesssim 1$, where a parsec is
$pc = 3.26$ light years (see e.g. Jacoby \etal 1992;
Fukugita, Hogan and Peebles 1993; Scott \etal 1994).

The CMB was produced at $t_{\rm rec} \simeq 3 \times 10^5$ yr, the
recombination epoch, when atoms became stable and replaced ions and electrons
in a plasma as the constituents of the universe. The Cosmic Background Explorer
(COBE) satellite (Mather \etal 1994) confirmed that the CMB has a blackbody
spectrum, and made the most accurate measurement of its temperature, $T_o =
2.726\pm 0.010$ K (95\% C.L.). Thus we know with great accuracy the number
density of relic photons (the CMB photons are the most abundant in the
universe, by several orders of magnitude, see e.g. Kolb and Turner 1990, p.
143)
$n_\gamma = (2 \zeta (3)/\pi^2) T_o^3 = 411$ cm$^{-3}$.

The cosmological abundance of ${}^4$He and of the trace elements D, $ {}^3$He
and ${}^7$Li are well accounted for in terms of nuclear reactions that occurred
at $t \simeq 10^{-2} - 10^2$ sec, the nucleosynthesis epoch, when the photon
temperature was $T \simeq 10 - 0.1 $ MeV (necessarily below the binding energy
of the light nuclei). This is the earliest available proof of the consistency
of the Big Bang model. The relative amount of the primordial light elements
synthesized is very sensitive to the relative abundance of baryons (protons and
neutrons) over photons, $\eta \equiv n_B/n_\gamma$ ($n$ is number density),
and to the expansion rate (and, thus, to the energy density of the universe
$\rho$,  since $H \sim \sqrt\rho$) at the epoch of nucleosynthesis. Recent
analysis (Walker \etal 1991, Smith \etal 1993) find that $\eta$ must be in a
small range $\eta = (2.8-4) \times 10^{-10}$ (95\%C.L.) to fix three very
different abundances: ${}^4$He, D + ${}^3$He and  ${}^7$Li. This implies a
bound on the energy density of baryons, $\Omega_B = \rho_B/\rho_c$, through the
relation $\eta = 2.68 \times 10^{-8} \Omega_B h^2$, which is
$0.01 \leq \Omega_B \leq
0.015$. It is convenient to express energy densities $\rho$ in units of the
critical density $\rho_c = 3 H_o^2/8\pi G = 10.5 h^2$ keV cm$^{-3}$ (defined
to be such that for $\rho_{\rm T0TAL} \leq \rho_c$ the universe will expand
forever and for $\rho_{\rm TOTAL} > \rho_c$ the universe will eventually
recollapse), i.e. $\Omega \equiv \rho/\rho_c$. Attempts to avoid the
nucleosynthesis upper bound on $\Omega_B$  have been largely unsuccessful so
far (Malaney and Mathews 1993). These include, inhomogeneities in the baryon
density (Applegate, Hogan and Scherrer 1987), late decaying particles
(Dimopoulos \etal 1988) and generating entropy after nucleosynthesis (Bartlett
and Hall 1991). We mention later a  new idea involving a decaying $\nu_\tau$
(Gyuk and Turner 1994).

An increase in the expansion rate of the universe leads to an earlier freeze
out of the weak interactions that determine the ratio of neutrons over protons
and, consequently, to a larger value of this ratio and to overproduction of
${}^4$He (since as a first approximation all neutrons end up in ${}^4$He
nuclei). Thus, the observational upper bound on the primordial ${}^4$He
abundance provides an upper bound on the total $\rho$ at the moment of
nucleosynthesis, that is expressed in terms of the allowed number of light
neutrino families, $N_\nu \lesssim 3 + \delta N_{\nu}$. Walker \etal 1991
found $\delta N_{\nu} =0.3$, but this number changes often, with the
incorporation of new measurements of the  cosmic abundances of
elements\footnote{$^*$}{ Loopholes of this bound exist but they require rare
specific neutrino properties, such as excess lepton numbers of the order of the
photon number density $n_\gamma$, namely $(n_{\nu_i} - n_{\nu^c_i})/ n_\gamma
\simeq O(1)$, (while one expects this leptonic asymmetry to be of the same
order of magnitude of the baryonic asymmetry $\eta \simeq O(10^{-10})$
(Olive \etal 1991), or a heavy tau neutrino with mass in the MeV range,
decaying with
a short lifetime ($\lesssim 30$ sec) into $\nu_\mu$ or, better, $\nu_e$ and
other invisible particles (a Majoron, for example), as mentioned later in this
section (Kawasaki \etal 1994 and Dodelson, Gyuk and Turner 1994a).}. Even if
the total width of the $Z^o$ measured at LEP insures that $N_\nu = 3$, the
nucleosynthesis bound is still useful. This is because it limits any extra
contribution  beyond those of three  left handed relativistic neutrinos to
$\rho$ during nucleosynthesis, to be equivalent at most to $\delta N_{\nu}$ of
a neutrino species.

Both CMB and nucleosynthesis provide a plethora of bounds on neutrinos, some of
which we will mention below. Other bounds are provided by the present cosmic
energy density and by structure formation  in the universe (namely the
formation of galaxies and galaxy aggregates).

May be the most important cosmological constraint on stable neutrinos is the
mass bound. Light neutrinos $(m_\nu <$ 1 MeV) with SM interactions are kept in
equilibrium with charged leptons and photons in the cosmic plasma (due to weak
interactions) until a temperature $T \simeq$ 1 MeV. At these temperatures the
rate of weak processes becomes smaller than the expansion rate of the universe
and neutrinos ``decouple'' or ``freeze out''. Their number (per comoving
volume, i.e. a volume increasing due to the Hubble expansion) becomes constant
afterwards. Taking into account that
the number of photons is increased when e$^+$e$^-$ annihilate (at $T
\lesssim m_e$, when e$^+$e$^-$ can no longer be formed), due to entropy
conservation, the present number density of neutrinos (plus antineutrinos) per
species is $n_{\nu_{\i}} = (3/11)n_{\gamma} = 102$ cm$^{-3}$.

If neutrinos are massive, their energy density is $\rho_{\nu_{\i}} = $
$m_{\nu_{\i}} n_{\nu_{\i}}$, therefore (Gerstein and Zeldovich 1972, Cowsik and
McClelland 1972)
$$
\Omega_\nu h^2 = \sum_{\i=1}^3  {m_{\nu_{\i}}\over 92\ {\rm eV}}~~.
\eqno(5.1)
$$
This bound assumes that only left-handed neutrinos are ever in equilibrium in
the primordial plasma (what is correct, even in the case of Dirac neutrinos for
Dirac masses below a few keV, see e.g. Kolb \etal  1991).
The best upper bound on $\Omega_o h^2$ does not
come from estimates of $\Omega_o$ and $h$, but from $t_o$. Due to the relation
between $t_o$, $H_o$ and $\Omega_o$ (that depends on the nature of the content
of the universe: matter, radiation or a cosmological constant, namely vacuum
energy),  a lower bound on the age of the universe translates into an upper
bound on $\Omega_oh^2$. Thus, $t_o \gsim 1.3 \times 10^{10}$ yr provides the
bound $\Omega_o h^2 \lesssim 0.4$ for a matter dominated universe (and zero
cosmological constant, see Kolb and Turner 1990, figure 3.3), thus
$\sum\limits_{\i=1}^3 m_{\nu_{\i}} \lesssim 37$ eV.  Only for small values of
$h$,  $0.4 \leq h \leq 0.5$, it can be  $ 0.25 \lesssim \Omega_o h^2 \lesssim
0.4$ (see the same figure). Thus, if the more popular lower bound of $h \simeq
0.5$ is taken then $\Omega_o h^2 \lesssim 0.25$ and consequently,
$\sum\limits_{\i=1}^3 m_{\nu_{\i}} \lesssim  23$ eV.  The presence of a
cosmological constant  $\Lambda$  would relax these constraints a bit (because
$\Omega_o h^2$ can be larger  given the same $t_o$, see e.g. Kolb and Turner
1990). Neutrinos with mass close to these upper bounds could dominate the mass
density and provide, therefore, the Dark Matter (DM).

It is by now well established that the dominant form of matter in the universe
is only detectable through its gravitational effects (and, because of  this, it
is called Dark Matter). Most measurements of the DM at large scales obtain
$\Omega_{DM} \simeq 0.2$ to $0.3$, but some obtain close to 1 (see e.g. Kolb
and  Turner 1990, Peebles 1993). Because all other known contributions to
$\Omega_o$ are much smaller, the total amount of DM in the universe is
responsible for the ultimate fate of our universe, expansion forever or
recollapse (if $\Omega_o \leq 1$ or $\Omega_o > 1$, respectively). The nature
of the DM is one of the most important entirely open questions in physics.  DM
candidates are classified as cold or hot according to the galaxy formation
scenarios derived from them. Galaxies are assumed to be formed  through
gravitational instability, from tiny  inhomogeneities in the energy density
(${\delta\rho/ \rho} < 10^{-4}$ at recombination), that leave an imprint in the
CMB, recently found by COBE. Very different mechanisms result if the DM is
relativistic (hot DM, HDM) or non-relativistic (cold DM, CDM) when galaxy size
inhomogeneities could first start collapsing at $T \simeq$ 1 keV. Massive
neutrinos (with $m <$ 1 keV) could be HDM. Simulations of structure formation
with pure HDM fail to fit the data, because galaxies form too late (by
fragmentation of the much larger structures that form first).

Pure CDM accounts for the bulk of the known data, but seems to fail
in detail. The COBE measurements of anisotropies in the CMB
provide a measurement of  density inhomogeneities
(Smoot \etal 1992, Gorski et al 1994).
In the context of models with $\Omega=1$ and primordial scale-invariant
 fluctuations (the simplest assumptions), once the
normalization given by COBE is imposed on the  spectrum of density fluctuations
predicted
by pure CDM at large scales, the spectrum has too much power on smaller scales,
namely the scales of galaxy clusters (Wright \etal 1994). A possible
acceptable solution explored at present is mixed DM (MDM or HCDM). Recent
simulations suggest an admixture of $\Omega_\nu \simeq 0.30$ of HDM in
neutrinos in a universe dominated by CDM (Wright \etal 1994, Nolthenius \etal
1993, Bonometto \etal 1993, Klypin \etal 1994). These simulations use
$\Omega_o = 1$, and $h = 0.5$, thus requiring $\sum\limits_{\i=1} ^3
m_{\nu_{\i}} \simeq$ 7 eV.

Up to now we mentioned neutrinos lighter than 1 MeV. Neutrinos with $m_{\nu}
\gsim 1$ MeV would decouple while they are non-relativistic and their density
is thus reduced by a Boltzman factor $\Omega_\nu \sim {\rm exp}(-m_\nu/T)$. In
this case the bound in the energy density $\Omega_o h^2$ is
satisfied for $m_\nu
\gsim$ few GeV (Lee and Weinberg 1972, Hut 1977, Sato and Kobayashi 1977,
Vysotsky, Dolgov and Zeldovich 1977). The mass bounds differ for a  Dirac or a
Majorana neutrino, they are higher for the latter (Kolb and Olive 1986, see
also Kolb and Turner 1990). Only a $4^{th}$ generation neutrino could be that
heavy, but the LEP bound excludes its existence unless $m_\nu > M_Z/2
\simeq$ 45 GeV. These heavy neutrinos would only have a small density
$\Omega_\nu < 10^{-2}$.

These constraints apply to neutrinos with only standard interactions (i.e. to
neutrinos that are non-standard only because they have a non-zero mass).
Neutrino masses could be in the forbidden range mentioned above, namely 30 eV
$< m_\nu < $ few GeV, if they have interactions beyond the SM that allow for
faster annihilation or decay.

Unstable neutrinos with $m < 1$ MeV whose relativistic decay products dominate
the energy
density of the universe  until the present must have a lifetime
$$
\tau \leq (92\ {\rm eV}/m_\nu)^2 (\Omega_o h^2)^2 t_o~~,
\eqno(5.2)
$$
 to insure
that the energy density of the decay products, $\Omega_{DP}$, is not too
large, i.e.
$\Omega_{DP} \leq \Omega_o$ (Dicus, Kolb and Teplitz 1977, Pal 1983, Kolb
1986)\footnote{$^*$}{This bound is obtained with a simple calculation that
assumes all neutrinos decay suddenly at t=$\tau$. A proper  calculation shows
that this assumption overestimates the $\Omega_{DP}$ by  about 15\% (Turner
1985). Thus the bound (5.2) should be actually higher by a factor of  1.3.}.
The bound in (5.2) and the corresponding bound for heavier masses are shown
in figure 1 with a continuous contour line.
Since neutrinos and photons are almost in equal numbers, the  neutrino energy
density $\rho_\nu = n_\nu m_\nu$ becomes dominant over the  radiation density
$\rho_{\rm rad} \simeq n_\gamma T$  as soon as $m_\nu > T$, i.e.  neutrinos
matter-dominate the energy density of the universe as soon  as they become non
relativistic (actually at T$\simeq  0.1 m_\nu$).  Thus, if neutrinos decay
while non-relativistic into relativistic  decay products, i.e. $\rho_\nu =
\rho_{DP}$ at $t=\tau_\nu$, these products  radiation-dominate the universe.
Because in (5.2) the universe is assumed to be radiation dominated until the
present, the constraint $t_o > 1 \times 10^{10}$ yr (1.3 $\times 10^{10}$ yr)
requires  $\Omega_o h^2 < 0.3~(0.1)$. If the universe is assumed to become
dominated  by matter at some time after the neutrino decay, the bound on the
lifetime is more restrictive than (5.2)  (and the bound in figure
1)\footnote{$^{**}$}{Notice this is in  contradiction with the equivalent
bounds
derived  in Kolb and Turner 1990 and presented in their figure 5.4,  that we
believe are not correct.}. In fact,   the massive neutrino has to  decay
earlier for the energy of their decay products to have a longer  time to
decrease, as $T^4$, and become subdominant before the present  with respect to
a matter density component that  decreases slower,  as $T^3$.   This is
actually required by structure  formation arguments. Because the growth of
density fluctuations is suppressed in the period of radiation domination of the
decay products, this period should finish at most when structure in ordinary
matter could start forming,  i.e. at recombination  $t_{\rm rec} \lesssim
10^{-5} t_o$, when atoms  become stable. This argument replaces $t_o$ by
$t_{\rm rec}$ in (5.2), thus yielding a bound more stringent by a factor
$10^{-5}$ (Steigman and Turner 1985)\footnote{$^{***}$}{For a more accurate
version of this bound based on numerical calculations and using the present
constraints from large scale structure and CMB see White, Gelmini and Silk
1994.}. This bound is shown for $m <$ 1 MeV in figure 1 with  a dashed
contour line.

 A neutrino with mass in the keV to MeV range, with a lifetime close to the
lower limit just obtained, could  not only be harmless but actually
can help in the
structure formation. It could actually be what CDM needs to find perfect
agreement with data (White, Gelmini and Silk 1994, Bardeen, Bond and Efstathiou
1987, Bond and Efstathiou 1991). The main effect of the decaying neutrino is to
delay the onset of the matter domination by the CDM, by adding to the radiation
density the contribution of the (relativistic) decay products. Also a heavier
neutrino, a $\nu_\tau$ with mass between 1 and 10 MeV could do it (Dodelson,
Gyuk and Turner 1994b), provided the lifetime is $\lesssim 10^2$ sec
to avoid excessively
perturbing nucleosynthesis (Dodelson, Gyuk and Turner 1994a, Kawasaki \etal
1994)

 In fact, a non-relativistic neutrino present
during nucleosynthesis, necessarily a tau-neutrino,
 may contribute more to $\rho$ than a  relativistic
species. Therefore, the bound on the additional number of relativistic
neutrinos  mentioned above, $\delta N_{\nu} < 0.3$, forbids the mass ranges
0.2 MeV$ - $33 MeV (for a Dirac neutrino) or  0.4 MeV$ - $30 MeV (for a
Majorana neutrino), if the neutrino is present during nucleosynthesis, i.e. for
$\tau > 10^2$ sec (Dodelson, Gyuk and Turner 1994a, Kawasaki \etal 1994,
Dolgov, Kainulainen and Rothstein 1994, Kolb and Scherrer 1992, Kolb
\etal 1991). This bound is shown in figure 1 with a
dot-dashed contour line.
These nucleosynthesis bound combined with the
laboratory upper limits (31 MeV, see section 4) nearly
exclude\footnote{$^{*}$}{The preliminary bound from Beijing of 29 MeV,
 if confirmed, would
already have closed the gap between both limits.}  a $\nu_\tau$ more massive
than about 0.4 MeV, if $\tau > 10^2$ sec (and if the $\nu_\tau$ does not have
additional annihilation channels besides those of the SM, so that its relic
density is the one computed with standard interactions).

For shorter lifetimes the bounds depend on the decay channel.
Radiative decays, i.e. into photons or
e$^+$e$^-$,  are excluded for all allowed neutrino masses, as can be seen in
figure 2. Bounds come from not allowing
distortions of the CMB (this excludes the gray region in figure 2, taken
from Kolb and Turner 1990) and from excessive entropy generation
or the dissociation
by the decay products of the  synthesized nuclei after nucleosynthesis.
Also stringent bounds come from astrophysics (section 9). Bounds obtained
from the supernova SN1987A exclude the hatched region in figure 2.

 The decay modes available are $\nu_\tau \rightarrow 3 \nu' $ and
$\nu_\tau \rightarrow \nu_e \phi$. The  mode $\nu_\tau \rightarrow \nu_e \phi$,
has been studied for all masses and lifetimes. For very short lifetimes $\phi$
is brought into equilibrium by decays and inverse decays and its contribution
to $\rho$ during nucleosynthesis  becomes unacceptably large (Kawasaki \etal
1994).

Let us mention two curious properties of these massive, fast decaying tau
neutrinos. A $\nu_\tau$ with  mass  $m >$1 MeV and short lifetimes $\tau <$ 10
sec, that decays into another neutrino (and other sterile particles), could
decrease the amount of ${}^4$He produced thus actually loosening the
nucleosynthesis bound on new light species. Available analyses differ slightly
in the range at stake (Dodelson, Gyuk and Turner 1994a, Kawasaki \etal 1994).
This possibility could become particularly useful if the observational data
would lead to a lower ${}^4$He abundance than now assumed (reducing the allowed
number of effective neutrino families to less than three). The second property
refers to a $\nu_\tau$ with mass (20--30) MeV decaying into $\nu_e$ (and
other invisible particles) with a  lifetime of (200--1000) sec, that has a
relic
density smaller than that of a  standard neutrino (i.e. this neutrino must have
larger than standard annihilation cross sections in the early universe). In
this case the bound on $\Omega_B$ from nucleosynthesis could be loosen (Gyuk
and Turner 1994). Moreover there is the possibility we already mentioned of a
massive short-lived neutrino of $m>$ 1 MeV helping in structure formation in
the early universe (Dodelson, Gyuk and Turner 1994b). Rejecting experimentally
a $\nu_\tau $ mass larger than 1 MeV would eliminate these possibilities.

As already mentioned, radiative decays $(\nu \rightarrow \nu' \gamma$,
 $\nu \rightarrow \nu' $e$^+$e$^-)$ cannot help in evading the cosmological
mass bound (see figure 2), i.e. cannot have
shorter lifetimes than those excluded in figure 1.
Only decays into neutral weakly interacting particles are allowed, namely
$\nu \rightarrow 3 \nu'$ and $\nu \rightarrow \nu' \phi$.
 It is difficult to find extensions of the SM in which the first decay is
fast enough while avoiding conflicts with cosmological as
well as laboratory bounds and  suppressing the related forbidden radiative
decays (for a review see e.g. Mohapatra and Pal 1991). The mode $\nu
\rightarrow \nu' \phi$, where $\phi$ is a Goldstone boson,  seems the most
promising, and finding a neutrino mass in the forbidden  cosmological range
would strongly suggest the existence of Goldstone bosons.

If neutrinos are ``cosmologically stable", i.e. $\tau \gg t_0$, the
radiative decay of a massive neutrino is not excluded by the above
mentioned bounds. A decaying DM neutrino with $m_\nu\simeq O(10$ eV) would
however produce a monochromatic UV line  (with $E_\gamma=m_\nu/2$ if
the daughter neutrino is massless) that could be observable with
satellite detectors. A neutrino of
$m_\nu\simeq 28$ eV and $\tau\simeq 10^{23}$ sec
has even been proposed  to account for the
observed ionization of the galactic and intergalactic hydrogen
(De R\'ujula and Glashow 1980, Sciama 1990).
Although this lifetime is much shorter than those resulting in the SM with the
addition of Dirac neutrino masses, it is surprising that it is just
at the right value in models in which both masses and radiative decays
(through a dipole transition) are generated by loop effects (Roulet
and Tommasini 1991, Gabbiani \etal 1991).

The nucleosynthesis bound on the effective number of relativistic neutrinos
$\delta N_{\nu}$ applies also to any new mechanisms that could bring sterile
particles into the primordial plasma, such as active-sterile neutrino mixings
(see sections 7 and 8) or a Dirac neutrino mass.
With respect to Dirac masses, recall from section 2 that if neutrinos are Dirac
particles, there exist right-handed chirality neutrinos ($\nu_R$, and also
left-handed chirality $(\nu^c)_L$) that in the SM do not have weak
interactions. Since the physical states of a neutrino are helicity
states, and helicity and chirality do not coincide for massive neutrinos, all
four helicity states have weak interactions. However, the admixture of the
``wrong" chirality (the interacting one) in a right-handed helicity neutrino is
of order $m/E$ (for $m<E$, $E$ is the neutrino energy), thus their interactions
are suppressed by a factor $(m/E)^2$ with respect to those of a left handed
helicity neutrino.  An upper bound on the abundance of right-handed helicity
neutrinos (given by $\delta N_{\nu}$ in this case) translates into an upper
bound on  their reaction rates and thus on the Dirac mass. If $m <
0.3$~MeV the
right-handed neutrinos decouple before the quark-hadron $QCD$ phase transition,
in which there is a large increase of the number of still interacting
particles. Thus the equilibrium number of right-handed neutrinos  becomes
relatively very small (Fuller and Malaney 1991, Enqvist and Uibo 1993).
However, there are also out of equilibrium processes in which right handed
neutrinos are produced, such as pion decays (Lam and Ng 1991).  Dolgov,
Kainulainen and Rothstein (1994) find  $m_{{\nu}_{\mu}} \lesssim 170$ keV
$[(\delta N_{\nu} -0.10)/0.20]^{1/2}$ and $m_{{\nu}_{\tau}} \lesssim
210$ keV$
[(\delta N_{\nu} -0.10)/0.20]^{1/2}$, if the temperature of the quark-hadron
phase transition is $T_{QCD}= 100$ MeV. These bounds are anyhow less
restrictive than the bound derived from the supernova SN1987A, that forbids
Dirac masses from $O(10$ keV) to 1 MeV (see section 9).
Notice that if the right
handed neutrino would be trapped inside the supernova by unknown interactions
beyond the SM (Babu, Mohapatra and Rothstein 1991) thus avoiding the SN1987A
bound (based on energy loss due to the escape of right handed neutrinos), the
same interactions would bring them into equilibrium during nucleosynthesis
(leading to an unacceptable $\delta N_\nu$=1). This conclusion is difficult to
obviate (Babu, Mohapatra and Rothstein 1992).

Let us mention last an upper bound on Majorana neutrino masses due to the
 persistence of a baryon asymmetry generated at temperatures higher than the
 electroweak phase transition, as most baryon generation models assume.
 The $L$ violating processes associated with
 the Majorana mass in conjunction with $B+L$ violating non-perturbative
 processes in the SM would erase any existing $B$ asymmetry
 unless the Majorana masses are small enough. A stringent upper
 bound of 0.1 eV was formerly derived, but recent re-considerations of this
 bound have loosen it considerably to around 10 keV
 (see Cline, Kainulainen and Olive 1993b).

\section{Neutrino oscillations}

\subsection{Neutrino mixing}

The phenomenon of neutrino oscillations (Pontecorvo 1958), appears
because the neutrino current (or interaction or flavour) eigenstates
$\nu_\alpha$, $\alpha = e,\mu, \tau$
 (namely, the neutrino states produced
in a weak decay in association with a given charged lepton flavour,
see (3.8)) are
generally superpositions of different neutrino mass eigenstates $\nu_i$,
$i=1,2,3$,
$$
\nu_\alpha=\sum_iU_{\alpha i}\nu_i \;.\eqno(6.1)
$$
Here the unitary matrix $U = K_\ell^{\dag}$, where $K_\ell$ is the
leptonic analog of the Cabibbo-Kobayashi-Maskawa quark
matrix $K_q$ (see (3.8) and (3.9)).
Due to the dissimilar propagation  of the neutrino mass eigenstates,
the flavour
content of the propagating neutrino  $\nu$ changes with time.
Since the neutrinos are usually detected
by means of charged current processes, sensitive to the
neutrino flavour, this oscillating behaviour may be observable. Moreover
this is an interference effect, sensitive therefore to very small neutrino
mass differences.

\subsection{Oscillations in vacuum}

To study the $\nu$ oscillations, assume that a $\nu_\alpha$ flavour
eigenstate (namely $\nu_{e}$, $\nu_{\mu}$ or $\nu_{\tau}$)
 is produced  at $x=0,\ t=0$.
Although the $\nu$ states are wave-packets with a certain spread in
momentum, it is sufficient to consider just the plane wave solutions,
since
limitations due to the finite coherence length of the wave-packets
in general do not affect the cases of interest (Nussinov 1976,
Kayser 1981).
We know that the space-time
dependence of a free mass eigenstate $\nu_i(t)$ of momentum
$p_i$ and energy $E_i=\sqrt{p_i^2+m_i^2}$ is
$$
\nu_i(t) = {\rm exp}[i(p_ix-E_it)]~~~\nu_i.\eqno(6.2)
$$
It is convenient to take the propagating neutrinos with a common
definite momentum $p$ (the same conclusion result if we take e.g. a
common energy and different momenta (Winter 1981) or the wave-packets
themselves),
so that for the ultra-relativistic neutrinos $E_i\simeq p+m_i^2/(2E)$
and the neutrino state is then
$$
\nu(t)={\rm
exp}[ip(x-t)]\sum_iU_{\alpha i} {\rm exp}\left(-i{m_i^2\over
2E}t\right)\ \nu_i \;.\eqno(6.3)
$$
The probability with which the neutrino
produced as $\nu_\alpha$ is converted into $\nu_\beta$ after
travelling a distance $x=t$ results
$$
\eqalign{
P(\nu_\alpha\to\nu_\beta)=& \left\vert
 \langle\nu (t)\vert\nu_\beta\rangle\right\vert
^2=\left\vert \sum_iU_{\alpha i}{\rm exp}
\left(-i{m_i^2\over 2E}t\right)U^*_{\beta i}\right\vert^2 \cr =&
{\rm Re}\sum_{i,j}U_{\alpha i}U^*_{\alpha j}U^*_{\beta i}U_{\beta j}
{\rm exp}\left[-i{m_i^2-m_j^2\over 2 E}x\right] .
}\eqno(6.4)
$$
We see that in order to have
$P(\nu_\alpha\to\nu_\beta)\neq\delta_{\alpha \beta}$ it is necessary
that at least two neutrinos be non-degenerate.

These probabilities satisfy some important relations. $CPT$ invariance
implies
$P(\nu_\alpha\to\nu_\beta)=P(\bar\nu_\beta\to\bar\nu_\alpha)$, while
for two flavour mixing, and only if $CP$ is conserved  also for mixing with
three or more flavours, one has
$P(\nu_\alpha\to\nu_\beta)=P(\nu_\beta\to\nu_\alpha)$ (Cabibbo 1978),
because these
probabilities are related by the replacement $U\leftrightarrow U^*$ in
eq. (6.4).
These relations are valid when neutrinos propagate in the vacuum.
When interactions with matter affect the neutrino propagation they may
no longer hold because the medium itself is generally not symmetric
under $CP$ and $CPT$. Finally, the unitarity of $U$
implies the probability conservation relation
$$P(\nu_\alpha\to\nu_\alpha)=1-\sum_{\beta\neq \alpha}
 P(\nu_\alpha\to\nu_\beta).\eqno(6.5)$$

We will hereafter consider the case of
mixing between just two neutrino flavours, $\nu_\alpha$ and
$\nu_\beta$,
$$\pmatrix {\nu_\alpha\cr\nu_\beta}=
\pmatrix {\cos\theta&\sin\theta\cr
-\sin\theta&\cos\theta}\pmatrix {\nu_1\cr\nu_2}\ .
\eqno(6.6)$$
To be more general, $\nu_\beta$ can be taken here as one of the three
known neutrinos or eventually as a new light singlet neutrino $\nu_s$ not
participating in the weak interactions (sterile
species). For a sterile neutrino the oscillation formalism is still valid but
the predictions for observations by both charged and neutral
current processes are affected.

The probability of neutrino conversion then results
$$
P(\nu_\alpha\to\nu_\beta)=
\sin^22\theta\ \sin^2{\Delta m^2\over 4E}x~~,
\eqno(6.7)$$
where $\Delta m^2\equiv m_2^2-m_1^2$. Clearly the probability of
observing the same flavour in the neutrino beam is just
$P(\nu_\alpha\to\nu_\alpha)=1-P(\nu_\alpha\to\nu_\beta)$. We see that
these probabilities oscillate with an amplitude proportional to the
neutrino mixing factor $\sin^22\theta$. Noting that
$$
{4 E\over \Delta m^2}=0.8 \;{\rm m}{E\;[{\rm MeV}]\over
\Delta m^2\;[{\rm eV}^2]}~~,
\eqno(6.8)$$
we also see that the oscillation length is macroscopic for the
typical energies of reactor ($1-10$ MeV) or accelerator ($10^{-2}-10^2$
GeV) neutrinos, at least for the cosmologically allowed values of the
masses of stable neutrinos ($<10^2$ eV).

When considering an experiment, it is important that the neutrino
source is not monochromatic but instead has in general a broad energy
spectrum. Furthermore, both the region of neutrino production and the
detector have finite sizes. These facts make the oscillatory term  in
the conversion probability to be usually averaged out, leading just
to $P(\nu_\alpha\to\nu_\alpha)=1-\sin^22\theta/2$. However, if $\Delta
m^2$ is so small that the oscillation length becomes much larger than
the baseline $d$ between the production point and the detector, i.e.
$\Delta m^2\;[eV^2]\ll E\;[MeV]/d\;[m]$, the oscillations do
not have enough time to develop and no effect can be seen. Clearly
going to larger baselines allows to test smaller $\Delta m^2$ values,
as long as the reduction in the $\nu$ flux $(\propto d^{-2}$) does not
become the limiting factor. Actually, there is intense activity at
present to develop long-baseline ($10-10^4$ km) experiments using
accelerator neutrinos, to increase the $\Delta m^2$ sensitivity.

The main sources of artificial neutrinos used in oscillation
experiments are:\hfil\break
\noindent - $\bar\nu_e$ from $\beta^-$ decays of fission products in
reactors; \hfil\break
\noindent - Fluxes of $\nu_\mu,\ \bar\nu_\mu$ and $\nu_e$, in
comparable amounts and with energies of tens of MeV,
from decays of stopped $\pi^+$ at low energy accelerators;\hfil\break
\noindent - Beams of $\nu_\mu$ or $\bar\nu_\mu$, with a small
contamination at the percent level of $\nu_e$ and $\bar\nu_e$, from
decays in flight of $\pi$ and $K$ produced in high energy
accelerators.

In addition to the experiments using artificial sources,
 studies of $\nu$ oscillations can be performed using
natural sources of neutrinos, such as atmospheric neutrinos produced
by cosmic rays in the upper atmosphere or solar neutrinos produced in
the fusion reactions inside the sun. In table 3 we show the typical
energies, the distances involved and the minimum $\Delta m^2$ values
leading to testable oscillations, for the different kinds of neutrino
sources just mentioned.

\bigskip
\tabcaption{Neutrino energies $E$, baseline distances $d$ and minimum
testable mass-square differences $\Delta m^2$ for reactor, accelerator
(existing short-baselines and proposed long-baselines), atmospheric
and solar neutrinos.}
\ialign{#\hfil&&\hglue 2pc plus2pc minus1pc#\hfil\cr
\br
&reactor&\hfil accelerator\span\omit&atmospheric&solar\cr
&&short-base.&long-base.\cr
\mr
$E$ [MeV]&$\lesssim 10$&\hfil$30-10^5$\span\omit&$10^3$&$\lesssim 14$\cr
$d$ [m]&$10-300$&$10^2-10^3$&$10^4-10^7$&$10^4-10^7$&$10^{11}$\cr
$\Delta m^2$
[eV$^2$]&$10^{-2}$&$10^{-1}$&$10^{-4}$&$10^{-4}$&$10^{-11}$\cr
\br}
\bigskip

As we will see in sections 7 and 8, studies of solar and atmospheric
 neutrinos provide some indications in favour of neutrino
oscillations. On the contrary, the searches of oscillations using
reactor and accelerator neutrinos have provided up to the present
no compelling evidence for oscillations, therefore excluding the
ranges of $\Delta m^2$ and
$\sin^22\theta$ explored, as we turn now to discuss.

\subsection{Oscillation experiments}

There are essentially two strategies for detecting neutrino
oscillations starting from a beam of neutrinos of a given flavour
$\nu_\alpha$: \hfil\break
\noindent $i)$ to measure the survival probability
$P(\nu_\alpha\to\nu_\alpha$) looking for an eventual reduction in the
$\nu_\alpha$ flux, i.e. the so called $disappearance$ experiments;\hfil\break
\noindent $ii)$ to try to directly observe in the detector the
interactions due to a different neutrino flavour, in the so called
$appearance$ experiments.

Usually a disappearance experiment has two detectors (or one that can
be moved), and the comparison of the measurements nearer to the source
with those farther away from it
are used to search for any possible reduction
in the $\nu_\alpha$ flux due to oscillations. This experimental set-up
implies that there is not only a minimum testable value of $\Delta
m^2$ (corresponding to oscillation lengths much larger than the
distance from the source
to the far detector), but also a maximum testable $\Delta m^2$,
since for large enough mass differences  oscillations would have
already been averaged in the nearby detector.

Appearance experiments look appealing because only a few
events above background are enough to establish an oscillation pattern,
while large statistics are required for a meaningful signal in a
disappearance experiment. There are however situations in which an
appearance experiment cannot be performed, like in reactor
experiments,
 where the CC interactions of the $\mu$ or $\tau$ neutrinos are
kinematically forbidden, or for the study of oscillations into sterile
species, that only show up in a disappearance experiment.

\subsection{Present situation}

In figure 3 we show the constraints obtained on $\Delta m^2$ and
$\sin^22\theta$, under the assumption of two flavour mixing, arising from
the unsuccessful searches of oscillations of the type
$\nu_e-\nu_\mu$ (figure $3a$) and
$\nu_\mu-\nu_\tau$ (figure $3b$).
 We have only depicted
the experiments giving the stronger constraints although several more
exist. The bound are slightly relaxed if mixing among three flavours
is allowed (Bl\"umer and Kleinknecht 1985).

For oscillations of e-neutrinos, the best bounds on $\Delta
m^2$ result from reactor experiments because of their small energies.
The $\bar\nu_e$ are here detected by the inverse $\beta$ reaction
$\bar\nu_ep\to e^+n$. Since these are disappearance experiments, the
resulting bounds apply to oscillations into any type of neutrinos, and
they actually represent the best experimental constraints on
$\nu_e$-$\nu_\tau$ oscillations.
At
the Goesgen reactor in Switzerland (Zacek \etal 1986)
the fluxes measured at three
distances (37.9, 45.9 and 64.7 m) were compared among themselves and
in addition they were compared to the flux expected from the knowledge
of the $\nu$ source and of the detector efficiency (this explains why
the excluded region extends also to large $\Delta m^2$ values). The
experiment of the Kurchatov group (Vidyakin \etal 1990),
in the site of three nuclear reactors in Moscow,
used one detector at 57 m from the first and second reactors and 231
m from the third. These comparatively large distances allowed them
to exclude down to $\Delta m^2\simeq 8\times 10^{-3}$ (for maximum
mixing) from the comparison of the rates measured with all the
different possible combinations of reactors on and off.

Although accelerator experiments have not reached such low values of
$\Delta m^2$, $\nu_\mu\leftrightarrow \nu_e$ appearance experiments
have achieved sensitivities to sin$^22\theta$ values much smaller
($\sim 3\times 10^{-3}$) than those reached at reactors ($\gsim
10^{-1}$). The best bounds on $\sin^22\theta$ are from the BNL
experiments E776 (Borodovsky \etal 1992) and E734 (Ahrluder \etal 1985)
and from the Serpukhov experiment SKAT (Ammosov \etal 1988). Also
shown in figure $3a$ are the constraints from the Los Alamos Meson
Physics Facility (LAMPF) (Durkin \etal 1988)
 and the bubble chamber BEBC at CERN (Angelini \etal 1986).

Regarding the $\nu_\mu-\nu_\tau$ oscillations, searches
looking for $\nu_\mu$ disappearance were performed at Fermilab CCFR
experiment and by the CDHS (Dydak \etal 1984) and CHARM (Bergsma \etal
1984)
collaborations at CERN. The
best constraints  on $\sin^22\theta_{\mu\tau}$ were obtained by the
Fermilab E531 experiment (Ushida \etal 1986) with an emulsion
detector sensitive to the
appearance of a $\tau$ track. Recently,  also the CHARM II
collaboration (Gruw\'e \etal 1993) was able to constrain
the $\nu_\tau$ appearance by means
of the study of the $\tau\to\pi\nu_\tau$ decay mode in their
fine-grained calorimeter. These results are shown in figure $3b$.

There is also an important astrophysical constraint on
$\nu_e-\nu_{\mu,\tau}$ oscillations, to be discussed in section 9,
arising from the non-disruption of $r$-processes in supernovae, that
excludes $\Delta  m^2$ values bigger than $\sim 4$ eV$^2$ for
$\sin^22\theta\gsim 10^{-5}$.

\subsection{Future prospects}

While one may say that in the past oscillation experiments have been
performed just looking blindly into the attainable ranges of $\Delta
m^2$ and $\sin^22\theta$, the situation looks quite different for the
future. In fact, there are two main physics issues that motivate the
forthcoming experiments.

First, there is the dark matter problem, that may be accounted for, at
least partially, if the heavier of the three neutrinos has a mass in
the range between a few eV and a few tens of eV
(see section 5). This is most
likely a $\nu_\tau$, and an analogy with the mixing in the quark
sector would suggest that its most significant mixing is with
$\nu_\mu$ (the nearest generation). Thus, this argument has strongly encouraged
$\nu_\mu\to\nu_\tau$ appearance searches sensitive to tiny mixing angles
for $\Delta m^2\gsim$ eV$^2$ (Harari 1989).
Two such experiments, CHORUS and NOMAD (di Lella 1993), are now running
at CERN. They are expected to take data in the period 1994--1995, and
to be sensitive
down to $\sin^22\theta_{\mu\tau}\sim 3\times 10^{-3}$.
CHORUS (de Jong \etal 1993) has an emulsion target with scintillating
fibers to
assist in the location of the $\tau$ tracks while NOMAD
selects kinematically the $\tau$ decays based on the
correlation between the resulting hadronic shower and the missing
$p_T$ carried by the produced $\nu_\tau$.
An approved  proposal (P803) related to the main
injector project at Fermilab, using 1 ton of emulsion target, will
extend the sensitivity to $\nu_\tau$ appearance down to
$\sin^22\theta\sim 6\times 10^{-5}$ by the end of the decade (Kodama 1990).

The second motivation comes from the observations performed with
atmospheric neutrinos that, as will be discussed in the section 8,
may suggest a mixing between $\nu_\mu$ and $\nu_\tau$ (or $\nu_e$) with
$\sin^22\theta\gsim 0.4$ and $\Delta m^2$ somewhere in between
$10^{-1}$ and $10^{-3}$ eV$^2$. The need to clarify this issue has
encouraged the realization of long-baseline experiments, where a
very intense $\nu$ beam is pointed towards a large far away detector
(Pantaleone 1990, Bernstein and Parke 1991).
Several proposals of this kind have been studied, including sending a
beam from: Fermilab main injector to
SOUDAN (baseline 710 km) (Allison
\etal 1991);
BNL to three new Cerenkov detectors at 1, 3 and 20 km (Mann and
Murtagh 1993); CERN SPS to Gran Sasso ICARUS detector (730 km)
(Revol 1993) or
to NESTOR (1680 km) (Resvanis 1993); CERN to Superkamiokande
(8750 km) (Rubbia 1993); KEK to  Superkamiokande (250 km) (Nishikawa 1992).

Since some of these experiments have detectors with the ability to
measure separately neutral and charged current events, a strategy
somehow in between the usual appearance and disappearance technics
can be used
to study oscillations among active neutrino species.
This strategy consists in using the NC events as
a measure of the total flux (including the flavours resulting from the
oscillations) and the CC events to measure the original flavour
content.
Another similar approach is to use the through-going muons produced in
the rock outside the detector as a measure of the $\nu_\mu$
content of the beam and the events contained inside the detector
(NC${}+{}$CC) as an indication of the total $\nu$ flux.

If the atmospheric $\nu$ problem were due to oscillations in the
$\nu_e-\nu_\mu$ channel (a possibility, however, that seems  now
excluded, see section 7),
these may also be studied using reactor $\bar\nu_e$
and large detectors at distances $\gsim 1$ km. There is a proposal to
use the San Onofre reactor in California with a 12 ton  detector at a
distance of 1 km from the reactor (phase I) and a kton detector at
a distance of 10-15 km (phase II) that will
cover all the relevant range. Another proposal is based on the use of
the Perry reactor and a kton detector at the IMB site that is 13 km
apart. Finally, there is a proposal to use the Chooz reactor in France
with a detector 1 km away.

Besides the proposed detectors just discussed, there are some ongoing
experiments that will provide results in the near future. The third
phase of the oscillation experiments at the Bugey reactor in France
may slightly improve the sensitivity of the Goesgen experiment. The
KARMEN detector at the spallation source ISIS in the Rutherford Appleton
Laboratory (Drexlin \etal 1990)
 is studying $\nu_\mu\to\nu_e$ and $\bar\nu_\mu\to\bar\nu_e$ appearance.
A very important improvement  in the $\nu_\mu\to\nu_e$ oscillations will be
achieved with the Liquid Scintillator Neutrino Detector (LSND) at
LAMPF (Whitehouse \etal 1991).

The approximate sensitivities expected in the experiments discussed in this
section are  shown in figures $4a$ ($\nu_e\to\nu_\mu$) and
$4b$ ($\nu_\mu\to
\nu_\tau$),
together with the region excluded at present (continuous lines).
Also indicated
are the regions relevant for the dark matter problem and for
oscillations of atmospheric neutrinos.

\section{Solar Neutrinos and Oscillations in Matter}\ssf
\subsection{The Solar Neutrino Problem}
The most important natural source of low-energy neutrinos ($E_\nu\leq
18$ MeV) reaching the earth, except at the moment of a galactic
supernova explosion, is our own sun. In fact, 98\% of the solar energy
production arises from the fusion chain that converts four protons
into a $^4$He nucleus, two positrons (by charge conservation) and two $\nu_e$
(by lepton number conservation), releasing about 27 MeV and with a
maximum $\nu$ energy of 0.42 MeV (for these so called $pp$ neutrinos)
\footnote{$^*$}{Since
the solar energy flux at the earth is $\simeq 8.5\times 10^{11}$ MeV
cm$^{-2}$ s$^{-1}$, the integrated $pp$ neutrino flux must be $\simeq
6\times 10^{10}$ cm$^{-2}$ s$^{-1}$.}. The proton-proton chain also
gives rise to the production of the heavier nuclei Li, Be and B,
although with negligible effects on the overall energy production. The
$^8$B decays give rise to an important continuum neutrino spectrum
extending up to $E\simeq 14$ MeV, while the $^7$Be produces, by
electron capture, two neutrino lines at energies 0.86 MeV (90\%) and
0.38 MeV (10\%). Another line at 1.44 MeV arises from $p+e+p\to d+\nu$ ($pep$)
while the $^3{\rm He}+p$ ($hep)$ reaction gives a small continuous spectrum up
to 18.7 MeV. Finally, in the CNO (carbon, nitrogen and
oxygen) chain, essentially a $^{12}$C catalyzes the fusion of 4
protons into $^4$He to produce the remaining 2\% of the solar energy
output. The decays of the $^{13}$O, $^{13}$N and $^{17}$F produced in
the intermediate steps lead to other continuous neutrino spectra. All
this is shown in figure 5 (Bahcall and Ulrich 1988).

Although the $pp$ neutrino flux is well established theoretically, the
other ones can depend significantly  on the solar modelling and the
input data used in it. For instance, the $^8$B and $^7$Be neutrino
fluxes are sensitive functions of the solar core temperature $T_c$
$$
\Phi (^8{\rm B})\propto T_c^{18}\ ,\ \ \ \Phi(^7{\rm Be})\propto
T_c^8\eqno(7.1)
$$
(while $\Phi(pp)\propto T_c^{-1.2}$, actually decreases with increasing
$T_c$ due to the conservation of the total solar luminosity).
 Hence, a change on the
radiative opacities (namely the cross-sections per gram) that determine the
energy
transport in the sun,
can affect these fluxes significantly. The opacities
themselves depend on the abundances of heavy elements and in
particular on that of Fe, that contributes sizably to the number of
free electrons in the plasma. Making different estimates of the latter
affect the $^8$B neutrino predictions by as much as
10\%\footnote{$^*$}{Another possible way of modifying the
predictions of $T_c$ is to consider non-standard solar models, e.g.
invoking large magnetic fields in the solar core, significant
turbulence, etc.}. Another source of uncertainties comes from the
nuclear cross sections involved. In particular, the $^7$Be($p,
\gamma$)$^8$B cross section is known at the energies that are relevant in the
sun  from an extrapolation of the values measured in the
laboratory at higher energies, and this procedure may also cause
discrepancies among different $^8$B neutrino
flux estimates (see e.g. Dar and Shaviv 1994).

Solar neutrinos have been observed by four experiments up to now, with
the thresholds indicated in figure 5. The pioneering Cl radiochemical
experiment (Davis \etal 1990), based on the reaction $^{37}{\rm
Cl}+\nu_e\to ^{37}{\rm Ar}+e$ ($E^{thr}_\nu=0.81$ MeV), has detected
solar neutrinos for more than 20 years. It is sensitive to the $^8$B
as well as the $^7$Be, CNO and $pep$ neutrinos. The Kamiokande
water Cerenkov detector (Hirata \etal 1991a-b) has been able to observe
in real time the $\nu e$ scattering, having an analysis threshold of 7
MeV that makes this device sensitive only to the $^8$B neutrinos. The
directional information of the events makes this detector the first
`neutrino telescope' able to obtain a `picture' of an extraterrestrial
neutrino source (the sun). Finally, there are two radiochemical
gallium experiments, SAGE at Baksan (Abazov \etal 1991) and
GALLEX at Gran Sasso (Anselmann \etal 1994). They use the reaction
$^{71}{\rm Ga+\nu_e\to
^{71}Ge+e}$, whose low threshold ($E^{thr}_\nu$=0.23 MeV) makes them
sensitive to
the $pp$ neutrinos in addition to the more energetic $^7$Be, $^8$B and
CNO ones. In table 4 we show the rates measured in these four
experiments, as well as the theoretical predictions based on the
standard solar models of Bahcall and Pinsonneault (1992) and
Turck-Chi\`eze and Lopes (1993).

\bigskip\tabcaption{Solar neutrino measurements and theoretical
expectations within the Standard Solar Model of Bahcall and
Pinsonneault (1992), SSM-BP, and Turck-Chi\`eze and Lopes (1993)
SSM-TCL. } \ialign{#\hfil&&\hglue 2pc plus2pc minus1pc#\hfil\cr \br
Experiment& Measurement&SSM-BP&SSM-TCL\cr \mr $^{37}$Cl
[SNU]&$2.23\pm0.23 \;^{\rm a}$ &$8\pm 1$  & $6.4\pm 1.4$\cr Kamioka
$\left({\rm Observed\over SSM-BP}\right)$& $0.50\pm 0.04\pm
0.06\;^{\rm b}$&$1\pm 0.14$&$0.77\pm 0.17$ \cr GALLEX [SNU]&$79\pm
10\pm 6\;^{\rm c}$& $131.5 ^{+ 7}_{-6}$& $122.5\pm 7$\cr SAGE
[SNU]&$73^{+18}_{-16} {}^{+5}_{-7}\;^{\rm d}$& \hfil ''& \hfil ''\cr \br}
\tabnote{$^{\rm a}$ Davis 1992, $^{\rm b}$ Suzuki 1993, $^{\rm c}$
Anselmann \etal 1994, $^{\rm d}$ Abdurashitov \etal 1994.}  \bigskip

The most intriguing feature of these four measurements is that they
all give values significantly below the theoretical predictions. This
constitutes the so called solar neutrino problem. The explanation of
this deficit calls either for modifications in the solar models or for
new properties of the neutrinos, such as mixings or magnetic moments,
that could convert the electron neutrinos into less detectable species
in their journey from the center of the sun to the earth
(radiochemical experiments are only sensitive to $\nu_e$, while the
$\nu -e$ cross section relevant for Cerenkov detection is 7 times
smaller for $\nu_{\mu,\tau}$ than for $\nu_e$, and vanishes for
sterile species). Also important is that the observed Ga rates are not
less than the $\sim 70$ SNU expected (reliably)  from the $pp$
neutrinos alone, since this would otherwise exclude any explanation
not invoking new neutrino properties. However, if all experimental
results are taken at face value, one should note that an explanation
based on a reduction of the core temperature is not possible, since it
would reduce much more the $^8$B flux than the $^7$Be one, contrarily
to what results from the comparison of Kamiokande with Cl data.

Turning to the explanation in terms of new neutrino properties, the
simplest one is to invoke oscillations in vacuum of the $\nu_e$
(Gribov and Pontecorvo 1967). In view of the typical $\nu$ energies
($\sim {\rm MeV}$) and the earth-sun distance (1 A.U.${}\simeq
1.5\times 10^{11}$ m), values of $\Delta m^2\gsim 10^{-11}$ eV$^2$ are
required for a significant $\nu_e$ suppression (see eq. (6.10) and
(6.11)). However, for $\Delta m^2\gsim 10^{-10}$ eV$^2$ oscillations are
completely averaged. In this case, the reduction factor results the same
 in all experiments, and it is larger
than 1/2 for the mixing among two flavours. It is only for $10^{-11}\ {\rm
eV^2}<\Delta m^2<10^{-10}$ eV$^2$ that the oscillation length is of
the order of 1 A.U. so that neutrinos of different energies are
converted in different amounts. With this `just so' oscillations
(Barger, Phillips and Whisnant 1981, Glashow and Krauss 1987), that
also require sizeable mixing angles, it becomes then possible to fit
simultaneously all three types of experiments (Barger \etal
1992, Krastev and Petcov 1992).

A more beautiful possibility is that of explaining the solar neutrino
problem using the enhancement of the oscillations induced by the
effects of the solar medium, that we now turn to discuss.
\subsection{Neutrino Oscillations in Matter}
 When neutrinos propagate
through matter, a crucial differentiation among the neutrino flavours
appears due to the fact that $\nu_e$ (unlike $\nu_\mu$ or $\nu_\tau$),
have
charged current interactions due to $W$ exchange with the electrons.
As noted by Wolfenstein (1978), these interactions can affect the
pattern of neutrino oscillations and the effect can be significantly
enhanced (Mikheyev and Smirnov 1985) when a resonance crossing takes
place, in the so called MSW effect (Bethe 1986, Rosen and Gelb 1986,
Parke 1986, Haxton 1986).

The charged current interaction can be written  as
$H_{CC}=(4G_F/ \sqrt{2})J^{+ {\mu}}_{\ell} J^{+ \dagger}_{\ell \mu}$
(see (2.3) and (3.8)) for energies much lower
 than the $W$ boson mass.
Because there are electrons  in normal matter
but not muons or taus, we are interested only in the e $\nu_e$
 interactions.
After a Fierz rearrangement to write them as a product of charged
conserving currents, we get
$$
H_{CC}= {G_F\over \sqrt{2}}\bar e\gamma^\mu(1-\gamma_5)e\bar\nu_e
\gamma_\mu(1-\gamma_5)\nu_e ~~.\eqno(7.2)
$$
For an unpolarized medium at rest the only non-vanishing
component of the electron current is the temporal part of the vector
current, that is nothing but the electron density $N_e$. Hence,  an
`effective' potential energy $\langle H_{CC}\rangle \equiv \langle
e\nu \vert H_{CC}\vert e\nu \rangle\simeq \sqrt{2}G_FN_e$ will affect
the propagation phase of the $\nu_e$\footnote{$^*$}{This can
also be thought of as inducing an index of refraction for $\nu_e$.}.
Although the neutral currents are also present, they can be omitted,
since they affect in equal amounts all active flavours giving no net
effect to the neutrino oscillations (they are relevant however for
oscillations into sterile species).

For definiteness we will consider the case of mixing between two
active neutrinos, $\nu_e$ and $\nu_\alpha$ ($\alpha=\mu$ or $\tau$)
$$
\pmatrix{\nu_e\cr\nu_\alpha}=
R_\theta\pmatrix{\nu_1\cr\nu_2}\ ,\ \ \ \ {\rm with}\
\ \ R_\theta=\pmatrix{
c\theta&s\theta\cr -s\theta&c\theta\cr}\eqno(7.3)
$$
($s\theta\equiv \sin \theta$, etc.). The evolution of the flavour
content along the neutrino path can be simply obtained from the
equation
$$
i{d\over dx}\pmatrix{\nu_e\cr\nu_\alpha}={1\over 2E}{\bf M}^2
\pmatrix{\nu_e\cr\nu_\alpha}.\eqno(7.4)$$
The matrix ${\bf M}^2$ is
$${\bf M}^2={1\over 2}\left[ R_\theta\pmatrix{ -\Delta m^2&0\cr
0&\Delta m^2\cr}R_\theta^T+2E\pmatrix{\langle H_{CC}\rangle
&0\cr 0&-\langle H_{CC}\rangle
\cr}\right] \eqno(7.5)
$$
The first term in the r.h.s. is the usual one, already appearing in
vacuum oscillations, and the second term arises from the
$\nu_e-e$ coherent forward scattering. We have subtracted a piece
proportional to the identity matrix (what changes just a common phase)
to put ${\bf M}^2$ in a more symmetric form.

It proves very convenient to define the matter eigenstates as
$$
\pmatrix{\nu_m^1\cr\nu_m^2}=R^T_{\theta_m}
\pmatrix{\nu_e\cr\nu_\alpha}\;,\eqno(7.6)
$$
 where $R_{\theta_m}$ diagonalizes the matrix ${\bf M}^2$, i.e.
$$
R^T_{\theta_m}{\bf M}^2R_{\theta_m}={1\over
2}\pmatrix{-\Delta_m&0\cr 0& \Delta_m\cr}.\eqno(7.7)
$$
Here $\Delta_m=\Delta m^2\sqrt{(a-c2\theta )^2+s^22\theta}$ with
$a=2E\langle H_{CC}\rangle/\Delta m^2$. The matter mixing angle
$\theta_m$ entering in $R_{\theta_m}$ satisfies
$$
s^22\theta_m={s^22\theta\over (c2\theta-a)^2+s^22\theta}.\eqno(7.8)
$$
Even if the vacuum mixing is very small, there will then be maximum
mixing in matter ($\theta_m=\pi/4$) in the `resonance region'
corresponding to an electron density such that $a=c2\theta$. In matter
with varying density, the width of the resonance corresponds to the
densities for which $\vert a-c2\theta \vert = \vert s2\theta\vert$. It
is important that if we were considering antineutrinos, the sign of
$\langle H_{CC}\rangle$ would be reversed and no resonance would
appear (for $\Delta m^2>0$). Hence, the presence of electrons in the
medium (and no positrons) results in very different oscillation
patterns for neutrinos and antineutrinos.

The $\nu_m$ evolution is determined by
$$
i{d\over dx}\pmatrix{\nu^1_m\cr \nu^2_m\cr}=\pmatrix{-\Delta_m
/4E&-i{d\over dx}\theta_m\cr i{d\over dx}\theta_m&\Delta_m/4E\cr}
\pmatrix{\nu^1_m\cr \nu^2_m\cr}~~,\eqno(7.9)
$$
with
$${d\theta_m\over dx}={1\over 2}{s2\theta\over (a-c2\theta
)^2+s^22\theta}{da\over dx}~~.\eqno(7.10)
$$

We see that the matter states are the propagation eigenstates in
a medium with constant density (d$\theta_m/$d$x\simeq 0$), since no
transitions between them can occur in this case. When the density of
the medium varies along the neutrino path, 
 transitions between matter mass eigenstates
are induced by a nonzero $d\theta_m/dx$. This is the
case in the sun, where the densities fall nearly exponentially in
the radial direction. The transitions between matter mass eigenstates
are usually
negligible unless the neutrinos are near the resonance layer, for
which the diagonal elements in (7.9) are minimum and $d\theta_m/dx$
is enhanced. If $P=\vert\langle\nu^1_m\vert\nu^2_m\rangle\vert^2$ is
the probability of $\nu^1_m\to\nu^2_m$ conversion in the resonance
crossing, the averaged probability to detect an electron neutrino that
has crossed  a resonance is
$$P_{\nu_e\nu_e}={1\over 2}+\left( {1\over 2}-P\right)c2\theta
\times c2\theta_m~~ ,\eqno(7.11)$$
where $\theta_m$ is the matter mixing angle corresponding to the point
where the $\nu_e$ was produced, while $\theta$ is the vacuum
angle\footnote{$^*$}{This is obtained by projecting at the production point
the $\nu_e$ into matter eigenstates, following then the outward
propagation in the sun including an eventual jump between matter
eigenstates at the resonance crossing, and using the fact that the
flavour content of those states changes smoothly with electron density
until they match the vacuum mass eigenstates when they leave the
sun.}.

Under the assumption that the electron density varies linearly in the
resonance layer, the probability of level crossing at resonance is
found to be
$$P=e^{-\pi\gamma/2},\eqno(7.12)
$$
where the adiabaticity parameter $\gamma$ is
$$
\gamma\equiv {\Delta_m\over 4E\vert d\theta_m/dx\vert}
={\Delta m^2\over 2E}{s^22\theta\over c2\theta}{1\over \vert d{\rm
ln}N_e/dx\vert_r}\eqno(7.13)
$$
(the sub-index $r$ stands for the resonance value).

In the adiabatic case, i.e. $\gamma\gg 1$, the off-diagonal terms in
(7.9) can be neglected even at resonance, so that $P=0$ and the
probability of having a $\nu_e$ after adiabatic conversion is
$P_{\nu_e\nu_e}=(1+c2\theta \times c2\theta_m)/2$. What happens is
that if the production point is at densities larger than the resonance
value, a $\nu_e$ is mainly the $\nu^2_m$ eigenstate
($\theta_m\simeq \pi/2$) and remains so during the adiabatic
propagation. Thus, it becomes essentially $\nu_\alpha$ when it comes out from
the sun. An almost complete flavour conversion results.
The resonance condition $a=c2\theta$ can be written as
$$
\Delta m^2 c2\theta \simeq    {E\over 10~\rm MeV}  {\rho\over
\rho_0}10^{-4}{\rm eV^2}~,\eqno(7.14)
$$
where $\rho_0\simeq 150$ gr/cm$^3$ is the central density of the sun
and we have used $N_e=Y_e\rho/m_p$ (with
$Y_e=N_e/(N_n+N_p)\sim0.7-0.8$ in the sun). We then see that the solar
neutrinos will meet a resonance only for $\Delta m^2\lesssim 10^{-4}$
eV$^2$.

It could be that in the resonance crossing the propagation
becomes non-adiabatic (namely, $\gamma\lesssim 1$),
making the flavour conversion
less efficient. This happens when the vacuum mixing angle is small so
that the resonance becomes very narrow. In the extreme non-adiabatic
case ($\gamma\ll 1$), corresponding to\footnote{$^*$}{For $r>0.1
R_\odot$, $N_e\propto {\rm exp}(-10.5 r/R_\odot)$.}
$$
\Delta m^2s^22\theta\ll 2E\left\vert{1\over N_e}{dN_e\over
dr}\right\vert_{r}
\simeq 6\times 10^{-8}\left({E\over 10~{\rm MeV}}\right){\rm eV}^2~~,
\eqno(7.15)
$$
the $\nu$ flavour content is essentially unchanged in the resonance crossing
and no enhancement in $P(\nu_e\to\nu_\alpha)$ is found.

The survival
probabilities depend strongly on the neutrino energy, as is shown in
figure 6, in which the effect of the conditions of crossing a resonance
and that the transition not be extremely non-adiabatic are clearly
seen. For instance, for $s2\theta=0.1$ as in the figure, only the
high energy neutrinos ($E\gsim 10$ MeV) would be suppressed if $\Delta
m^2\sim 10^{-4}$ eV$^2$ (adiabatic case) while instead for $\Delta
m^2\sim 10^{-6}$ eV$^2$ it would be mainly the $pp$ neutrinos and not
the high energy ones to be converted (non-adiabatic case).

An agreement with the deficit observed at the Davis experiment, with
adiabatic neutrino evolution, is obtained in two regions of the
$\Delta m^2-s^22\theta$ plane. The first is
for  $\Delta m^2\simeq
10^{-4}$ eV$^2$, if the resonant layer is not too narrow, i.e.
$s2\theta \gsim 10^{-2}$. The second is  the so-called ``large angle
solution", $s^22\theta\sim 0.7$ and $10^{-4}$ eV$^2>\Delta m^2>10^{-8}$
eV$^2$, for which the MSW oscillations start approaching the
large-mixing vacuum oscillations. A non-adiabatic solution exists for
$\Delta m^2s^22\theta\simeq 10^{-7.5}$ eV$^2$, closing a
`triangle' in the $\Delta m^2-s^22\theta$ plane. Due to the different
energy dependence of the reduction factor in the three regimes,
experiments with other thresholds, or capable of measuring the neutrino
spectra, can distinguish among the three solutions. The results of the
Kamiokande detector in fact disfavor the $\Delta m^2\simeq 10^{-4}$
eV$^2$  adiabatic solution, that would deplete too much the high
energy $^8$B neutrinos,  while the gallium experiments disfavor
$\Delta m^2$ values below $10^{-6}$ eV$^2$, that would result into an
excessive conversion of $pp$ neutrinos. The combination of all the
experiments, then, points out  to two possible solutions, a non-adiabatic
one with $s^22\theta\sim 8\times 10^{-3}$ or a large angle solution
with $s^22\theta\sim 0.7$, both with $\Delta m^2\sim 10^{-5}$ eV$^2$,
as is shown in figure 7, taken from Hata and Langacker (1993b) (see
also Hata and Langacker 1994).

For $\nu_e$ oscillations into sterile species $\nu_s$, the solution is
slightly different because the $\nu_s$ do not have NC interactions.
Hence, the matter effects in the sun as well as the cross
section for detection in Kamiokande are affected. As a result
 (Hata and Langacker 1994), the
large angle solution (that would anyhow also conflict with
nucleosynthesis (Shi \etal
1993)) is absent for $\nu_e-\nu_s$ oscillations
and the small angle solution is slightly shifted.

There are two interesting cases in which the neutrino propagation
through the earth may be affected by matter effects. In these cases
no MSW effect, namely the crossing of a resonance layer,
really takes place since the density varies only
slightly, from 3 to 5.5 gr/cm$^3$ in the mantle and from 10 to 13
gr/cm$^3$ in the core.
Anyhow, oscillations of neutrinos that are
resonant can be significantly enhanced since the matter mixing becomes
maximal.
The first is the case  of the accelerator neutrinos in the planned
long-baseline experiments and of the atmospheric neutrinos that
pass through the earth before
reaching the detectors. They would be resonant in the mantle if $\Delta
m^2\simeq 3\times 10^{-3}$ eV$^2$ for energies $\simeq 10$ GeV. Hence,
matter effects can extend the sensitivity to smaller
$s^22\theta_{e\alpha}$ values in this mass range (Carlson 1986,
Akhmedov, Lipari and Lusignoli 1993, Fiorentini and Ricci 1993).
However, very long-baselines are required to see any effect because
the wavelength of the oscillations in matter, $\lambda_m\equiv 4\pi
E/\Delta_m$, becomes
$$
\lambda_m\simeq {3\times 10^6 \rm m\over tg 2\theta (\rho/5\
gr\;cm^{-3})}\eqno(7.16)
$$
in the resonance (taking $Y_e=0.5$). This is of the order of or larger
than the radius of the earth.
The second case in which matter effects in the earth are relevant
is that of the $^8$B neutrinos observed by Kamiokande
($E\simeq 10$ MeV). The resonance in the earth in this case is achieved
for $\Delta m^2\simeq
3\times 10^{-6}$ eV$^2$, and can give rise  to a day--night effect in this
mass range if $s
2\theta$ is not too small (if $\lambda_m$ is not too large, Baltz and Weneser
1987,
Bouchez \etal 1986). The non-observation
of this effect (Hirata \etal 1991a) leads to the exclusion
of the region shown in figure 7 (see also Hata and Langacker 1993a
and references therein).

\subsection{Other Solutions}
 Many alternative solutions have been
proposed to explain the solar neutrino deficit. One of them is based
 on the interaction of the neutrinos with the magnetic
fields of the sun, that, in the convective, zone increase in strength in
periods of high activity (Cisneros 1971, Voloshin, Vysotsky
and Okun 1986).
 The resulting effect would be to flip the
neutrino chirality to produce either a sterile state (with Dirac type
moments) or just a different active flavour (with Majorana type
$\vert\Delta L\vert=2$ transition moments).
This solution became
especially attractive to account for an anticorrelation with the
solar activity of the rates, observed in the Davis experiment but not
confirmed by Kamiokande.

The required
magnetic moments  are $\mu\simeq 10^{-11}\mu_B$ ($\mu_B\equiv e/2m_e$ is
the Bohr magneton). Although these values are not in conflict with
direct experimental bounds, $\mu< 1.08\times 10^{-9}\mu_B$ (Review of
Particle Properties 1994),
they would result in excessive stellar cooling
(Raffelt 1990), or conflict
with SN1987A in the case of Dirac type
transitions (see e.g. Barbieri and Mohapatra 1988), as we will see in
section 9.

 Another
difficulty is that it is not easy to find models providing magnetic
moments of the required size keeping the neutrino masses small, since
a loop diagram contributing to $\mu$ gives a mass term when the photon
is removed, unless some particular symmetries are invoked (Voloshin
1988). Furthermore, transition type spin precession is quenched when
the mass difference between states is $\Delta m^2\gsim 10^{-7}$
eV$^2$. However,  matter effects could
compensate the mass terms and lead to an enhanced `spin-flavour'
precession in a resonance crossing (Akhmedov 1988, Lim and Marciano
1988). Recent analysis of the experimental data in this context can be
found in Krastev (1993), Akhmedov, Lanza and Petcov (1993) and Pulido
(1993).

The usual MSW picture is also modified in the presence of other
non-standard properties in the neutrino sector. For instance, if
mixing with heavy singlet neutrino species occurs, the $Z$-mediated
interactions can become non-universal in flavour. Then, it may even
become possible to have a resonance for massless neutrinos, by
compensating the effect of the charged-current
with the neutral-current $\nu_e$-interactions (Valle 1987). However, due to
experimental constraints on the weak boson couplings, this may only be
achievable in media such as neutron stars, but not in the sun.

More
interesting for the solar neutrino problem are new flavour
changing neutrino interactions that appear in supersymmetric or GUT
extensions of the standard model. They may induce neutrino oscillations even
for vanishingly small vacuum mixings (Roulet 1991), and their resonant
amplification can result in sizeable effects,  even for small
(experimentally allowed) couplings. In this type of scenarios, the
resonance conversion in the massless case may take place in the
presence of large new diagonal couplings of $\nu_\tau$ (Guzzo \etal
1991) (for a general discussion see Barger \etal (1991)).
Finally, the possibility that non-universal
gravitational interactions (or new flavour-dependent long range
forces) affect the solar neutrino oscillations (Halprin and Leung 1991),
and the possibility of matter induced neutrino
 decay into another neutrino and a Majoron in the sun (see e.g. Berezhiani
 and Rossi 1993) have also been considered.

\subsection{Future Prospects}
Several new solar  neutrino detectors
are in construction or under study at present. The Sudbury Neutrino
Observatory in Canada (Aardsma \etal 1987), using 1 kton of heavy
water, will be able to detect, besides the scattering off electrons,
the reactions $\nu_xd\to pn\nu_x$ and $\nu_e d\to ppe$.
 By observing
the $n$ capture, the deuterium
disintegration ($E^{th}_\nu=2.2$ MeV) will allow to measure the NC rates
 due to all active neutrino flavours.
The inverse $\beta$ decay will be useful to reconstruct the
neutrino spectrum accurately above $\simeq 5$ MeV. The comparison of the
NC and CC rates can give a clear signal of oscillations, independently
of any solar model input. The spectral shape can help to
distinguish among different solutions. This experiment (starting in
1995) and the Superkamiokande water Cerenkov (Totsuka 1990), with 22
kton fiducial mass (starting in 1996), will be able to study with
large statistics the $^8$B neutrino fluxes.

The study of the Be line is the main objective of the Borexino
experiment (Arpesella \etal 1992). It consists of 100 ton fiducial mass
of  ultrapure
liquid scintillator with extremely low radioactive background, in order
to be able to detect the $\nu e$ scattering in the window $0.25{\rm\
MeV}<E<0.8$ MeV (90\% of the signal coming from $^7$Be). The events of
larger energies will also permit to study the $^8$B neutrino fluxes. A
signal that can be useful is the annual modulation of the Be flux due
to the eccentricity of the earth orbit. For `just so'
oscillations the modulation will differ from the simple $r^{-2}$ behaviour. A
counting test facility with 2 ton fiducial mass is now being built at
Gran Sasso. Another experiment, Icarus, a liquid argon TPC to be installed
at Gran Sasso
(3 modules of 5 kton each), could allow to study the $^8$B neutrinos
measuring separately CC and NC events. A 3 ton prototype built at
CERN has performed satisfactorily (Revol 1993). Two experiments to
detect $pp$ neutrinos are under study at present, both helium-based.
Heron (Lanou, Maris and Seidel 1987) would
detect the rotons produced by neutrino interactions in superfluid helium and
Hellaz (Arzarello \etal 1994) would consist of a helium gas TPC at high
pressure.

 These experiments, together
with increased statistics in the gallium experiments, should firmly
establish the nature of the solution to the solar neutrino problem,
 within the next ten years.

\section{Atmospheric neutrinos}

The cosmic ray nuclei that impinge on the atmosphere are stopped in
the upper layers by collisions with the air nuclei. These interactions
can produce pions and kaons that in turn decay giving rise to
neutrinos (e.g. $\pi^+\to\mu^++\nu_\mu$ and similarly for $\pi^-$ and
$K^\pm$). Also the muons so produced decay after loosing some energy
and produce more neutrinos via $\mu\to\nu_\mu+e+\bar\nu_e$ (muons
with energies below a few GeV decay before reaching the ground).
The resulting neutrino fluxes on earth are quite significant, and
were observed many years ago (Krishnaswamy 1971, Reines 1971) by
looking at horizontal and upgoing muons
produced in the rock surrounding deep underground detectors (downgoing
muons are dominated by those directly produced high in the atmosphere
by very energetic meson decays). In recent years, it became also
possible to observe contained events ($E_\nu\leq 2$ GeV), produced
directly by neutrino interactions inside large volume detectors (IMB,
Kamiokande, Fr\'ejus, NUSEX, SOUDAN).

Atmospheric neutrinos are now an important subject for all underground
neutrino detectors. They also constitute a background for proton decay
searches, and due to the long `baseline' distances involved ($10-10^4$
km), they are of particular interest for $\nu$ oscillation studies.

Several theoretical computations of the resulting $\nu$ fluxes have
been performed, both for the low energy neutrinos that lead to
contained events (Barr, Gaisser and Stanev 1989, Lee and Koh 1990,
Honda \etal 1990, Kawasaki and Mizuta 1991, Bugaev and Naumov 1989)
 and for those of higher energies that produce
throughgoing muons (Volkova 1980, Mitsui, Minorikawa and Komori 1986,
Butkevich, Dedenko and Zheleznykh 1989).
The different computations typically agree on the absolute values of the
$\nu_e+\bar\nu_e$ and $\nu_\mu+\bar\nu_\mu$ fluxes
only within 20-30\%. The reason is that they are
affected by the poor knowledge of the primary cosmic ray spectrum and
composition, the uncertainties in the meson production cross section,
the use of different calculational methods, etc. Also the resulting
absolute values of muon fluxes, that could in principle be used to
normalize the calculations, are poorly known experimentally.
Hence, the measurements of the absolute neutrino fluxes can
hardly give reliable information on neutrino oscillations.

This difficulty can be overcome by considering the ratio
$R=(\nu_\mu+\bar\nu_\mu)/(\nu_e+\bar\nu_e)$ of contained events of the
$\mu$ and e-type. Most uncertainties cancel in this ratio and, in
fact, the different calculations agree to within better than 5\% on the
value of this quantity. One naively expects $R=2$ at low energies, as
would result from the meson decay chain presented above.
Detailed evaluations of $R$ lead to
a value larger than 2 only by a few percent,  for $E_\nu\lesssim 1.5$ GeV.
When this ratio was measured some years ago by the Kamiokande
collaboration (Hirata \etal 1988),
a significant discrepancy with the expected value was
found. This discrepancy was later confirmed by the other
Cerenkov detector IMB
(Casper \etal 1991, Becker-Szendy \etal 1992b),
and recently also by the SOUDAN II tracking calorimeter (Kafka 1993).
No anomaly was found with the other detectors of this type,
Fr\'ejus (Berger \etal 1989) and NUSEX (Aglietta \etal 1989),
that are, however, statistically less significant
than the Cerenkov detectors. These experiments report the ratio of ratios
$$
R(\mu/e)\equiv{R_{obs}\over R_{MC}}~~, \eqno(8.1)
$$
where $R_{obs}$ is the observed ratio of $\mu$ to e-type events and
$R_{MC}$ the Monte Carlo results, that take into account also
experimental cuts and detector details. The values for this quantity
are given in the table 5.

\bigskip
\tabcaption{Ratio of observed and Monte Carlo ratios of $\mu$-type to
e-type contained neutrino events.}

\ialign{#\hfil&&\hglue 2pc plus2pc minus1pc#\hfil\cr
\br
Experiment& $R(\mu/e)=R_{obs}/R_{MC}$&reference\cr
\mr
Kamiokande&$0.60^{+0.07}_{-0.06}\pm 0.05^{\rm a}$&(Hirata \etal 1992)\cr
 & $0.67^{+0.08}_{-0.07}\pm 0.07^{\rm b}$&(Fukuda \etal 1994)\cr
IMB& $0.54\pm 0.05\pm 0.12$&(Becker-Szendy 1992b)\cr
Nusex&$0.99^{+0.35}_{-0.25}$&(Aglietta \etal 1989)\cr
Fr\'ejus&$1.06\pm 0.18\pm 0.15^{\rm c}$&(Berger \etal 1989)\cr
&$0.87\pm 0.16\pm 0.08^{\rm d}$\cr
Soudan II&$0.69\pm 0.19\pm 0.09$&(Kafka 1993)\cr
\br}
\tabnote{$^{\rm a}\;$Sub-GeV; $^{\rm b}\;$Multi-GeV
$^{\rm c}\;$Vertex contained; $^{\rm d}\;$Fully contained.}
\bigskip

In figure 8 we show the momentum distribution of e-type
($8a$) and $\mu$-type ($8b$) events
in the Kamiokande data (Beier \etal 1992,
Hirata \etal 1992), compared to a Monte Carlo using the
neutrino fluxes from Lee and Koh (1990). Similar results were also obtained
by IMB (Becker-Szendy \etal 1992)).

The deficit of $\mu$-type neutrinos (or excess of e-type, or
both combined), found in some of the detectors, constitutes the so
called atmospheric neutrino problem. If taken at face value, it would
suggest neutrino oscillations  of the type $\nu_\mu-\nu_\tau$,
$\nu_\mu-\nu_e$ or $\nu_\mu-\nu_s$.
 These would require large mixing angles (sin$^22\theta\gsim
0.4$) and, due to the typical $\nu$ energies (GeV) and large baselines
($\lesssim 12000$ km) they would require $\Delta m^2\gsim 10^{-3}$ eV$^2$.

Another feature of the data is that there is no evidence of zenith angle
dependence on $R(\mu/e$) at low energies, $E_\nu\leq$ GeV (Hirata \etal 1992),
suggesting
that oscillations should have already settled at baselines $\simeq 30$ km
for these energies. However, a significant angular dependence has been
observed in the recent analysis of multi-GeV events by Kamiokande
(Fukuda \etal 1994) which is consistent with the sub-GeV data in the
scenario of neutrino oscillations, because of the increase of the
oscillation length with energy. The
values of $\Delta m^2$ and sin$^22\theta$ consistent with the Kamiokande
results are shown in figure 9 for the $\nu_\mu-\nu_e$ channel ($9a$) and
$\nu_\mu-\nu_\tau$ channel ($9b$). Also shown are the bounds from
reactors and accelerators and from the negative Fr\'ejus
result, that for the $\nu_\mu-\nu_e$ option exclude the parameter
space allowed at present by Kamiokande.
 An explanation in
terms of $\nu_\mu-\nu_s$ oscillations would be in conflict with constraints
from primordial nucleosynthesis on the number of effective neutrino species
at $T\simeq$ MeV mentioned in section 5, $N_\nu \lesssim 3.3$ (Walker 1993),
since for the required values of $\Delta m^2$ and $\sin^2 2\theta$
 the sterile species $\nu_s$ would be brought into
equilibrium by the mixing itself (Dolgov 1981, Barbieri and Dolgov
1991, Kainulainen 1990, Enqvist \etal 1992, Shi \etal 1993)
and hence lead to $N_\nu\simeq 4$.

Since neutrino oscillations depend strongly on the neutrino energy, another
way to test the oscillation solution to the atmospheric  neutrino problem
is to compare the upward going
muons stopping inside the detector (typical energies $E_\nu\sim 10$ GeV)
with those going through the detector ($E_\nu\sim 100$ GeV).
This method is independent of the overall normalization of the neutrino
fluxes,  but depends on its spectral shape. The
analysis of the IMB data (Becker-Szendy \etal 1992a, Frati \etal 1993)
shows no evidence of an anomaly, what
excludes the region  with $\Delta m^2\sim
10^{-3}-10^{-2}$ eV$^2$ (since $d\simeq 10^4$ km) and large mixing angles
shown in figure $9b$.

Also the IMB-1 collaboration (Bionta \etal 1988) used the non observation of
an anomaly between the upgoing and downgoing $\nu_\mu$ ($E_\nu\sim$ GeV), to
constrain $\nu_\mu-\nu_\tau$ oscillations with $\Delta m^2$ around $10^{-4}$
eV$^2$ (IMB-1 curve in figure $9b$), for which only the $\nu_\mu$ from the
lower hemisphere would have oscillated sizably (and not those from above).

Finally, there have been attempts to directly compare the upward-going
muons with the theoretical expectations to further constrain the mixing
parameters (these fluxes are consistent with no oscillations
(Becker-Szendy \etal 1992a, Mori \etal 1991, Boliev \etal 1991)).
The uncertainties in the absolute neutrino fluxes and on the quark
structure functions entering the inelastic cross section for muon
production, make however those analysis less reliable (Frati \etal
1993). In conclusion, the atmospheric neutrino anomaly could be
explained in terms of oscillations between $\nu_\mu$ and $\nu_\tau$,
with $\Delta m^2\simeq 10^{-2}$ and $\sin^22\theta\simeq 0.5$.

For the future, the results of the Soudan II experiment will
significantly improve since their preliminary data (based on the first
1.5 kt y) still have large
statistical uncertainties. This is important because none of the
previous tracking calorimeters (Fr\'ejus and Nusex) observed an
anomaly in the atmospheric neutrino data. Also relevant in this
respect is the planned exposure of a Cerenkov detector to e and
$\mu$ beams at KEK, to check for possible systematic effects due to
the $\mu/e$ identification capability of these devices, that was never
tested before. In addition, the new detectors
 Superkamiokande, SNO, Dumand, Amanda and NESTOR, that are under
construction, will provide much more information on this issue, in
particular from the detailed study of the zenith angle dependence of
the $\nu$ fluxes (Doncheski, Halzen and Stelzer 1992)
 as well as the other methods mentioned before.
As we discussed in the section 6, also the proposed
long-baseline oscillation experiments can help to reliably settle the
issue of whether the atmospheric neutrinos oscillate or not.

\section{Neutrinos from Supernovae and Other Stars}

Neutrinos play an important role in stellar evolution.  Neutrinos produced in
the interior of stars can just stream out, except at the large densities of
collapsing supernovae, where they are trapped for some time. While in ordinary
main-sequence stars, such as our sun, the photon luminosity is larger than the
neutrino luminosity, for stars whose central temperature is higher than 10 keV
(O and Si burning stars and beyond) neutrino emission is the main form of
energy loss. Thus, stars can test any non-standard neutrino property that
increases the neutrino energy loss rate from stars or the neutrino energy
transfer inside stars, thus changing their standard course of evolution (for a
review see Raffelt 1990a). For example, an increase of energy loss accelerates
the evolution of stars, shortening their lifetimes.

Nowhere the effect of neutrino cooling is more dramatic than in a supernova
explosion. The observation of the neutrino burst from the supernova SN1987A
provided a confirmation of the theoretical understanding of the stages of the
gravitational collapse of a type II (core collapse) supernova (see e.g.
Burrows 1990 or Hillebrandt and Hoflich 1989). Most of the energy liberated in
the collapse of the iron core of the progenitor star into a neutron star, i.e.
$\simeq 3 \times 10^{53}$ ergs, is emitted in neutrinos in about 10~sec. While
the collapse is well understood, the actual explosion which ejects the outer
layers of the original star is not. Because only $\simeq 1\%$ of the total
energy
liberated in the collapse goes into the explosion, relatively very small
effects have to be well understood to account for it theoretically. Therefore,
supernovae explosions are very difficult to obtain in simulations. This
difficulty is sometimes called the ``supernova problem". Neutrino properties
that could transfer some of the neutrino energy to the outer layers of the star
have been repeatedly advocated as a solution to this problem.

A surprising wealth of information on neutrinos have been obtained from
SN1987A, considering that only 19 neutrino events were observed (by two large
water Cerenkov detectors, IMB, Bionta \etal 1987 and Bratton \etal 1988, and
Kamiokande II, Hirata \etal 1987 and 1988, through the reaction $\nu_e^c + p
\to n + e^+$).  Let us mention some of the bounds obtained (for a review see
e.g. Raffelt 1990a and 1990b).

The simple arrival of (electron anti-) neutrinos from the SN1987A, provides a
bound on any effect that could have removed them from the burst before arriving
to earth, such as decay, scattering, or interaction with magnetic fields. The
lower bound obtained on their lifetime  is $\tau_{\nu_e} > 6 \times 10^5$ sec
($m_{\nu_{e}}$/eV). Due to the possibility of a difference between
current and mass eigenstate neutrinos this bound does not  exclude that part of
the electron neutrinos emitted in the sun could decay before arriving to the
earth (Frieman, Haber and Freese 1988), explaining in
this way the solar neutrino problem. However decaying neutrinos do not provide
a  satisfactory fit to the solar neutrino observations (Acker and Pakvasa
1994).
The neutrinos arriving from SN1987A where not removed by scattering, for
example, with a background of relic Majorons, thus  the neutrino majoron
coupling (see section 3) must be $(g_{\rm eff})_ {\nu_e \nu_e J} < 10^{-3}$
(Kolb and Turner 1987). The arrival of neutrinos provides also a bound on
their charge. If neutrinos would have a non-zero electric charge
$Q_{\nu_{e}}$, their
trajectories would have been bent in the galactic magnetic field. The
observation of the neutrino burst gives a limit $Q_{\nu_{e}}\lesssim 3 \times
10^{-15}e$ (Barbiellini and Cocconi 1987)  (while the best limit from the
neutrality of atoms is $Q_{\nu_{e}} \lesssim 10^{-21}e$, Bauman et al 1989).

An upper bound on $m_{\nu_{e}}$ is due to the (non) dispersion of the
neutrino emission time. The neutrino signal was spread over $\simeq
10$ seconds and the emission time could not have been much longer.
Massive neutrinos from the star arrived to earth with a delay time
(time of arrival minus time of emission) $\Delta t = d / v \simeq d
[1+ (1/2) (m_\nu/E_\nu)^2]$ where $d \simeq 50$ kpc$=1.6 \times 10^5$
lyr is the distance travelled and $E_\nu$ is the neutrino energy. If
all neutrinos (with equal mass $m_\nu$) had been emitted at the same
time, the more energetic ones should have arrived earlier. This is not
what happened (energies and times were not so correlated). Thus, by
increasing the assumed mass $m_\nu$, the inferred dispersion of the
emission times increases and for some mass value becomes unacceptably
large. This argument yields the upper bound $m_{\nu_{e}} < 23$ eV (at
$95\%~ C.L.$ ignoring systematic uncertainties, Loredo and Lamb
1989). This bound (as those of section 4.1) is independent of the
Dirac or Majorana nature of the neutrinos. There are no bounds on the
other flavour neutrino masses because the bulk of the events observed
were $\nu_e^c$, due to detection cross-section arguments.

The number and energy of the detected events correspond well with what
is expected, leaving not much room for new sources of energy loss.
Also a new mechanism for energy loss would have speed up the cooling,
thus shortening the pulse with respect to what was seen. For example
Majoron models (singlet or mostly singlet Majorons, the only viable
ones, see section 3)  are constrained by these argument, that reject a
region in $m_\nu$-$V$ space, for $V$, the scale of spontaneous
symmetry breaking, in the range few GeV $\lesssim V \lesssim$ 1 TeV
(Choi and Santamaria 1990). If neutrinos are Dirac particles, their
mostly sterile right-handed helicity components can be produced due to
their $m/E$ admixture in the left-handed chirality component (the
interactive component). Subsequently they escape, because their cross
sections are too small to trap them, increasing the cooling rate of
the neutron star. A bound $m_\nu^D < O(10$ keV) has been obtained
(Raffelt and Seckel 1988, Burrows, Gandhi and Turner 1992, Mayle \etal
1993 and references therein) from preventing a shortening of the
duration of the observed cooling neutrino pulse. This bound applies
only to neutrinos lighter than 1 MeV, that can be copiously produced,
because the production of heavier neutrinos within the star is
approximately suppressed by a Boltzman factor (Sigl and Turner 1994).
The upper bound of $O(10$ keV) is hard to improve due to the
difficulty in evaluating the effect of the very fast multiple
scattering of nucleons inside the supernova on the emission of the
right-handed neutrinos (Raffelt and Seckel 1991 and 1992). This
difficulty  affects most bounds (also on other particles) based on
energy loss in SN1987A. Sterile right handed neutrinos could also have
been produced due to new interactions with an effective coupling
constant $G_R$ (the equivalent of the usual Fermi constant $G_F$). The
above cooling arguments yield $G_R \lesssim 10^{-4}G_F$ (Raffelt and
Seckel 1988, 1991, 1992).

 If neutrinos are Dirac particles and they are unstable, the $\nu_R$
that escaped from the  inner, hotter,  regions of the supernova, could
decay into energetic ($E\simeq 100$ MeV) left-handed neutrinos that
would have been seen as extra events if 1 keV $\lesssim m_\nu \lesssim
300$ keV and $ 10^{-9} {\rm sec}(m/1$ keV$) \lesssim \tau \lesssim 5
\times 10^7 {\rm sec} (m/1$ keV), thus these ranges are forbidden for
Dirac neutrinos (Dodelson, Frieman and Turner 1992). Following a
similar argument, one finds  a more dubious  bound that would apply to
both Dirac and Majorana neutrinos in the same mass range, emitted from
the ``neutrino sphere",  the boundary of the region where neutrinos
are temporarily trapped (so these are interacting neutrinos),  if
electron neutrinos are produced in the decay. Then, lifetimes $3\times
10^{5}$ sec (keV$/m) \lesssim \tau \lesssim 2 \times 10^{10}$ sec
(keV$/m)$ are rejected (Gelmini, Nussinov and Peccei 1992, Mohapatra
and Nussinov 1992, Soares and Wolfenstein 1989).  This bound is less
meaningful because neutrinos are emitted from the neutrino sphere with
 energies of $O(10$ MeV), consequently the energy of the decay
products is smaller than before, thus reducing the expected number of
events and their  energies, what makes these events more difficult to
distinguish from background events. These bounds on lifetimes apply to
invisible decay modes, in which the decay products do not include
photons or e$^+$e$^-$, i.e. $\nu \to 3 \nu'$ or $\nu \to \nu' \phi$,
where $\phi$ is a Goldstone boson, a Majoron in most models (see
section 3).
Much more restrictive bounds apply to radiative decays, as shown below.

Turning now to the electromagnetic properties of neutrinos, a magnetic or
electric dipole moment $\mu_\nu$ would also allow for helicity flips into
otherwise inert $\nu_R$ of Dirac neutrinos in the electromagnetic scattering of
the trapped $\nu_L$ with charged particles. Thus, the same arguments on energy
loss from SN1987A impose, for Dirac neutrinos only,  $\mu_\nu <
10^{-12}\mu_B$, with the uncertainty mentioned above  (Barbieri and Mohapatra
1988, Lattimer and Cooperstein 1988, Goldman et al 1988) for  neutrinos that
can be copiously produced in the supernova, namely $m\lesssim 1$~MeV. Here
$\mu_B= e/2 m_e$ is the Bohr magneton.
Other bounds, valid for $m_\nu \lesssim 10$ keV, arise from energy
loss  due to the plasmon decay $\gamma \rightarrow \nu_i^c\nu_j$ in
red-giants and in white dwarfs. These are respectively $\mu_\nu < 3
\times 10^{-12} \mu_B$ (Raffelt 1990c, Raffelt and Weiss 1992) and
$\mu_\nu < 1 \times 10^{-11}\mu_B$ (Blinnikov 1988, Raffelt 1990a).
These two  bounds apply to both Dirac (direct or diagonal and
transition) and Majorana (transition) neutrino dipole moments
(`diagonal' and `transition' refer to flavour space, see section 2).
Note that these bounds would exclude the spin-flip solution to the
solar neutrino problem (see section 7).

For comparison, let us mention here the cosmological bound based on the
effective number of relativistic neutrinos during nucleosynthesis, $N_\nu <
3.4$ (Olive \etal 1990, see section 5). It requires $\mu_\nu \lesssim 5
\times 10^{-11} \mu_B$ for Dirac neutrinos (Morgan 1981, Fukugita and Yazaki
1987), because otherwise the mostly inert right-handed helicity components
could be brought in equilibrium in the early universe  due to the
electromagnetic interactions of left-handed neutrinos. On the other hand, the
laboratory bounds for magnetic moments (Review of Particle Properties 1994 and
references therein) are $\mu_{\nu_e} < 1.08 \times 10^{-9} \mu_B$,
$\mu_{\nu_\mu} < 7.4 \times 10^{-10} \mu_B$ and $\mu_{\nu_\tau} < 5.4 \times
10^{-7} \mu_B$. This  latter bound on $\mu_{\nu_\tau}$ applies only to a direct
moment (Cooper-Sarkar \etal 1992). Using the same data (Cooper-Sarkar
\etal 1985),
Babu, Gould and Rothstein (1994) found a  laboratory bound for
transition moments $\mu_{\nu_\tau {\rm tran}} < 1.1  \times 10^{-9}
($MeV$/m_{\nu_\tau}) \mu_B$, that is more restrictive for an MeV tau neutrino.
In the simplest extension of the SM, where only $\nu_R$ and Dirac masses
$m_\nu$ are added (section 3.2), the predicted neutrino magnetic moment is much
smaller than any of these bounds, $\mu_{\nu}\simeq 3 \times 10^{-19} (m_\nu
/$eV$) \mu_B$ (Lee and Schrock 1977, Fujikawa and Shrock 1980).

The non-observation (by the Solar Maximum Mission Satellite, SMM) of a
$\gamma$-ray burst in association with the SN1987A neutrino burst, forbids
radiative neutrino decays $\nu \to \gamma \nu'$ with lifetimes
$\tau_{\rm rad}
> 2 \times 10^{15}$ sec $(m_\nu/$eV) for $m_\nu < 20$ eV. For larger masses,
the photons produced would be spread out in time and the upper bound weakens
to $\tau_{\rm rad} \gsim 3 \times 10^{16}$ sec for
20 eV $ \lesssim m_\nu \lesssim 100$ eV and
$\tau_{\rm rad} > 0.8 \times 10^{18} ({\rm eV}/m_\nu)$ for 100 eV
$\lesssim m_\nu \lesssim$ 1 MeV (Kolb and Turner 1989, Raffelt 1990a and 1990b
and references therein). The mass-lifetime region excluded by the SMM bounds
is shown in the hatched area of figure 2. These bounds on $\tau_{\rm rad}$
translate into very
restrictive bounds on transition moments,  for all neutrino types and flavours
(through the relation $\tau_{\rm rad}= 8 \pi/ {\mu_\nu}^2 {m_\nu}^3$,
valid  for
negligible final neutrino mass). These bounds are  $\mu_{\nu {\rm tran}}< 1
\times 10^{-8} ($eV$/m_\nu)^2 \mu_B$ for $m_\nu \lesssim$ 20 eV and $\mu_{\nu
{\rm tran}}<5 \times 10^{-10}(1$ eV$/m_\nu)\mu_B$ for 20 eV $\lesssim
m_\nu\lesssim$ 1 MeV (Raffelt 1990a).

The mass vs. lifetime bounds for visible  decay modes of neutrinos heavier than
1 MeV (only possible for $\nu_\tau$) have been recently reanalysed (Sigl and
Turner 1994, Babu, Gould and Rothstein 1994). These neutrinos have another
visible decay mode besides $\nu_\tau \to \nu' \gamma$, namely $\nu_\tau \to
\nu' $e$^+$e$^-$. Even if this is the dominant decay, the SMM bound can be
applied, because photons will be produced with branching ratio $10^{-3}$
through bremsstrahlung  (Sigl and Turner 1994)  or by other processes
(Mohapatra, Nussinov and Zhang 1994). The SMM bound only  applies  if neutrinos
decay outside the supernova, i.e. for $\tau > 100$ sec.  For shorter lifetimes,
the decay products would be trapped inside and bounds  result from their effect
on the  supernova energetics. The combinations of both bounds exclude a massive
 $\nu_\tau$ of lifetime $ 10^{-6}$ sec $\lesssim \tau \lesssim 10^8$ sec
decaying into visible modes (Sigl and Turner 1994). The excluded region for
$m_\nu > 1$ MeV taken from Sigl and Turner (1994) is included in the hatched
area of figure 2. Laboratory bounds
(Cooper-Sarkar \etal 1985, Babu, Gould and Rothstein 1994) exclude these modes
for  even shorter lifetimes, $ \tau\lesssim 0.1 (m_\nu/$MeV) sec
\footnote{$^*$}{Very unlikely short lifetimes $\tau \lesssim 2 \times 10^{-12}
(m_\nu/$MeV) would be allowed only for the mode $\nu_\tau \to \nu_s \gamma$,
where $\nu_s$ is a sterile neutrino.}. So, if the dominant decay modes are
visible, all lifetimes shorter than $\simeq 10^8$ sec are forbidden by the
combination of SN1987 bounds and laboratory bounds. If we add the
nucleosynthesis bounds (see section 5, Dodelson, Gyuk and Turner 1994, Kawasaki
\etal 1994) that forbid $\nu_\tau$ heavier than $\simeq 0.3$~MeV if the
lifetime is shorter than $\simeq 100$~sec, the combination of the three bounds
reject a $\nu_\tau$ heavier than 0.3~MeV that decays dominantly into visible
modes for any lifetime.

The MSW effect (see section 7) could also happen within a supernova. The
oscillation of $\nu_\mu$ or $\nu_\tau$ into $\nu_e$ outside the neutrino sphere
could have important consequences. The reason is that the emitted $\nu_\mu$ and
$\nu_\tau$ are more energetic than the emitted $\nu_e$. This is because
$\nu_\mu$ and $\nu_\tau$ only interact through neutral currents with the
surrounding matter. This means, they have a larger mean free path than $\nu_e$
(that also interact through charged currents) and they consequently emerge from
deeper and, thus, hotter layers of the  supernova core. Thus, if a large
fraction of the $\nu_\tau$ (or $\nu_\mu)$ are transformed into energetic
$\nu_e$ (that, as we just said, have larger interaction rates with the
surrounding matter), the increased energy given to outer layers of the star may
help explode them (Fuller et al 1992). However, the same flavour conversion may
preclude nucleosynthesis through $r$-processes in supernovae. Heavy elements
are believed to be synthesized through rapid neutron capture  processes
($r$-processes) in the neutron-rich outer layers of an exploding  supernova.
The more energetic electron-neutrinos, resulting from a MSW  conversion of
outgoing $\nu_\mu$ or $\nu_\tau$ before reaching the region where
$r$-processes should occur, would reduce the amount of neutrons in the material
of the crucial layers (through $\nu_e$ + n $\to$  p + e$^-$). The environment
would then become proton-rich and this would stop $r$-processes entirely.
Because there is no other known production site for the elements that should
be produced though $r$-processes in supernovae, this effect must be avoided.
This excludes the region of the $\Delta m^2 - \sin^22\theta$, for
$\nu_e-\nu_\mu$
and $\nu_e-\nu_\tau$ oscillations, corresponding to
approximately $\Delta m^2 \gsim 4$ eV$^2$  for  $\sin^2 2\theta
\gsim 10^{-5}$ (Qian \etal 1993). This is an important region, because it
corresponds to a $\nu_\tau$ or $\nu_\mu$ with a mass suited to be  hot dark
matter (between a few eV and a few tens of eV, see section 5), if $\nu_e$ is
much lighter.

The IMB and Kamioka cerenkov detectors have shown the feasibility of observing
the neutrinos from supernova explosions, and as we just discussed their few
events allowed to obtain significant information on neutrino properties. In the
future, with the construction of Superkamiokande and also especially of the SNO
detector sensitive to neutral currents, the detection of a supernova within our
galaxy (there are $\sim 2$ galactic supernova explosions per century!) may
allow to constrain the masses of $\nu_\mu$ and $\nu_\tau$ if they are larger
than $\simeq 25$ eV (Minakata and Nunokawa 1990, Seckel, Steigman and Walker
1991, Krauss et al 1993, Burrows, Klein and Gandhi 1992), and maybe even down
to 15 eV with new type of detectors (Cline \etal 1994). It may also become
possible to detect the background of neutrinos from all past supernovae, that
should be the dominant neutrino flux at energies just beyond those of the solar
neutrinos (20--30 MeV).

Finally, let us mention the under-water (ice) neutrino telescopes under
construction at present with energy thresholds between a few and 10 GeV.
They will observe the high-energy neutrinos that should be emitted in
comparable numbers with $\gamma$ rays in the most energetic galactic
and extragalactic processes, such as the shock acceleration in Active
Galactic Nuclei.
These experiments, DUMAND, AMANDA, Baikal and NESTOR, with an area of
approximately 0.02 km$^2$, can be considered the prototypes for the  final
desirable full size experiments of 1 km$^2$ (for a review see Halzen 1993).

 \section {Concluding Remarks}

The hints for non-zero neutrino masses in solar and atmospheric neutrinos,
and  the possibility of confirming or rejecting them in the near
future,
make the subject of neutrino masses a particularly exciting
field of research at present.
So much so that one could indulge in the actually premature exploration
of the consequences of the confirmation of both. One could even wonder
about their compatability with  other possible desirable mass values,
such as masses of order 1-10 eV for neutrinos to account for part of the
Dark Matter in the Universe, or Majorana masses within the reach of present
neutrinoless double beta decay experiments, i.e. $<m_{\nu_e}> \simeq$ 1 eV.

The MSW solution to the solar neutrino problem would require $m_{\nu_i}^2-
m_{\nu_e}^2 \simeq O(10^{-6})$~eV$^2$, while the solution to the atmospheric
neutrino deficit requires
$m_{\nu_j}^2- m_{\nu_\mu}^2 \simeq O(10^{-2})$~eV$^2$. An obvious simultaneous
solution for both requires $m_{\nu_\tau}\simeq 10^{-1}$~eV,
$m_{\nu_\mu}\simeq 10^{-3}$~eV and $m_{\nu_e} \ll m_{\nu_\mu}$, with
$\nu_i =\nu_\mu$ and $\nu_j =\nu_\tau$. These mass values can be
obtained in a see-saw model with a large
right-neutrino Majorana mass $M\simeq 10^{11}$~GeV (see section 3).

Incorporating either the solar neutrino solution or the atmospheric neutrino
solution with either DM neutrinos or $\langle m_{\nu_e}\rangle
 \simeq$ 1 eV is easy in see-saw models with smaller values of $M$
 (see e.g. Peltoniemi, Tommasini and Valle 1993).
 A solution incorporating
both the solar and the atmospheric neutrino solutions and either DM neutrinos
or $\langle m_{\nu_e}\rangle \simeq$ 1 eV requires
quasi-degenerate neutrinos, where the
mass differences necessary to account for the oscillations are much smaller
than the masses themselves (Caldwell and Mohapatra 1993, Peltoniemi and
Valle 1993).
 Models of this type, possibly with extra inert neutrinos, can even incorporate
all four mass requirements
mentioned. In a possible solution the three active neutrinos are almost
degenerate.
This can be obtained if the see-saw mechanism determines the mass differences
but the left-neutrino Majorana masses give
the masses themselves (see section 3).

The confirmation of a see-saw mechanism is by no means the only
possible outcome of neutrino mass experiments in the near future.
We may uncover the Majorana nature
of neutrinos in future neutrinoless double beta decay experiments
if the effective $\nu_e$ mass is not much smaller than 1 eV
(section 4.2). Still, we may find neutrino masses in the range forbidden by the
present energy density of the universe (in direct mass
searches on with a galactic supernova explosion),
what would be a strong indication of the existence of Goldstone
bosons (typically singlet or mostly singlet Majorons, see section 3),
necessary for the neutrinos to decay or
annihilate fast in the early universe (section 5). Or even
neutrinoless double beta decay with the emission of a boson, that
would be an even more direct indication of Goldstone bosons.
Even if these results are possible, they are unlikely. We do
expect instead a resolution of the solar and atmospheric neutrino
problems within the next decade.
These are certainly exciting times for neutrino physics.

\ack
We would like to thank the editors of this journal for their patience.
G.G. was supported in part by the U.S. Department of Energy under Grant
DE-FG03-91ER 40662 TaskC.

\references
\refjl{Aardsma G \etal (SNO) 1987}{\PL}{194B}{321}
\refjl{Abazov A I \etal (SAGE) 1991}{\PRL}{67}{3332}
\refjl{Abdurashitov J N \etal 1994}{\PL}{328B}{234}
\refjl{Abela R \etal 1984} {\PL}{146}{149}
\refjl{Aglietta N \etal (NUSEX) 1989}{\sl Europhys. Lett.}{8}{611}
\refjl{Ahrens L A \etal (BNL E734) 1985}{\PR D}{31}{2732}
\refjl{Acker A and Pakvasa S 1994}{\PL}{B320}{320}
\refjl{Anselmann \etal 1994}{\PL}{B327}{377}

\refjl{Akhmedov E 1988}{\PL}{213B}{64}
\refjl{Akhmedov E \etal 1993a}{\PL}{B299}{90}
\refjl{Akhmedov E, Berezhiani Z and Senjanovic' G 1992}{\PRL}{69}{3013}
\refjl{Akhmedov E \etal 1993b}{\PRL}{D47}{3245}
\refjl{Akhmedov E, Lipari P and Lusignoli M 1993}{\PL}{300B}{128}
\refjl{Akhmedov E, Lanza A and Petcov S T 1993}{\PL}{303B}{85}
\refjl{Albrecht H \etal (ARGUS) 1992}{\PL}{B292}{221}
\refbk{Allison W W M \etal 1991}{Fermilab proposal P822}{}
\refjl{Ammosov V V \etal (SKAT) 1988}{\ZP C}{40}{487}
\refjl{Angelini C \etal (BEBC) 1986}{\PL}{179B}{307}
\refjl{Anselmann P \etal (GALLEX) 1993}{\PL}{314B}{445}
\refjl{Arpesella C \etal 1992}{\NP B (Proc. Suppl.)}{28A}{486}
\refbk{Arzarello \etal 1994}{preprint LPC-94-28}
\refbk{Assamagan K \etal 1994}{preprint PSI-PR-94-19}
\refjl{Aulak C S and Mohapatra R N 1983} {\PL} {119B}{136}
\refjl{Avignone F T \etal 1991}{\PL}{B256}{556}
\refjl{Babu K S 1988} {\PL} {203B}{132}
\refjl{Babu K S, Gould T and Rothstein 1994} {\PL}{B 321}{140}
\refjl{Babu K S and Mohapatra 1991} {\PRL}{66}{556}
\refjl{Babu K S, Mohapatra and Rothstein 1991}{\PRL}{67}{545}
\refjl{\dash 1992}{\PR}{D45}{3312}
\refjl{\dash 1993} {\PRL}{70}{2845}
\refjl{Bahcall J N and Pinsonneault M H 1992}{\sl Rev. Mod.
Phys.}{64}{885}
\refjl{Bahcall J N and Ulrich R K 1988}{\sl Rev. Mod.
Phys.}{60}{297}
\refjl{Baltz A J and Weneser J 1987}{\PR D}{35}{528}
\refjl{Balysh A \etal 1994}{\PL}{B322}{176}
\refjl{Bamert P and Burgess C 1994}{\PL}{B329}{289}
\refjl{Barbiellini G and Cocconi G 1987}{Nature}{329}{21}

\refjl{Barbieri R and Dolgov A 1990}{\PL}{B237}{440}
\refjl{\dash 1991}{\NP B}{349}{743}
\refjl{\dash 1991}{\NP}{B349}{743}
\refjl{Bardeen J, Bond J and Efstathiou G 1987}{\it Astrophys. J}
{321}{5068}

\refjl{Barbieri R, Ellis J and Gaillard M K }{\PL}{B90}{249}
\refjl{Barbieri R and Mohapatra R 1988}{\PRL}{61}{27}

\refjl{Barger V, Phillips R J N and Whisnant K 1981}{\PR D}{24}{538}
\refjl{\dash 1991}{\PR D}{44}{1629}
\refjl{\dash 1992}{\PRL}{69}{3135}

\refjl{Barr G, Gaisser T K and Stanev T 1989}{\PR D}{39}{3532}
\refjl{Bauman J \etal 1989}{Nucl. Inst. Methods}{A284}{130}
\refjl{Beck \etal 1993}{\PRL}{70}{2853}
\refjl{Becker-Sendy R \etal (IMB) 1992a}{\PRL}{69}{1010}
\refjl{Becker-Sendy R \etal (IMB) 1992b}{\PR D}{46}{3720}
\refjl{Beier E W \etal 1992}{\PL}{283B}{446}
\refbk{Belesev A I \etal (Troistk) 1994}{preprint INR-862/94}
\refjl{Berezinski V and Valle J W F 1993}{\PL}{B318}{366}

\refbk{Berezhiani Z G and Rossi A 1993}{\it Venice 92, Proc. Neutrino
Telescopes}{p 123}

\refjl{Berezhiani Z G, Smirnov A YU and Valle J W F 1992}{\PL}{B291}{99}
\refjl{Bernatowicz \etal 1992}{\PRL}{69}{2341}
\refjl{Berger Ch \etal (FREJUS) 1989}{\PL}{227B}{489}
\refjl{Bergsma F \etal (CHARM) 1984}{\PL}{142B}{103}
\refjl{Bergkvist K E 1972}{\NP}{B39}{317 and 371}

\refjl{Bernstein R H and Parke S 1991}{\PR D}{44}{2069}
\refjl{Bertolini S and Santamaria A 1988}{\NP}{B310}{714}
\refjl{Bethe H A 1986}{\PRL}{56}{1305}
\refjl{Bilenky S M and Petcov S T 1987} {\RMP}{59}{671}
\refjl{Bionta R M \etal (IMB) 1987}{\PRL}{58}{1494}
\refjl{Bionta R M \etal (IMB) 1988}{\PR \ D}{38}{768}
\refbk{Blinnikov S I 1988}{ITEP-preprint No 19, unpublished}

\refjl{Bl\"umer H and Kleinknecht K 1985}{\PL}{161B}{407}
\refbk{Boliev M M \etal (BAKSAN) 1991}{\sl Proc. of the 3$^{rd}$
International Conf. on Neutrino Telescopes,}{ Venice, Italy, Ed. Baldo
Ceolin M, p. 235}
\refjl{Bond J and Efstathiou G 1991}{\PL}{B265}{245}
\refjl{Bonometto S \etal 1993}{\it Astroph-9311018}{}

\refjl{Bonvicini G 1993 {\sl Neutrino 92}}{\NP  Proc.
Suppl.} {31}{65}
\refjl{Boris \etal 1987}{\PRL}{58}{2019}
\refjl{Borodovsky L \etal (BNL E776) 1992}{\PRL}{68}{274}
\refjl{Bouchez J \etal 1986}{\ZP C}{32}{499}
\refjl{Brahmachari B \etal 1993}{\PR}{D48}{4224}
\refjl{Bratton C \etal (IMB) 1988} {\PR}{D37}{3361}
\refjl{Bugaev E V and Naumov V A 1989}{\PL}{232B}{391}
\refjl{Burgess C P and Cline J M 1993} {\PL}{B298}{141}
\refjl{\dash 1994}{\PR}{D49}{5925}
\refjl{Burrows A 1990}{Annu. Rev. Nucl. Sci}{40}{181}
\refjl{Burrows A, Gandhi R and Turner M S 1992}{\PRL}{68}
\refjl{Burrows A, Klein D and Gandhi R 1992}{\PR}{D45}{3361}

\refjl{Butkevich A V, Dedenko L G and Zheleznykh 1989}{\sl Sov. J.
Nucl. Phys.}{50}{90}
\refjl{Cabibbo N 1963}{\PRL}{10}{531}
\refjl{Cabibbo N 1978}{\PL}{72B}{333}
\refjl{Caldwell D and Mohapatra R N 1993}{\PR}{D48}{3259}
\refjl{\dash 1994}{\PR}{D50}{3477}

\refjl{Carlson E D 1986}{\PR D}{34}{3364}
\refjl{Carone C 1993}{\PL}{B308}{85}
\refjl{Casper D \etal (IMB) 1991}{\PRL}{66}{2561}
\refjl{Chikashige Y, Mohapatra R N and Peccei R D 1981}
{\PL}{98B}{265}
\refjl{\dash 1980}{\PRL}{45}{1926}
\refbk{Choi K \etal 1989},{unpub. preprint CMU-HEP89-22}
\refjl{Choi K and Santamaria A 1991}{\PL}{B267}{504}
\refjl{\dash 1990}{\PR}{D42}{293}

\refjl{Cinabro D \etal (CLEO)}{\PRL}{70}{3700}
\refjl{Cisneros A 1971}{\sl Astrophys. Space Sci.}{10}{87}
\refjl{Cline D \etal 1994}{\PR}{D50}{720}
\refjl{Cline J 1992}{\PRL}{68}{3137}
\refjl{Cline J, Kainulainen K and Olive K 1992}{\PL}{280}{153}
\refjl{\dash 1993a}{Astropart. Phys 1}{387}
\refjl{\dash 1993b}{\PRL}{71}{2372}
\refjl{Cooper-Sarkar A M \etal (WA66 coll) 1985}{\PL}{160B}{207}
\refjl{\dash 1992}{\PL}{280B}{153}
\refjl{Cowsik R and McClelland J 1972}{\PRL}{29}{669}

\refjl{Curran S C \etal 1949}{\PR}{76}{853}
\refjl{Cvetic M and Langacker P 1992}{\PR}{D46}{2759}

\refjl{D'Ambrosio G and Gelmini G 1987} {\it Zeit. Physics}{C35}{461}
\refbk{Darriulat P 1994}{Summary talk,{\it Int. Conference on High Energy
Physics}, Glasgow}

\refbk{Dar A and Shaviv G 1994}{Preprint TECHNION-PH-94-5}{}
\refbk{Davis R \etal 1990}{\it Proc. of the 21st International
Cosmic Ray Conference}{Ed. Protheroe R J (University of Adelaide
Press, Adelaide) p 143}

\refbk{de Jong M \etal (CHORUS) 1993}{CERN preprint}{CERN-PPE/93-131}
\refbk{Demarque P, Deliyannis C and Sarajedini A 1991}{Observational
Tests of Cosmological Inflation}{eds. T. Shanks \etal, Kluwer,
Dordrecht, p. 111}

\refbk{Dassie D. \etal 1994}{(NEMO)}{Preprint LAL-94-46}
\refjl{De R\'ujula A and Glashow S 1980}{\PRL}{45}{942}
\refjl{Di Lella L 1993}{\NP \sl (Proc. Suppl.)}{31}{319}
\refjl{Dicus O, Kolb E and Teplitz V 1977}{\PRL}{169}
\refjl{Dimopoulos \etal 1988}{\PRL}{60}{7}

\refjl{Dodelson S, Frieman J and Turner M 1992}{\PRL}{68}{2572}
\refjl{\dash 1994a}{\PR}{D49}{5068}
\refjl{\dash 1994b}{\PRL}{72}{3754}

\refbk{Dolgov A D, Kainulainen K and Rothstein I,} {preprint FERMILAB-Pub
-94/1994}
\refjl{Dolgov A 1981}{\sl Sov. J. Nucl. Phys.}{33}{700}
\refjl{Doncheski M A, Halzen F and Stelzer T 1992}{\PR \ D}{46}{505}
\refjl{Drexlin G \etal (KARMEN) 1990}{\NIM}{A289}{490}
\refjl{Durkin L S \etal (LAMPF E645) 1988}{\PRL}{61}{1811}
\refjl{Dydak F \etal (CDHSW) 1984}{\PL}{134B}{281}
\refjl{Ejiri H \etal 1994}{\NP (Proc. Suppl.)}{B35}{372}
\refjl{Elliot S R \etal 1991}{J. Phys. G}{17}{S145}
\refjl{Elliot S R \etal 1992}{\PR}{C46}{1535}
\refjl{Enqvist K, Kainulainen and Thomson M 1992}{\NP}{B373}{498}

\refjl{Enqvist K \etal 1992}{\NP}{B373}{191}
\refjl{Enqvist K, Uibo 1993}{\PL}{B301}{376}
\refbk{Fiorentini G and Ricci B 1993}{Proc. Neutrino Telescopes
Conf.}{Venice, p. 69}
\refjl{Fujikawa K and Shrock R 1980}{\PRL}{45}{963}
\refjl{Fukuda Y \etal (KAMIOKANDE) 1994}{\PL}{335B}{237}
\refjl{Fukugita M, Hogan and Peebles P 1993}{Nature}{366}{309}

\refjl{Fukugita M and Yazaki S 1987}{\PR}{D36}{3817}

\refjl{Fuller G and Malaney R 1991}{\PR}{D43}{3136}

\refjl{Fuller G, Mayle R, Meyer B and Wilson J 1992}{Astrophysics
J.}{389}{517}

\refjl{Frati W \etal 1993}{\PR D}{48}{1140}
\refjl{Frieman J, Haber H and Freese K 1988}{\PL}{200B}{115}
\refjl{Gabbiani F, Masiero A and Sciama D 1991}{\PL}{B259}{323}

\refjl{Gelmini G and Valle {JWF 1984}}{\PL}{142B}{181}
\refjl{Gelmini G, Nussinov S and Peccei 1992}{Int. J. Mod.
Phys.}{A7}{3141}

\refjl{Gelmini G, Nussinov S and Yanagida T 1983}{\NP}{B219}{31}
\refjl{Gelmini and Roncadelli M 1981}{\PL}{99B}{411}
\refjl{Gell-Mann M, Ramond and Slansky R (1979)}{``Supergravity''}
{(North Holland)}
\refjl{Georgi H Glashow S L and Nussinov S (1981)}{\PL}{99B}{411}
\refjl{Gerstein S and Zeldovich Ya 1972}{\it Zh. Eksp. Teor. Fiz Psi'ma}
{4}{174}
\refjl{Glashow S L 1991}{\PL}{256B}{218}
\refjl{Glashow S L and Krauss L M 1987}{\PL}{190B}{199}
\refjl{Goldman I \etal 1988}{\PRL}{60}{1789}
\refjl{Gorski \etal 1994}{\it Astrophys. J}{430}{L89}

\refjl{Gribov V N and Pontecorvo B M 1967}{\PL}{28}{493}
\refjl{Gruw\'e M \etal (CHARM II) 1993}{\PL}{309B}{463}

\refjl{Guzzo M, Masiero A and Petcov S T 1991}{\PL}{260B}{154}
\refbk{Gyuk G and Turner M 1994}{preprint FERMILAB-Pub-94/059-A}

\refjl{Halprin A and Leung C N 1991}{\PRL}{67}{1833}
\refbk{Halzen F 1993}{Madison preprint MAD/PH/785}
\refjl{Harari H 1989}{\PL}{216B}{413}
\refjl{Hata N and Langacker P 1993a}{\PR}{D48}{2937}
\refbk{\dash P 1993b}{U. of Pennsylvania preprint UPR-0581T}{}
\refjl{\dash P 1994}{\PR}{D50}{632}
\refjl{Haxton W C 1986}{\PRL}{57}{1271}
\refjl{Hillebrandt W and H\"oflich P 1989}{\RPP}{52}{1421}

\refbk{Hime A 1993}{Neutrino 92}{\NP}{(\sl Proc. Suppl.)}{31}~{50}

\refjl{Hirata K S \etal (KAMIOKANDE) 1988}{\PL}{205B}{416}
\refjl{Hirata K S \etal (KAMIOKANDE) 1991a}{\PRL}{66}{9}
\refjl{Hirata K \etal (KAMIOKANDE) 1987}{\PRL}{58}{1490}
\refjl{\dash}{\PR}{38}{448}

\refjl{Hirata K S \etal (KAMIOKANDE) 1991b}{\PR D}{44}{2241}
\refjl{Hirata K S \etal (KAMIOKANDE) 1992}{\PL}{280B}{146}
\refjl{Honda M \etal 1990}{\PL}{248B}{193}
\refjl{Holzschuch E 1992}{Rep. Prog. Phys.}{55}{1035}
\refjl{Holzschuch \etal}{\PL}{B287}{381}
\refjl{Hut P 1977}{\PL}{69B}{85}
\refjl{Ioannisian A and Valle J W F 1994}{\PL}{B332}{93}
\refbk{Jacoby \etal 1992}{Pub. Astron. Soc. Pacific}{104}~{599}

\refjl{Jeckelmann B \etal 1986}{\PRL}{56}{1444}
\refjl{Johnson R, Ranfone S and Schechter 1986}{\PL}{179B}{355}
\refjl{\dash 1987}{\PR}{D35}{282}
\refjl{Jungman G and Luty M 1991}{\NP}{B361}{24}

\refbk{Kafka T 1993}{\sl Talk at the TAUP 93 Conf.,}{ Gran Sasso, Sept.
1993}
\refjl{Kainulainen K 1990}{\PL}{244B}{191}
\refjl{Kawakami H \etal}{\PL}{B256}{105}

\refjl{Kawasaki M and Mizuta S 1991}{\PR D}{43}{2900}
\refjl{Kawasaki M \etal 1994}{\NP}{B419}{105}

\refjl{Kayser B 1981}{\PR D}{24}{110}
\refjl{\dash 1982}{\PR}{D 26}{1662}
\refjl{\dash 1984}{\PR}{D39}{1023}
\refbk{\dash 1988}{in CP violation,}{Ed C. Jarslkog, World Scientific,
Singapore, p. 334}
\refjl
{\dash 1985}{Comments Nucl. Part. Phys.}{14}{69}
\refbk{Klypin A \etal }{Preprint SCIPP-94-09}

\refjl{Kobayashi M and Maskawa T 1973}{Prog. Theo. Phys.}
{49}{652}
\refbk{Kodama K \etal 1990}{Fermilab proposal P803}{}
\refbk{Kolb 1986}{Proceedings of the 1986 Theoretical Advanced Studies
Institute (Santa Cruz, CA)}
\refjl{Kolb E W and Olive K 1986}{\PR}{D33}{1202}
\refjl{\dash E: 1986}{\PR}{D34}{2531}
\refjl{Kolb E W and Scherrer R 1982}{\PR}{D25}{1481}
\refjl{Kolb E W and Turner M S 1987}{\PR}{D36}{2895}
\refbk{\dash 1990}{The Early Universe}{(Addison-Wesley)}
\refjl{Kolb E W \etal 1991}{\PRL}{67}{533}
\refjl{\dash 1989}{\PRL}{62}{509}
\refjl{Konopinski E S and Mahmoud M 1953}{\PR}{92}{1045}

\refjl{Krastev P I and Petcov S T 1992}{\PL}{285B}{85}
\refjl{Krauss L M \etal 1992}{\NP}{B380}{507}

\refjl{Krishnaswamy M M \etal 1971}{\sl Proc. Roy. Soc.
London}{A323}{489}
\refjl{Lam W and Ng K W 1991}{\PR}{D44}{3345}
\refjl{Langacker P 1981}{Phys. Rept.}{72}{185}
\refbk{\dash 1988}{Neutrino Mass and Related Topics}{Editors S. Kato
and T. Oshima (World Scientific) }{p.35}
\refbk{\dash 1992}{Proceedings of Neutrino Telescopes, Venice, Italy}
\refjl{Langacker P and Polonsky N 1992}{\PR}{D47}{4028}
\refjl{Lanou R, Maris H and Seidel G 1987}{\PRL}{58}{2498}
\refjl{Lattimer J and Cooperstein 1988}{\PRL}{61}{23}
\refjl{Lee B W and Shrock 1977}{\PR}{D16}{1444}
\refjl{Lee B W and Weinberg S 1977}{\PRL}{39}{165}
\refjl{Lee D G and Mohapatra R}{\PL}{B329}{463}

\refjl{Lee H and Koh Y S 1990}{\NC }{105B}{883}
\refjl{Lim C S and Marciano W J 1988}{\PR D}{37}{1368}
\refjl{Lopez-Fernandez A \etal 1993}{\PL}{B312}{240}
\refjl{Loredo T J and Lamb D Q 1989}{Ann. N.Y. Acad. Sci.}{571}{601}
\refjl{Lusignoli M, Masiero A and Roncadelli M 1990}{PL}{B252}{247}
\refjl{Majorana E 1937}{\NC}{14}{171}
\refjl{Maier B 1994}{Nucl. Phys. (Proc. Suppl.)}{B35}{358}
\refjl{Malaney R and Mathews G 1993}{\it Phys. Repts.}{229}{145}

\refbk{Mann A and Murtagh M 1993}{Proposal for a Long-baseline
Neutrino Oscillation Experiment at the AGS}{}
\refjl{Masiero A and Valle J W F 1990}{\PL}{B251}{273}
\refjl{Mather J \etal 1994}{(Astrophys. J)} {420}{439}
\refjl{Mayle R, Schramm D N and Turner M S 1993}{\PL}{B317}{119}

\refjl{Mikheyev S P and Smirnov A Yu 1985}{\sl Sov. J. Nucl.
Phys.}{42}{913}
\refjl{Minakata H and Nunokawa H 1990}{\PR}{D41}{2976}

\refjl{Mitsui K, Minorikawa Y and Komori H 1986}{\NC }{9C}{995}
\refbk{Moe M K 1994}{\NP(Proc. Suppl.)}{\bf 35}~{386}
\refjl{Moe M K 1993}{Int. J. of Mod. Phys.}{E2}{507}
\refbk{Mohapatra R 1986}{Unification and Supersymmetry}{(Springer-Verlag)}
\refjl{Mohapatra R N and Nussinov S 1992}{Int.  Mod. Phys.}{A7}{5877}
\refjl{Mohapatra R N, Nussinov S and Zhang X 1994}{\PR}{D49}{3434}

\refbk{Mohapatra R N and Pal P 1991}{Massive Neutrinos in Physics and
Astrophysics}{(World Scientific)}
\refjl{Mohapatra R N and Parida M K 1992}{\PR}{D47}{264}
\refjl{Mohapatra R N and Senjanovic' G 1981}{\PR}{D23}{165}
\refjl{Morales A 1992}{\NP Proc. Suppl}{A28}{181}
\refjl{Morgan J A 1981}{\PL}{102B}{247}
\refjl{Mori M \etal (KAMIOKANDE) 1991}{\PL}{270B}{89}
\refjl{Nieves J 1982}{\PR D}{26}{3152}

\refbk{Nishikawa K 1992}{Preprint}{INS-Rep-924}
\refbk{Nolthenius R, Klypin A and Primack J 1993}{SCIPP-93-46}

\refjl{Nussinov S 1976}{\PL}{63B}{201}
\refjl{Olive K \etal 1990}{\PL}{236B}{454}
\refjl{Olive K \etal 1991}{\PL}{B265}{239}
\refjl{Pal P 1983}{\NP}{B227}{237}

\refjl{Pantaleone J 1990}{\PL}{246B}{245}
\refjl{Parke S J 1986}{\PRL}{57}{1275}
\refbk{Parke S J 1993}{Proc. Moriond Workshop on Neutrino Physics, }{ Villars}
\refjl{Particle Data Group 1986}{\PL}{B170}{1}
\refbk{Peebles P J E 1993}{Principles of Physical Cosmology}{(Princeton,
U. Press, N. Jersey)}

\refjl{Peebles P J E \etal 1991}{Nature}{352}{769}
\refjl{Peltoniemi J T and Valle J W F 1993}{\NP}{B406}{409}
\refjl{Peltoniemi J T, Tommasini D and Valle J W F 1993}{\PL}{B298}{383}

\refjl{Pontecorvo B 1968}{\sl Sov. Phys. JETP}{26}{984}
\refjl{Pulido J 1993}{\PR D}{48}{1492}
\refjl{Qian Y Z \etal 1993}{\PRL}{71}{1965}

\refjl{Raffelt G 1990}{\PRL}{64}{2856}
\refjl{\dash 1990a}{\it Phys. Rep}{198}{1}
\refjl{\dash 1990b}{\it Mod. Phys. Lett.}{A5}{2581}
\refjl{\dash 1990c}{\it Astrophys. J}{365}{559}
\refjl{Raffelt G and Seckel D 1988}{\PRL}{60}{1793}
\refjl{\dash 1991}{\PRL}{67}{2605}
\refjl{\dash 1992}{\PRL}{68}{3116}
\refjl{Raffelt G and Weiss A 1992}{Astron. Astroph.}{264}{536}

\refjl{Reines F \etal 1971}{\PR}{D4}{80}
\refjl{Reiss D 1982}{\PL}{B115}{217}
\refjl{Renzini A 1993}{16th Texas Symposium Eds Akerlof and Srednicki
Ann. N.Y. Acad. Sci.}{688}~{124}
\refjl{Review of Particle Properties 1994}{\PR}{D50}{No. 3}

\refbk{Revol J P 1993}{CERN preprint}{CERN-PPE/93-01}
\refbk{Resvanis L K \etal (NESTOR) 1993}{Preprint}{NE-93-24}
\refbk{Robertson R G H 1994}{Particle and Nuclear Astrophysics and
Cosmology in the Next Millennium}{Workshop, Snowmass, Colorado.}
\refbk{Robertson R G H and Knapp D A 1988}{Ann. Rev. Nucl. Part. Sci.38}{185}
\refjl{Robertson R G H \etal 1991}{\PRL}{67}{957}

\refjl{Rosen S P and Gelb J M 1986}{\PR D}{34}{969}
\refjl{Ross G G and Valle J W F 1985}{\PL}{151B}{375}
\refjl{Rothstein I Z, Babu K S and Seckel D 1993}{\NP}{B403}{725}

\refjl{Roulet E 1991}{\PR D}{44}{R935}
\refjl{Roulet E and Tommasini D 1991}{\PL}{B256}{218}
\refbk{Rubbia C 1993}{Preprint}{CERN-PPE/93-08}
\refjl{Sato and Kobayashi H 1977}{\it Prog.. Theor. Phys.}{58}{1775}

\refjl{Schechter J and Valle J W F 1981}{\PRL}{65}{1883}
\refjl{\dash 1982}{\PR}{D25}{283E}
\refjl{\dash 1982'}{\PR}{D25}{774}
\refjl{Schrock R E and Suzuki M 1982}{\PL}{B110}{250}
\refjl{Sciama D 1990}{Comm. Astrophys.}{15}{71}
\refbk{Scott  D \etal 1994}{Cf PA-TH-94-30 to
appear in}{Astrophysical Quantities, Ed. Cox A)}
\refjl{Seckel D, Steigman G and Walker T P 1991}{\NP}{B366}{233}

\refjl{Shi X \etal 1993}{\PR}{D48}{2568}
\refbk{Sigl G and Turner M S 1994}{Preprint FERMILAB-Pub-94/001-A}
\refjl{Smith M, Kawano L and Malaney R 1993}{\it Astrophys. J Suppl}
{85}{219}
\refjl{Smooth \etal 1992}{\it Astrophys. J}{396}{L1}
\refjl{Soares J M and Wolfenstein L 1989}{\PR}{D40}{3666}
\refjl{Steigman G and Turner M S 1985}{\NP}{B253}{375}

\refjl{Stockdale I E \etal (CCFR) 1985}{\ZP C}{27}{53}
\refbk{Suzuki Y 1993}{Talk at the TAUP93 Conference,}{ Gran Sasso,
Sept. 1993}
\refjl{Tomoda T 1991}{Rep. Prog. Phys.}{54}{53}
\refbk{Totsuka Y 1990}{\sl Proc. Int. Conf. Underground Physics
Experiments}{(Tokyo: Institute for Cosmic Ray Research)}{p 129}
\refjl{Turck-Chi\`eze S and Lopes I 1993}{\sl Astrophys. J.}{408}{347}
\refjl{Turkevich A L \etal 1991}{\PRL}{67}{3211}
\refjl{Turner M S 1985}{\PR}{D31}{681}

\refjl{Ushida N \etal (FNAL E531) 1986}{\PRL}{57}{2897}
\refjl{Valle J W F 1987}{\PL}{199B}{432}
\refjl{Vergados J D 1986}{Phys. Rep.}{133}{1}
\refjl{Vidyakin G S \etal 1990}{\sl Sov. Phys. JETP}{71}{424}
\refjl{Volkova L V 1980}{\sl Sov. J. Nucl. Phys.}{31}{1510}
\refjl{Voloshin M B, Vysotsky M I and Okun L B 1986}{\sl Sov.
J. Nucl. Phys.}{44}{440}
\refjl{Voloshin M B 1988}{\sl Sov. J. Nucl. Phys.}{48}{512}
\refjl{Vuilleumier J L \etal 1993}{\PR}{D48}{1009}
\refjl{Vysotsky M, Dolgov A and Zeldovich Ya 1977}{\it JETP
Lett}{26}{188}
\refjl{Walker T P \etal 1991}{Astrophys. J}{376}{51}
\refjl{Weinheimer Ch. \etal}{\PL}{B300}{210}
\refbk{Weinstein A and Stroynowski R 1993}{Annu. Rev. Nucl. Part.
Sci}{\bf 43}~{457}
\refbk{White M, Gelmini G and Silk J 1994}{UCLA/94/TEP 37}
\refjl{Wilczeck F 1982}{\PRL}{49}{1549}
\refbk{Wilkerson J F 1993}{``Neutrino 92''}{\NP B}{(Proc.
Suppl.){\bf 31}~}{32}

\refbk{Whitehouse D A \etal (LSND) 1991}{\sl Proc. of Particles and
  Fields '91}{Vancouver, Ed. Axen D \etal, World Scientific,
  Singapore, V 2, p 949}
\refjl{Wilczek F 1982}{\PRL}{49}{1549}

\refjl{Winter R G 1981}{\NC ~\sl Lett.}{30}{101}
\refjl{Witten E 1980}{\PL}{91B}{81}

\refjl{Wolfenstein L 1978}{\PR D}{17}{2369}
\refjl{\dash 1981}{\NP}{B186}{147}
\refjl{Wright E \etal 1994}{\it Astrophys. J}{420}{1}
\refjl{Wu X R \etal 1991}{\PL}{B272}{435}
\refbk{Yanagida T 1979}{Proc. of the Workshop on Unified Theories and
Baryon Number in the Universe, KEK, Japan}

\refjl{Zaceck G \etal 1986}{\PR D}{34}{2621}
\refjl{Zee A 1980}{\PL}{93B}{389}
\refjl{Zeldovich Y 1952}{Doklady Akad. Nauk SSSR}{86}{505}

\figures
\figcaption{Bounds on unstable neutrinos that apply on any decay mode.
The gray area is rejected. The
continuous contour shows the bound $\Omega_{DP} \leq \Omega_o$. The dashed
contour shows the structure formation bound only for experimentally allowed
masses. The dot-dashed contour shows the nucleosynthesis bound on the
allowed number of effective extra neutrino species $\delta N_\nu$
(see section 5).}
\figcaption{Bounds on unstable neutrinos whose main decay products include
photons or e$^+$e$^-$ pairs. The gray area, excluded by cosmological limits
(distortions of the CBR and other photon backgrounds), is reproduced from
the figure 5.6 of Kolb and Turner 1990. The hatched area is excluded by
the non observation of photons in coincidence with the observed neutrino
flux from the supernova SN1987A (see section 9).}
\figcaption{Present constraints on $\Delta m^2-\sin^22\theta$ from
reactor and accelerator searches of $\nu_e-\nu_\mu$ oscillations (a)
and for $\nu_\mu-\nu_\tau$ (b). For references see the text.}
\figcaption{Sensitivity of future oscillation searches together with the
present bounds and the regions suggested by atmospheric neutrinos and
to account for the dark matter. Figure $4a$ is for $\nu_e-\nu_\mu$
oscillations while $4b$ is for $\nu_\mu-\nu_\tau$. }
\figcaption{Standard solar neutrino spectrum (Bahcall and Ulrich 1988)
and experimental thresholds for the existing detectors.}
\figcaption{Survival neutrino probabilities vs. $E/\Delta m^2$ for
$\sin 2\theta=0.1$.}
\figcaption{MSW contours for the different experiments and preferred
values of $\Delta m^2-\sin^22\theta$ (shadowed). Also shown is the
region excluded by the non-observation of a day-night asymmetry by
Kamiokande.}
\figcaption{Spectrum of e-type and $\mu$-type atmospheric neutrinos
measured by Kamiokande. The histogram is the MC prediction using the
fluxes of Lee and Koh.}
\figcaption{Values of $\Delta m^2-\sin^22\theta$ suggested by
Kamiokande measurement of $R(\mu/e)$ atmospheric ratio
and its zenith angle dependence,
together with the bounds from accelerators and reactors for
$\nu_\mu-\nu_e$ (a) and $\nu_\mu-\nu_\tau$ (b) oscillations.
Also shown are the regions excluded by other searches with atmospheric
neutrinos: Fr\'ejus value of $R$, IMB-1 up/down $\nu_\mu$, and IMB
stopping/through-going muons.}

\vfill

\end

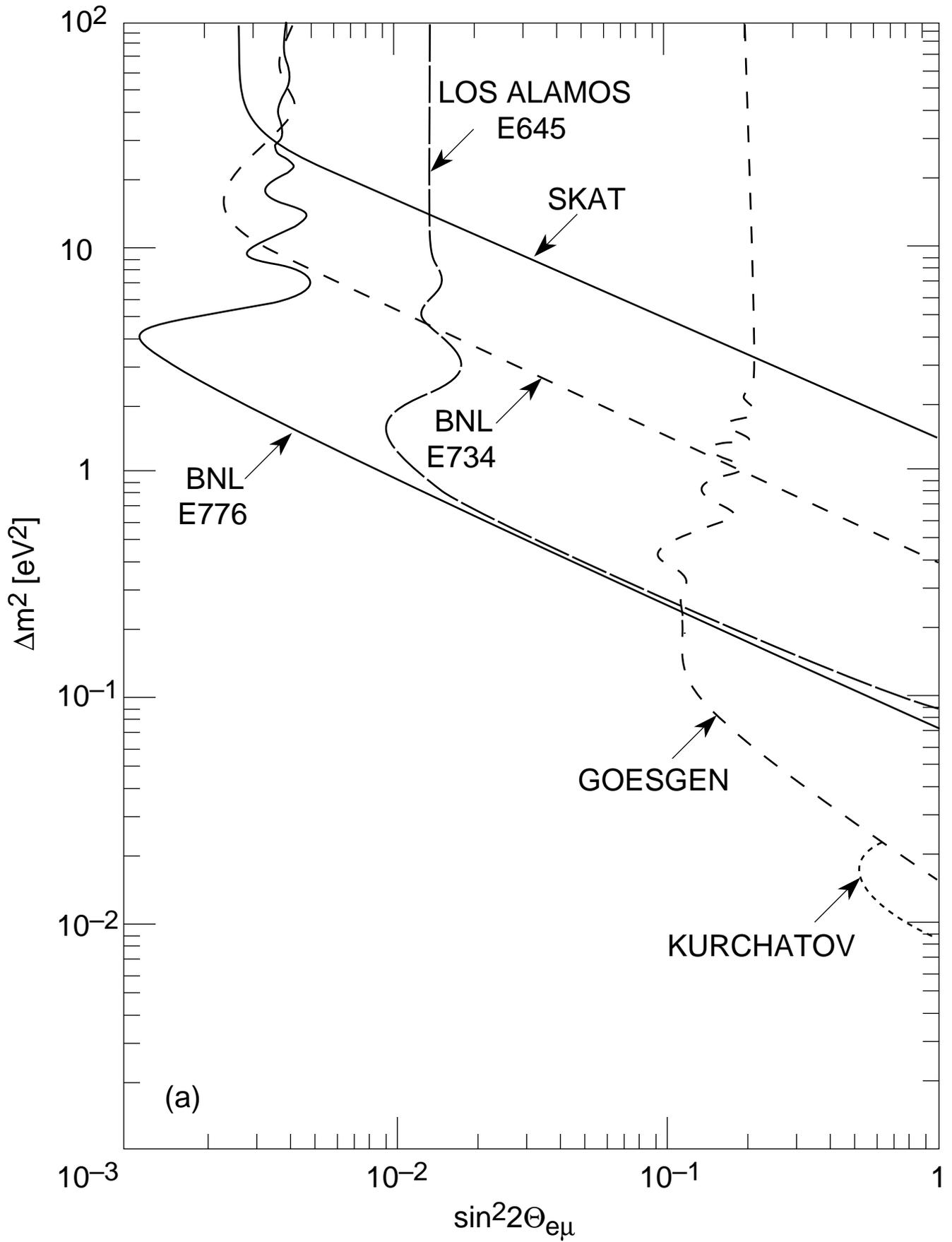
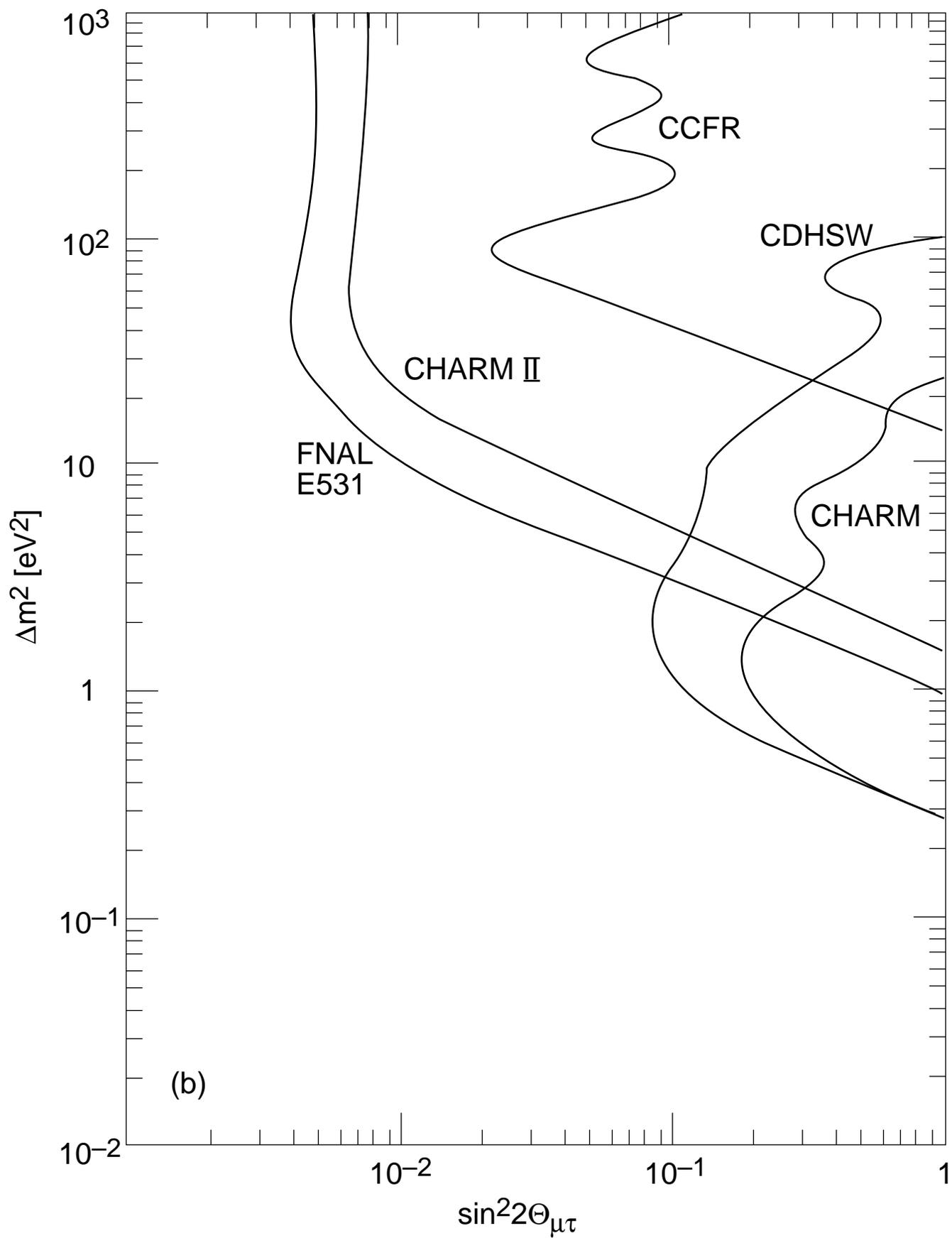
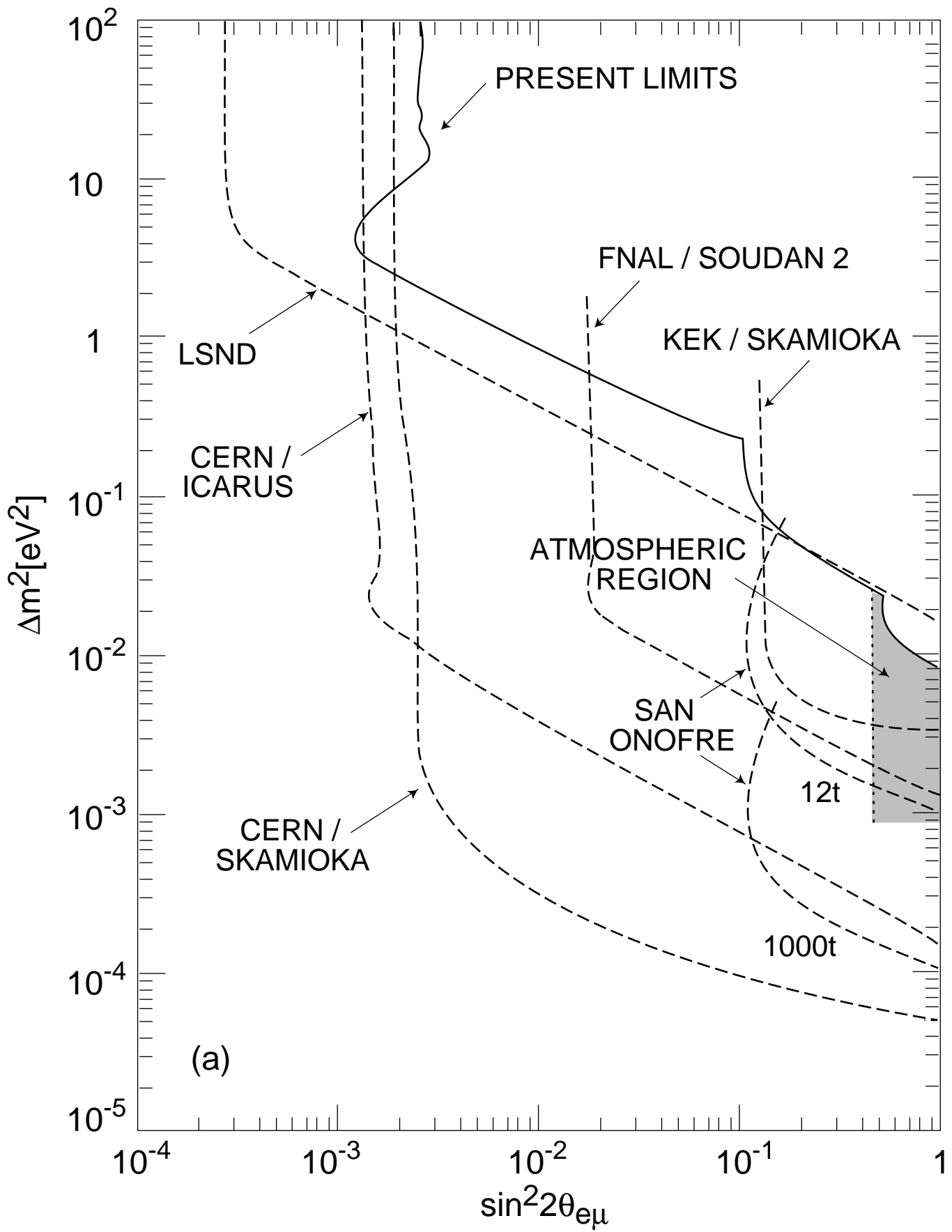
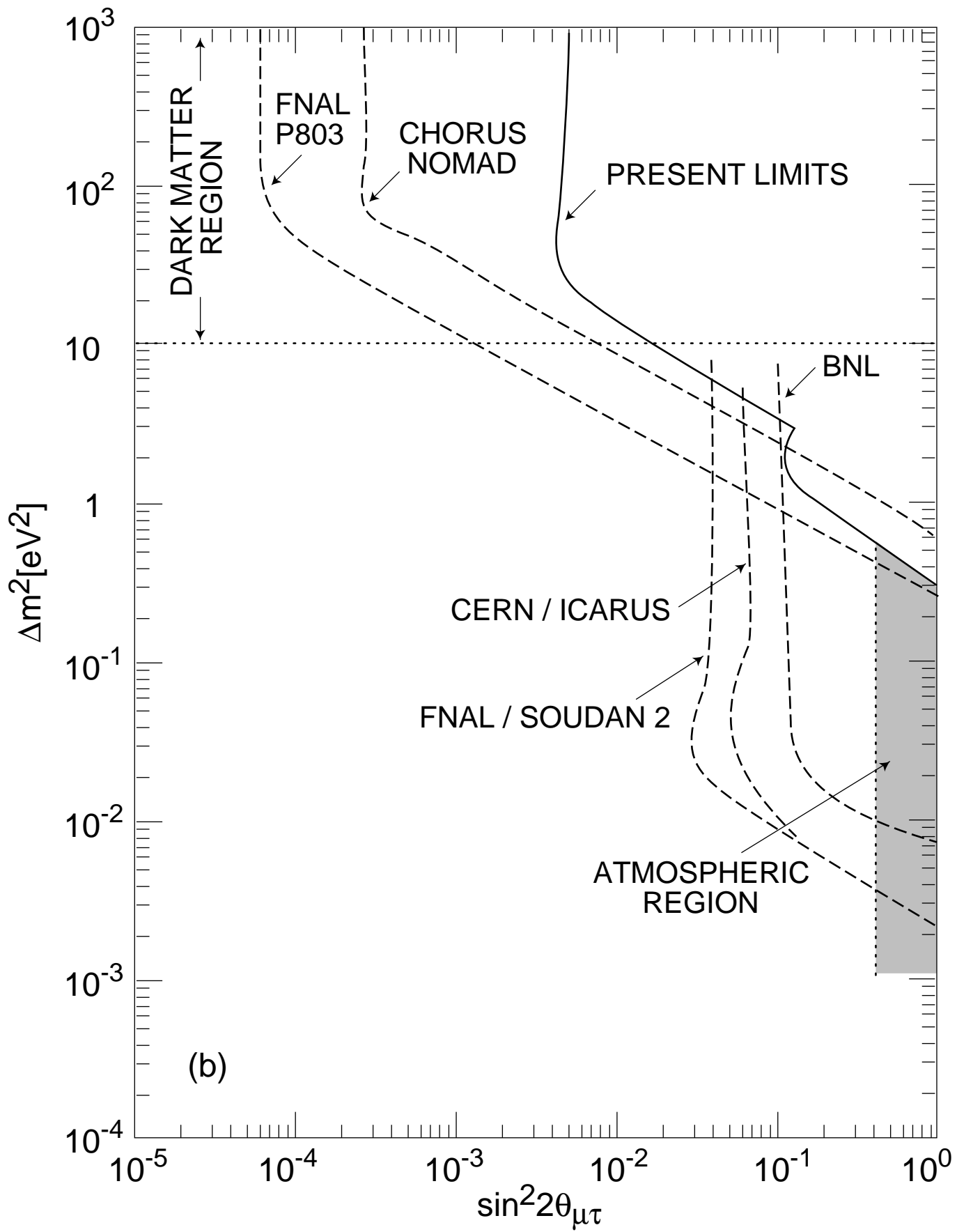
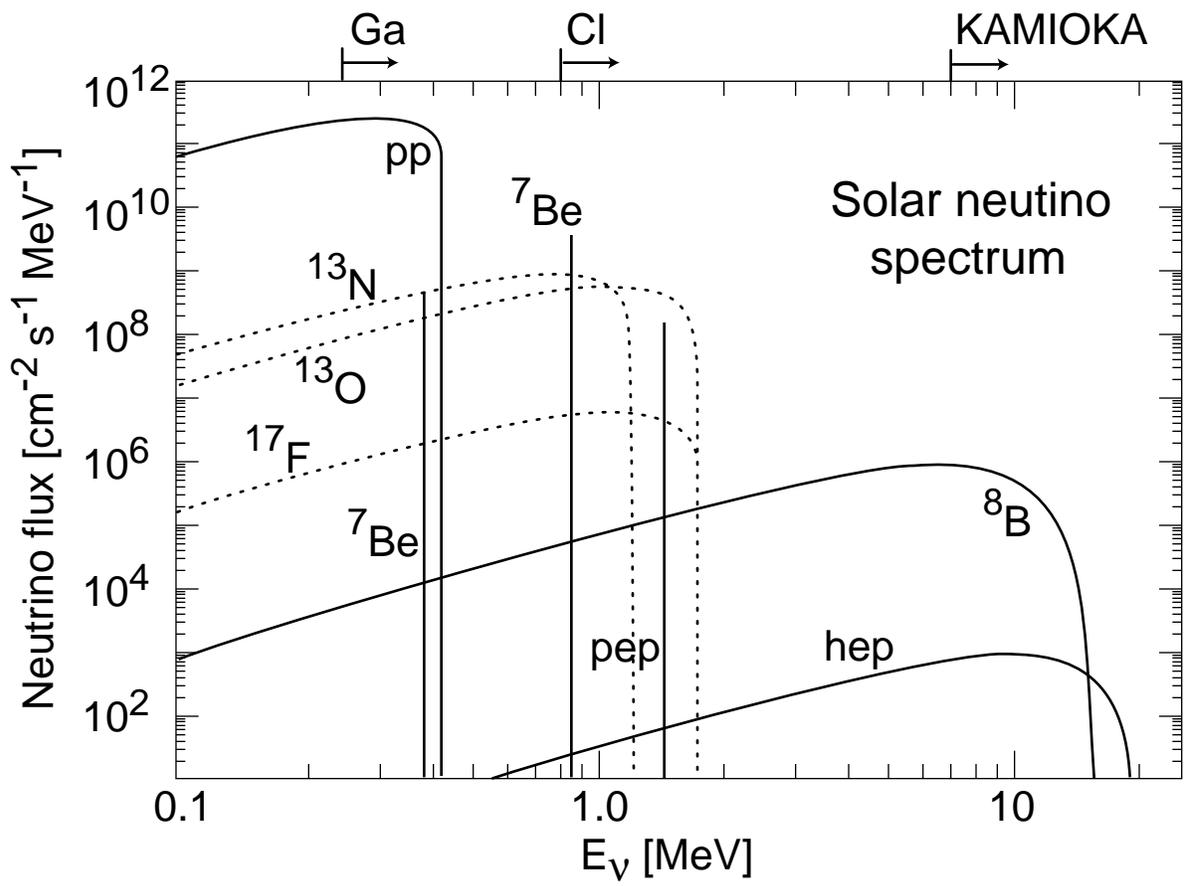
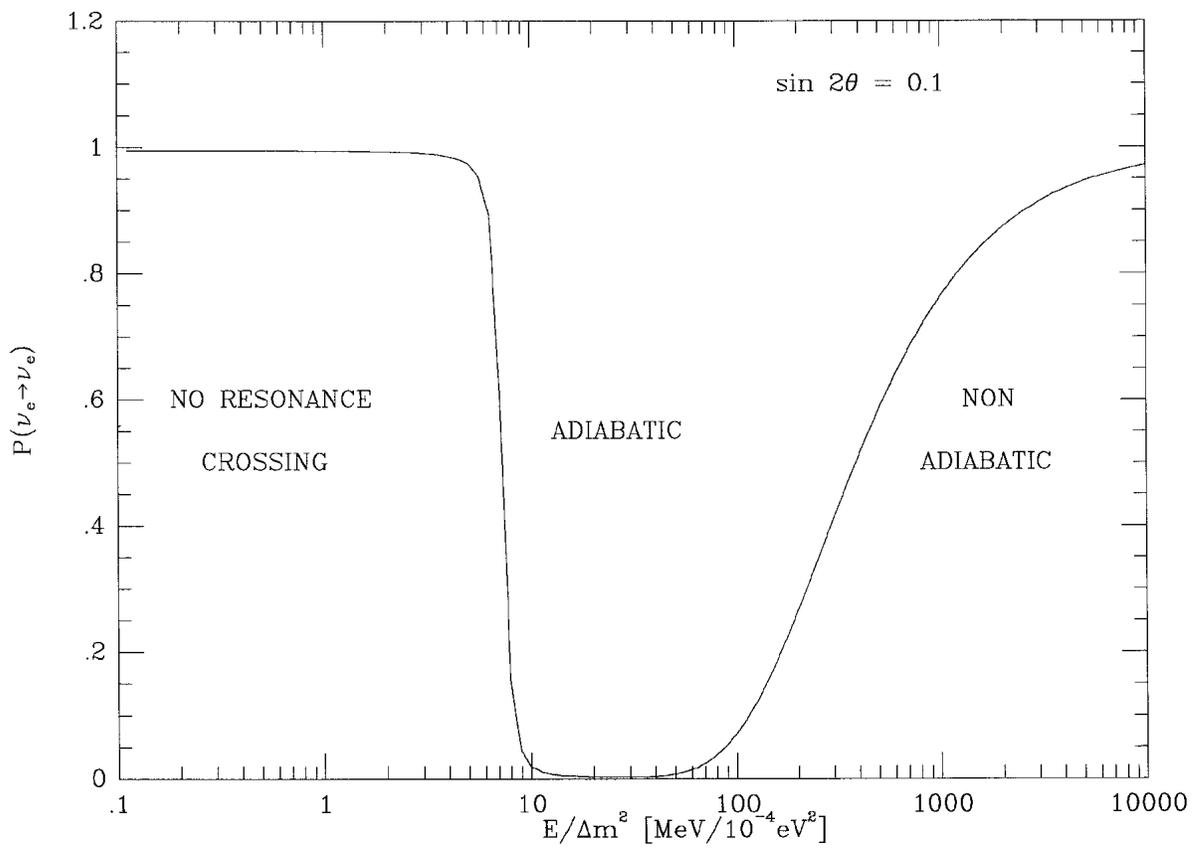
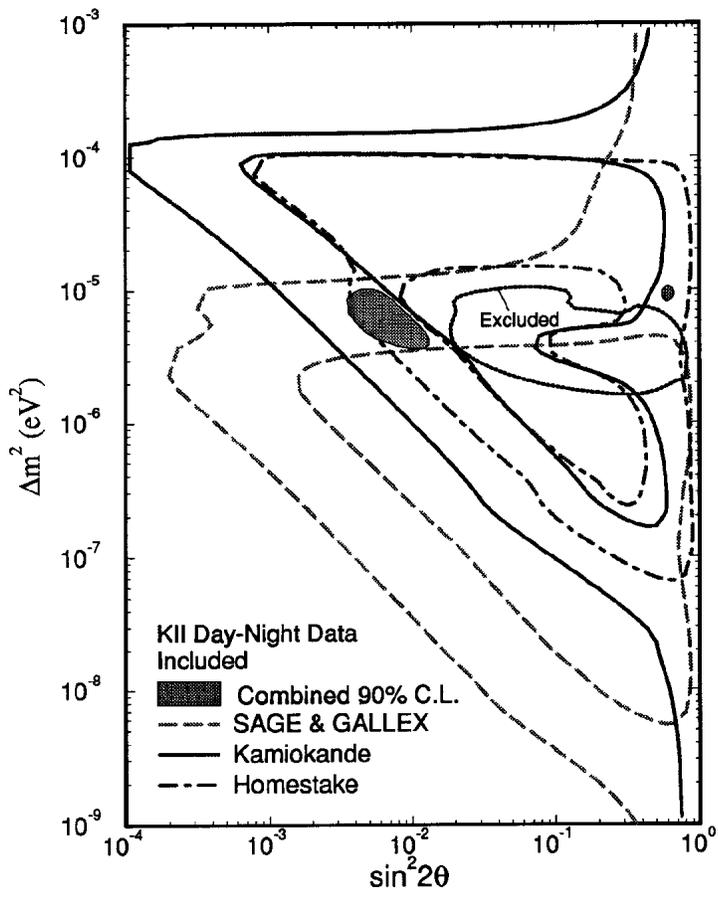
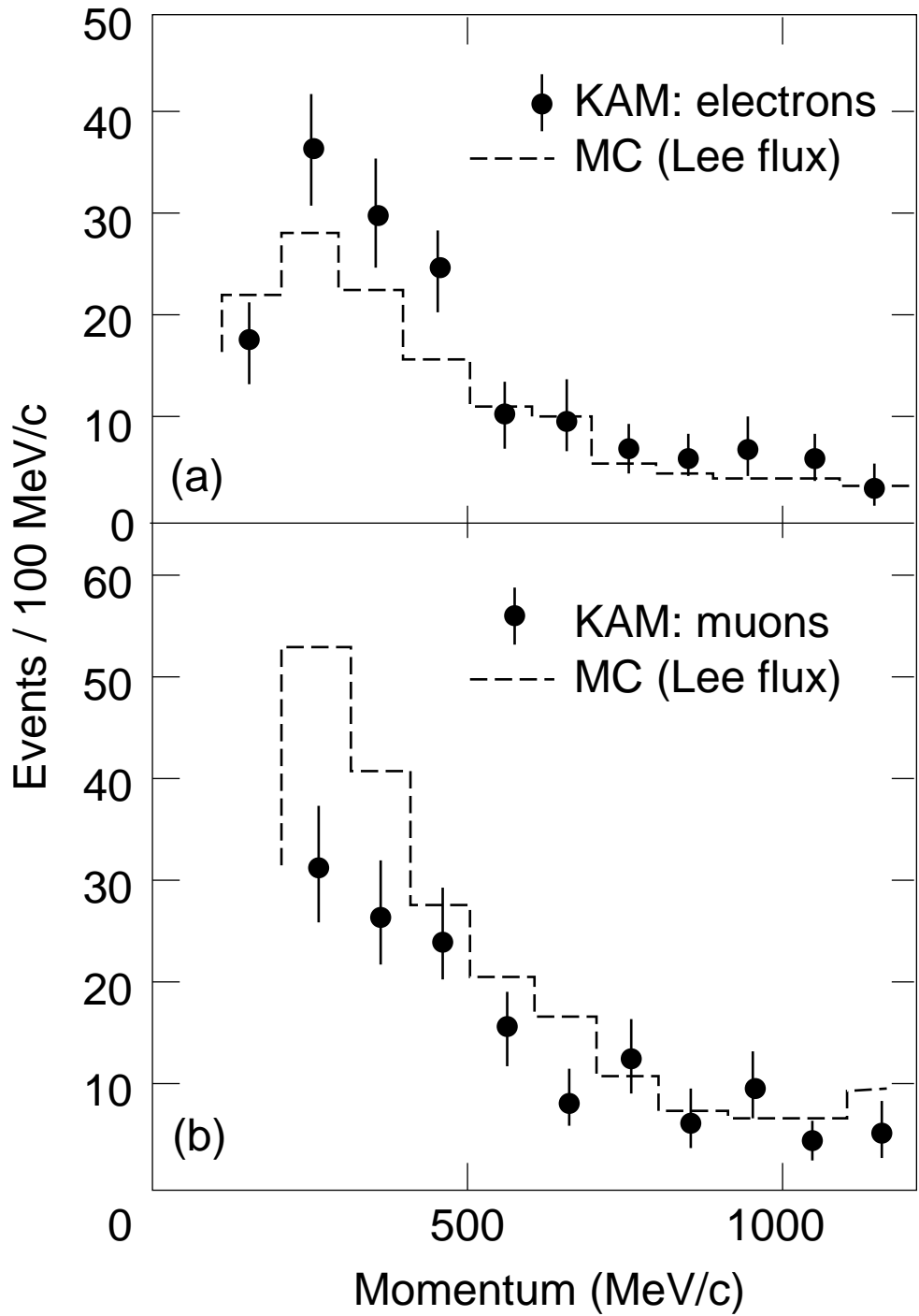
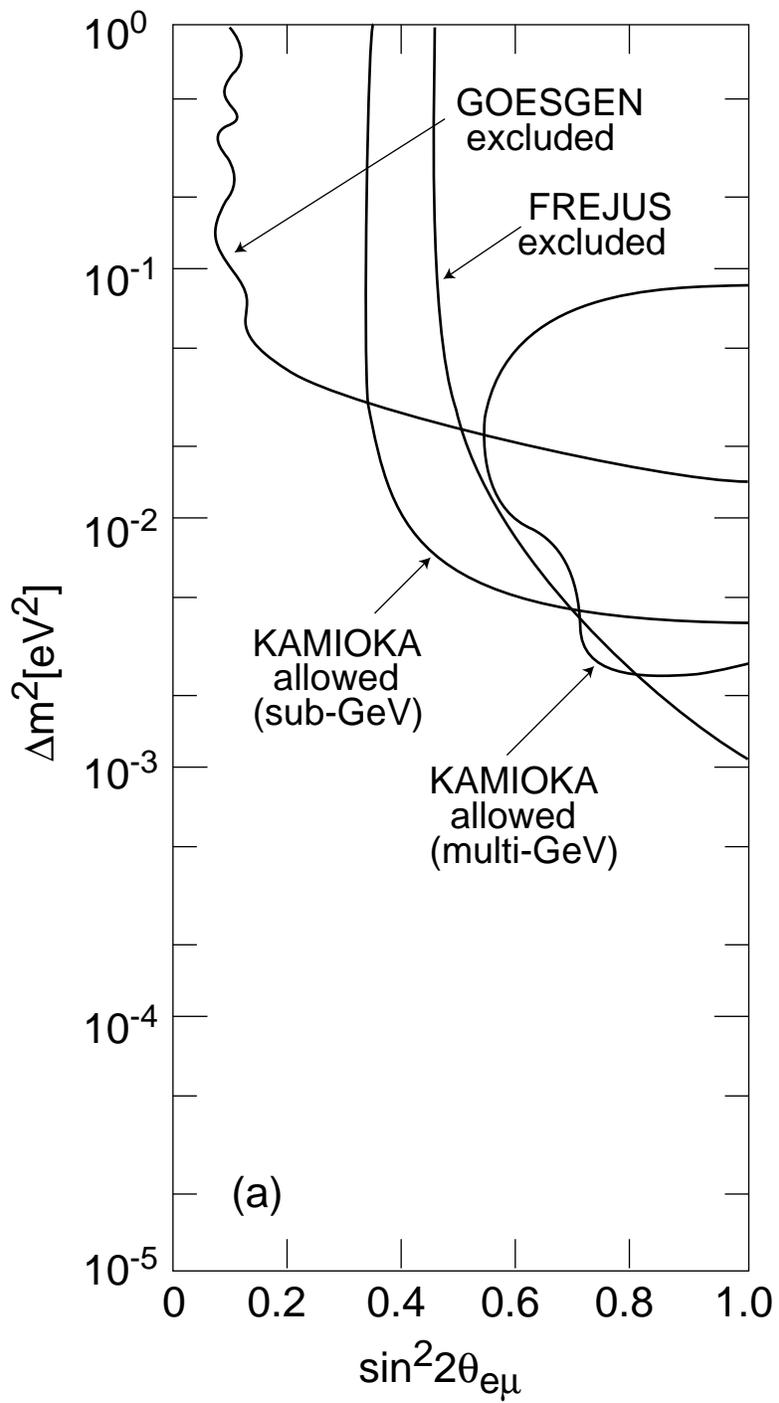
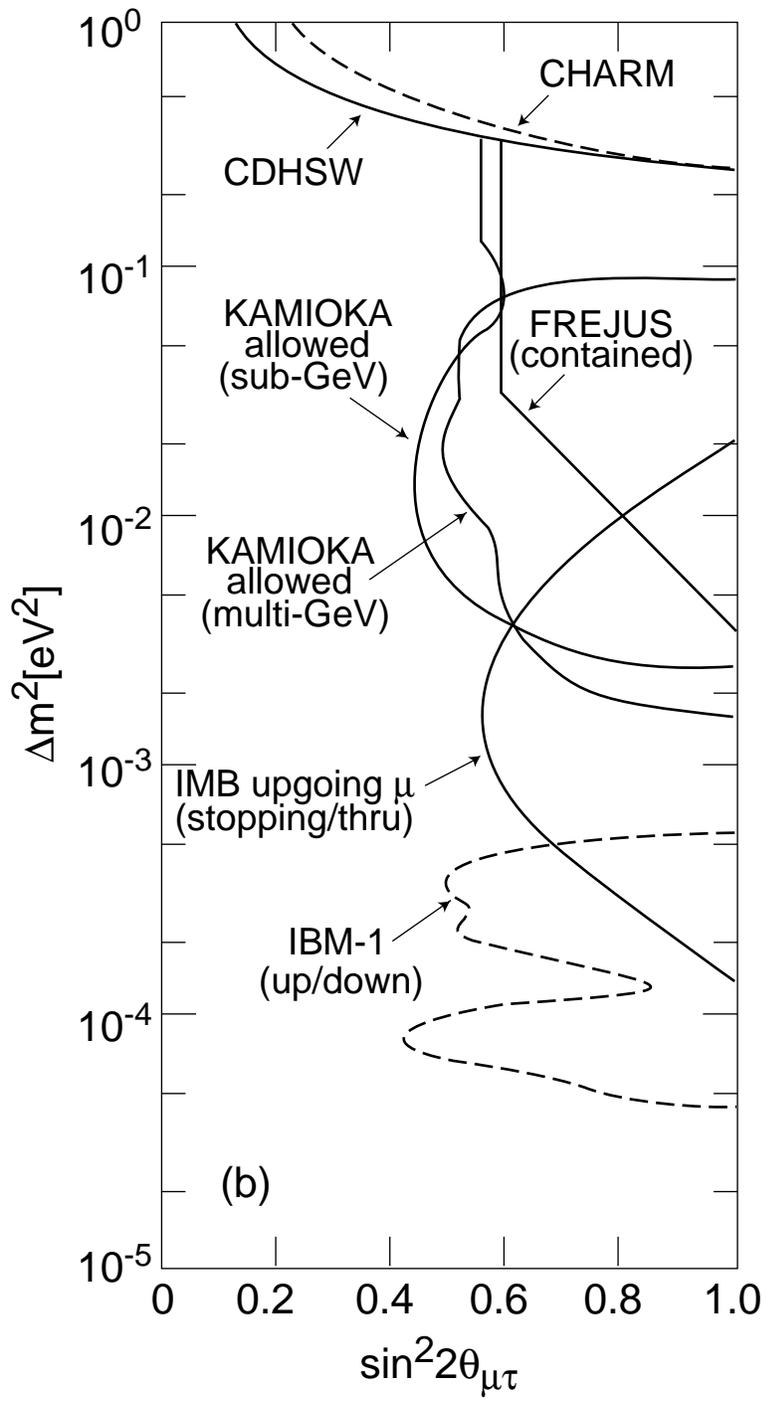
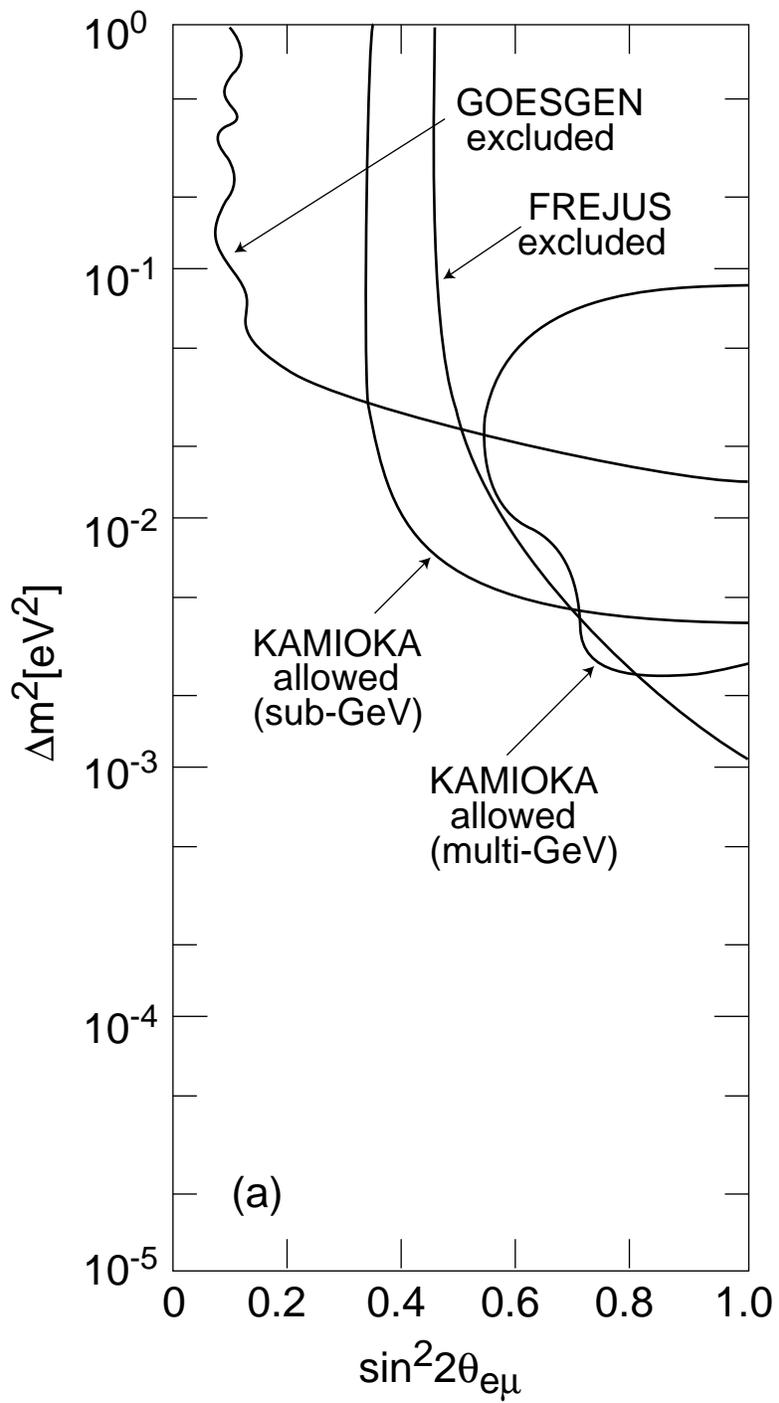
\bye